\documentclass[%
 reprint,
superscriptaddress,
 amsmath,amssymb,
 aps,
]{revtex4-2}

\usepackage{graphicx}
\usepackage{dcolumn}
\usepackage{bm}
\usepackage[table]{xcolor}
\usepackage{color, colortbl}
\usepackage[utf8]{inputenc}
\usepackage{gensymb}
\usepackage{xcolor}
\usepackage{ulem}
\usepackage{floatrow}
\usepackage{makecell}
\usepackage{tabularx,colortbl}
\usepackage{blindtext}
\usepackage{booktabs}

\begin{document}

\preprint{APS/123-QED}

\title{Electrical manipulation of magnetic anisotropy in a Fe$_{81}$Ga$_{19}$/PMN-PZT magnetoelectric multiferroic composite}

\author{W. Jahjah}
    \affiliation{Univ. Brest, Laboratoire d'Optique et de Magnétisme (OPTIMAG), EA 938, 29200 Brest, France}
    \affiliation{Cr Research Group, Department of Physics, University of Johannesburg, PO Box 524, Auckland Park 2006, South Africa}
\author{J.-Ph. Jay}
\email{jay@univ-brest.fr}
\author{Y. Le Grand}
\author{A. Fessant}
    \affiliation{Univ. Brest, Laboratoire d'Optique et de Magnétisme (OPTIMAG), EA 938, 29200 Brest, France}
\author{A.R.E. Prinsloo}
\author{C.J. Sheppard}
    \affiliation{Cr Research Group, Department of Physics, University of Johannesburg, PO Box 524, Auckland Park 2006, South Africa}
\author{D.T. Dekadjevi}
    \affiliation{Univ. Brest, Laboratoire d'Optique et de Magnétisme (OPTIMAG), EA 938, 29200 Brest, France}
    \affiliation{Cr Research Group, Department of Physics, University of Johannesburg, PO Box 524, Auckland Park 2006, South Africa}
\author{D. Spenato}
  \affiliation{Univ. Brest, Laboratoire d'Optique et de Magnétisme (OPTIMAG), EA 938, 29200 Brest, France}

\date{\today}

\begin{abstract}
Magnetoelectric composites are an important class of multiferroic materials that pave the way towards a new generation of multifunctional devices directly integrable in data storage technology and spintronics. This study focuses on strain-mediated electrical manipulation of magnetization in an extrinsic multiferroic. The composite includes 5 nm or 60 nm Fe$_{81}$Ga$_{19}$ thin films coupled to a piezoelectric (011)-PMN-PZT. The magnetization reversal study reveals a converse magnetoelectric coefficient $\alpha\textsubscript{CME,max}\approx2.7\times{10^{-6}}$ s.m$^{-1}$ at room temperature. This reported value of $\alpha\textsubscript{CME}$ is among the highest so far compared to previous reports of single-phase multiferroics as well as composites. An angular dependency of $\alpha\textsubscript{CME}$ is also shown for the first time, arising from the intrinsic magnetic anisotropy of FeGa. The highly efficient magnetoelectric composite FeGa/PMN-PZT demonstrates drastic modifications of the in-plane magnetic anisotropy, with an almost 90$\degree$ rotation of the preferential anisotropy axis in the thinner films under an electric field $E=10.8$ kV.cm$^{-1}$. Also, the influence of thermal strain on the bilayer's magnetic coercivity is compared to that of a reference bilayer FeGa/Glass at cryogenic temperatures. A different evolution is observed as a function of temperature, revealing a substrate thermo-mechanical influence which has not yet been reported in FeGa thin films coupled to a piezoelectric material.

\end{abstract}

\maketitle


\section{\label{sec:level1}Introduction }

The research on the magnetoelectric (ME) coupling saw a renaissance \cite{Spaldin05} considering multiferroic (MF) materials in 2005 \cite{Fusil2014}. MF materials \cite{eerenstein06} simultaneously show at least two ferroic orderings, such as ferroelectric (FE) and (anti-)ferromagnetic as well as ferroelastic. There have recently been a frenzy over MF materials in condensed matter physics due to their novel potential applications, including multifunctional devices such as spintronics, ME transducers, actuators, sensors, and multiple-state memories \cite{Matsukura15, Hu2011, Chiba2008, Heron2011}. Among multiferroic orderings, the coexistence of ferroelectricity and ferromagnetism is highly desired, as it allows electric-field control of magnetism without the need for magnetic fields, known as the converse ME effect (CME); the direct ME (DME) effect being the magnetic-field control of electric polarization \cite{Eerenstein2007}. Multiferroics may thus pave the way to faster, smaller, more energy-efficient data-storage technologies. The CME effect has been observed as an intrinsic effect in some oxide single-phase MF materials \cite{Zhao2006, hur2004} such as BiFeO$_3$ (BFO), one of the few rare multiferroics at room temperature, thus the most intensively investigated single-phase multiferroics \cite{Catalan09, tony12, Richy2018, Jahjah2018}.

However, most of the single-phase MF materials possess either low
permittivity or low permeability at room temperature and thus exhibit weak ME coupling which hinders their applications \cite{Palneedi2016}. In addition, these materials are often complicated to fabricate, potentially leading to electrical leakage problems due to structural defects and impurities \cite{Wang2010, Jahjah2018}. ME composites, on the other hand, also known as extrinsic multiferroics, consist of stacked magnetostrictive and piezoelectric phases and offer a wide range of materials, as well as a flexibility in fabrication \cite{Srinivasan2010, Nan2008}. These composites are a powerful tool to achieve, at room temperature, giant ME coupling response compared to those found in single-phase materials. Much efforts have been made to electrically control the magnetization via three main mechanisms: charge carrier \cite{vaz2012, Nan2014, zhou2015}, spin exchange (exchange bias coupling) \cite{Laukhin2006, yang2015_EBcontrolV, liu2011_EB_FeGa}, and strain-mediated coupling \cite{Thiele2007, Nan2014, Yang2009, alberca2015, Staruch2016, Wu2011, Zhang2014, Yang2015, Biswas2017}. Particularly, the strain-mediated mechanism has been found very appealing for further exploration \cite{cheng2018, Wang2010}. In strain-mediated composites, the ME coupling occurs when, in the case of CME, an applied electric field induces strain in the piezoelectric phase through the direct piezoelectric effect. This strain is transferred to the magnetostrictive phase and in turn induces inverse magnetostriction (Villari effect  \cite{Lacheisserie1993}), which translates to a change in magnetic properties.

In providing the strain, relaxor-PbTiO$_3$ (relaxor-PT) based ferroelectrics are widely used for their excellent piezoelectric properties. Rhombohedral Pb(Mg$_{1/3}$Nb$_{2/3}$)O$_3$-Pb(Zr,Ti)O$_3$ (PMN-PZT) single crystals are an example of ternary-system ferroelectrics that provide, near their morphotropic phase boundary (MPB), many advantages when compared to their binary counterparts PMN-PT (Pb(Mg$_{1/3}$Nb$_{2/3}$)O$_3$-Pb(Ti)O$_3$) and PZT-PT or PZT ceramics \cite{Zhang2007,Zhang2010}. They offer comparable piezoelectric coefficients ($d\textsubscript{33}=1000-2000$ pC.N$^{-1}$) and electromechanical coupling factors ($k\textsubscript{33}\geq$0.9) compared to binary crystals, while possessing double the coercive field values on the order of $E\textsubscript{c}=5$ kV.cm$^{-1}$, higher Curie temperature ($T\textsubscript{c}=130-170\degree$ C) and higher FE transition temperature ($T\textsubscript{RT}=90-160\degree$ C), significantly expanding the temperature range of usage for high-power applications \cite{Richter2008,Zhang2010}. The piezoelectric coefficient notation $d_{ij}$ refers to `\textit{j}' axis as the working deformation direction under an applied $E$ field along the polarization `\textit{i}' axis, in compliance with the IEEE standards for relaxor-based FE single crystals \cite{IEEEStandardRelaxorPiezo}.

Furthermore, it is possible to increase the strain amplitude by cutting and poling the crystal along particular crystallographic directions \cite{Palneedi2018, wang2009,Hwang2018,luo2006}. When the relaxor-PT crystals operate in the $\langle110\rangle$-poled longitudinal-transverse (L-T) mode (32 mode, vibration along $\langle001\rangle$), they possess very high $d_{32}$ and $k_{32}$ values. Meanwhile, as the crystals are poled and driven through the thickness rather than the length, the required electric field to drive them is much lower than that for the L-L 33 mode \cite{LuoZhang2014Piezo, IEEEStandardRelaxorPiezo}.  Consequently, a large in-plane anisotropic piezostrain could be achieved, with ${d_{32} = -1850}$ pC.N$^{-1}$ and ${d_{31} = 599}$ pC.N$^{-1}$ in the (011)-PMN-PZT case \cite{Patil2013,Kambale2013,Ryu2015, Palneedi2017,Bilgen2011}. This is a crucial driving mechanism to the observation of large anisotropic ME properties in combination with a magnetostrictive film. 

Little focus has been given to such ternary relaxor ferroelectrics in electrically controlling the magnetism. In particular, (011)-PMN-PZT single crystals have only recently been used in controlling magnetization through CME coupling in ME composites \cite{Lian2018-2}, whereas they have shown a wide range of fresh interesting results in power generation and energy harvesting based on the DME effect \cite{Park2010, Bilgen2011, Kambale2013, Patil2013,Ryu2015, Annapureddy2018,Palneedi2018, Hwang2018,Chu2018}. Some of these works \cite{Annapureddy2018,Palneedi2018} also involved galfenol or FeGa as the magnetostrictive phase considering its promising magnetic properties.

Regarding the magnetostrictive phase, FeGa thin films combine remarkable properties such as low
hysteresis, large magnetostriction, good tensile strength, machinability and recent progress in commercially viable methods of processing \cite{Clark2000,Atulasimha2011,jay2019}. Although FeGa magnetostrictive properties are comparatively lower than those of Terfenol-D (a terbium-iron-dysprosium alloy), gallium, when substituted for iron increases the tetragonal magnetostriction coefficient $\lambda_{100}$ over tenfold \cite{Clark2000}. Another advantage of FeGa alloys is the rare-earth free composition; the cost is thus reduced compared to the rare-earth alloys family that also has another drawback which is brittleness. 

FeGa has been the choice of magnetostrictive material for many studies, suggesting versatile proposals for the development of multifunctional devices exploiting both DME \cite{finkel2015,wang2011,dong2005} and CME effects \cite{liu2011_EB_FeGa, parkes2012, vaz2012, Xie2014, liu2014_review, Zhao2006, Ahmad2015, Phuoc2017, Zhang2018, Hu2015,lou2009}. Among these studies some have reported the dynamic self-biased effect, also called remanent CME, which is a desirable property recently seeked to control the magnetization using an electric field in ME devices without the need for the assistance of an external biasing magnetic field \cite{Zhang2014-selfbiasCME,ChulYang2011-selfbiasCME,Fitchorov2011a,Yang2015,Mandal2011,Lian2018-2,zhou2016}. Self-biased CME is indeed important for lower-energy consumption and more compact ME devices.

Nevertheless, the CME coupling has not yet been investigated in a ME composite that brings together two highly performant components in the ME extrinsic multiferroics research such as the (011)-PMN-PZT ferroelectric single crystals and the magnetostrictive polycrystalline Fe$_{81}$Ga$_{19}$ thin films.

Furthermore, the field of multiferroics covers aspects ranging from technological applications to fundamental research problems. The study of multiferroics increasingly influences neighbouring research areas, such as complex magnetism and ferroelectricity, oxide heterostructures and interfaces, etc \cite{fiebig16}. This lead us to shed light over the bonding relationship of both phases in the extrinsic MF, and show how a different type of strain such as thermal strain at low temperatures can act on a magnetostrictive material depending on the substrate's nature. This indeed has never been done before on the FeGa/PMN-PZT system. \\

In this contribution, we report on the magnetoelectric coupling in a Fe$_{81}$Ga$_{19}$/(011)-PMN-PZT composite through a systematic experimental study. For FeGa sample thicknesses $t\textsubscript{FM}=5$ and 60 nm, magnetization reversal loops $M(\mu_0H)$ ($\mu_0$ being the vaccum permeability) are first presented along the [100] and [0$\overline{1}$1] directions and for only two electric field $E$ values in section A. These measurements are then extended to bipolar $E$-field cycles to show the evolution of magnetic properties in section B. For more insight on anisotropy properties, we also present  in section C the azimuthal behaviors of FeGa through angular measurements of $M(\mu_0H)$ under $E=0$ kV.cm$^{-1}$ and $E$ $\textgreater$ 0 kV.cm$^{-1}$. To quantify the relative magnetization change upon applying an electric field on our ME composite, we report in section D a CME coupling coefficient $\alpha\textsubscript{CME}=\mu_0\Delta{M}/\Delta{E}$ (in s.m$^{-1}$) among the highest reported so far, and compare it to a comprehensive literature recap of reported $\alpha$\textsubscript{CME} values. We show as well an angular dependency of the $\alpha\textsubscript{CME}$ that is strongly related to the anisotropy properties of FeGa. In the final section E, low-temperature measurements ranging from 10 K to 300 K were carried out, in order to explore the thermo-magneto-mechanical effects by comparing the FeGa temperature-dependent magnetic coercivity's behavior on two different substrates: amorphous glass and the single-crystalline ferroelectric PMN-PZT.

\section{Experimental procedures}

\begin{figure*}
\includegraphics[width=0.9\textwidth]{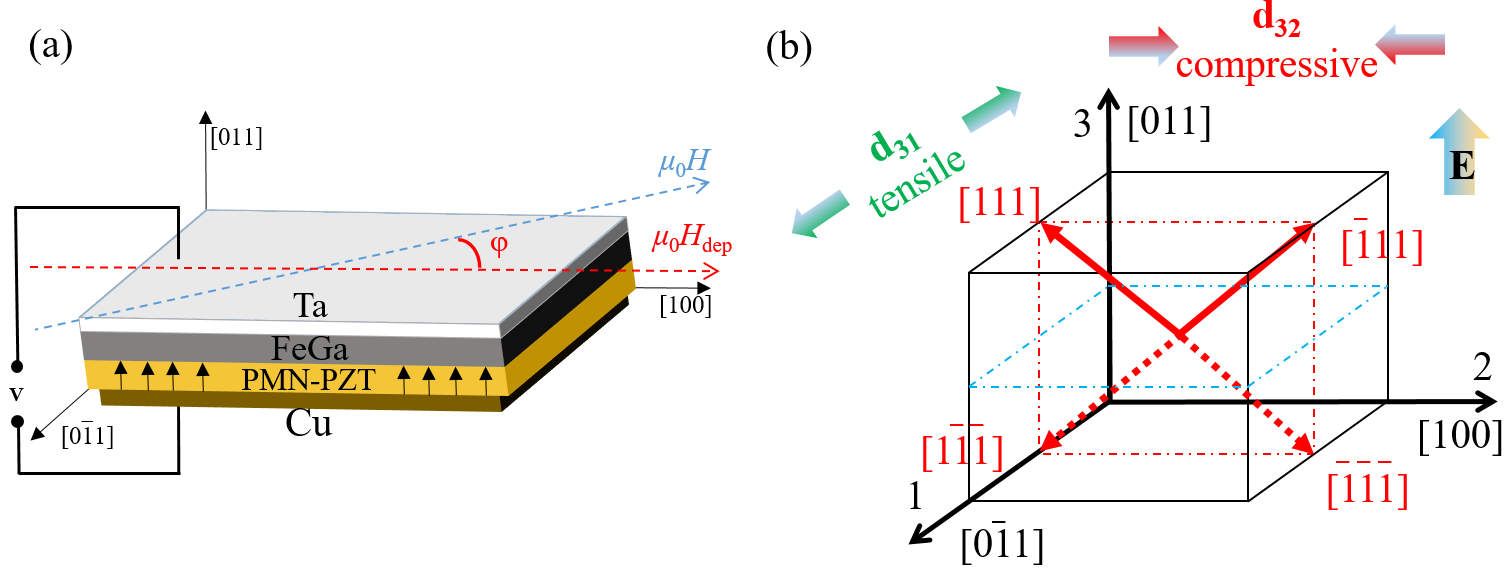}
\caption{\label{fig:fig1}(a) Schematic drawing of the ME composite consisting of the magnetostrictive FeGa and the (011)-oriented PMN-PZT beam. The configuration of measurement is also indicated using $\varphi$ angle between the applied magnetic field $\mu_0H$ (dashed blue axis) and the deposition field $\mu_0H$\textsubscript{dep} (dashed red axis along the [100] length of the beam). Black arrows inside the PMN-PZT layer represent the positive electric field direction. Drawing is not to scale.
(b) Sketch of the polarization vectors of the rhombohedral PMN-PZT unit cell in the (011)-oriented case. Also presented are the $d_{31}$ and $d_{32}$ modes of strain, the electric field $E$ poling direction along [011] with its plane (dashed blue square).}
\end{figure*}

Samples consisting of bi-layered ME composites are prepared by depositing the magnetostrictive FeGa thin films onto the piezoelectric (011)-PMN-PZT, using radio frequency (RF) magnetron sputtering. SSCG-grown (Solid State Crystal Growth) PMN-PZT rhombohedral single crystals are commmercially available  as \textit{CPSC160-95} from Ceracomp Co. Ltd., Korea \cite{ceracomp} The PMN-PZT slabs are (011)-oriented and poled along the thickness, thus, along [011], creating an in-plane anisotropic strain behavior which allows the L-T working mode. This means that the (011) plane undergoes an anisotropic deformation while applying an electric field parallel to the [011] poling direction. Figure \ref{fig:fig1} presents the (011)-PMN-PZT unit cell along with the corresponding working mode.

As presented in Fig.\ref{fig:fig1}b, poling the crystal along the non-polar direction [011] creates a macro-symmetric multi-domain structure \cite{Palneedi2018,Shanthi2008} Such an engineered domain state is more stable than the single-domain state and offers an almost hysteresis-free strain-$E$ behavior, because the two dipole orientations [$\overline{1}$11] and [111] are energetically equivalent and are equally populated under the [011] poling. When the (011)-oriented  crystal is actuated by $E$ parallel to [011], the two possible polar directions are expected to incline close to the $E$ direction in each domain, which results in an increased rhombohedral lattice distortion and a large piezoelectric response. Such a move strongly deforms the (011) plane (marked in dashed blue lines): it induces simultaneously a strong compressive strain along the [100] direction ($d_{32}$) and a tensile strain along the [0$\overline{1}$1] direction ($d_{31})$ \cite{Zhang2014} This results in different signs and magnitudes of the planar piezoelectric coefficients, i.e. $d_{32} = -1850$ pC.N$^{-1}$ and $d_{31} = 599$ pC.N$^{-1}$. Therefore the (011)-oriented PMN-PZT single crystal displays large anisotropic piezoelectric properties, with $\lvert d_{32} \rvert$ $\approx  3d_{31}$.  

Initially the PMN-PZT beams with dimensions $7^L\times5^W\times0.3^T$ mm$^3$ were cleaned with ethanol and acetone. A Fe$_{81}$Ga$_{19}$ polycrystalline target with a diameter of 3 inches was used in a \textit{Oerlikon Leybold Univex 350} sputtering system. The base pressure prior to the film deposition was typically 10$^{-7}$ mbar. Fe$_{81}$Ga$_{19}$ thin films  were deposited onto the PMN-PZT beams at room temperature using 100 W deposition power, and about 10 sccm Argon pressure. The FM FeGa thicknesses were $t\textsubscript{FM} = 5$ and 60 nm. The stack was capped \textit{in situ} with a 10 nm-thick Ta layer to protect the FeGa layer against oxidation. The growth was carried out under an in-plane magnetic field $\mu_0H$\textsubscript{dep} $\sim$ 30 mT (300 Oe) along the beam length, i.e. [100] direction (Fig.\ref{fig:fig1}a), in order to favor a preferential magnetic anisotropy direction. Other samples of FeGa/Ta with $t\textsubscript{FM} = 5$ and 60 nm were also deposited on Glass substrates (Schott D 263 TM \cite{schottglass}) which serve as reference samples.

For the electrical study, metallic electrodes must be available on both sides of the bi-layered ME composite. The 10 nm-thick Ta layer serves as a top electrode. For the bottom electrode, the samples were turned over and a 200 nm-thick Cu layer was deposited onto the bottom surface of the PMN-PZT beam using a mask of slightly smaller dimensions ($6.5^L\times4.5^W$ mm$^2$) than the PMN-PZT beam to avoid any short-circuit contact along the lateral edges with the FeGa. The final layered structure for our samples is Ta(10 nm)/FeGa($t$\textsubscript{FM})/PMN-PZT(0.3 mm)/Cu(200 nm) as shown in Fig.\ref{fig:fig1}a.  

Static magnetic measurements were performed to probe the magnetization reversal with a commercial \textit{Evico} MOKE microscope (Magneto-Otpical Kerr Effect) \cite{Evico}. An additional set-up within the MOKE apparatus was conceived to enable the application of an electric field across the sample thickness, thus allowing to perform magnetic measurements under an applied static electric field (Fig \ref{fig:fig1}a). The as-deposited samples Ta(10 nm)/FeGa($t$\textsubscript{FM})/PMN-PZT(0.3 mm)/Cu(200 nm) were held on a support inside the electromagnet. The support base contains a bottom electrode that contacts the bottom Cu layer; and the top Ta layer is in contact with a thin brass needle tip as the top electrode. Both electrodes are wired to a DC power supply (up to 300 V), and an ammeter together with a 10 M$\Omega$ protecting resistor were series-wound in the circuit to monitor the current during all the measurements. 

Furthermore, the coercive field temperature dependency was obtained from $M(\mu_0H)$ measurement using a \textit{Cryogenic} cryogen-free physical properties measurement platform with a vibrating sample magnetometer (VSM) inset \cite{CryogenicLtd}. The magnet was initially demagnetized after which the sample was cooled in zero applied magnetic field to the desired temperature. The $M(\mu_0H)$ was then measured using the low magnetic field option of the Cryogenic system.\\

\section{Results and discussion}

\subsection{Electrical tuning of the magnetization reversal}

\begin{figure}[h!]
    \centering
    \includegraphics[width=1\textwidth]{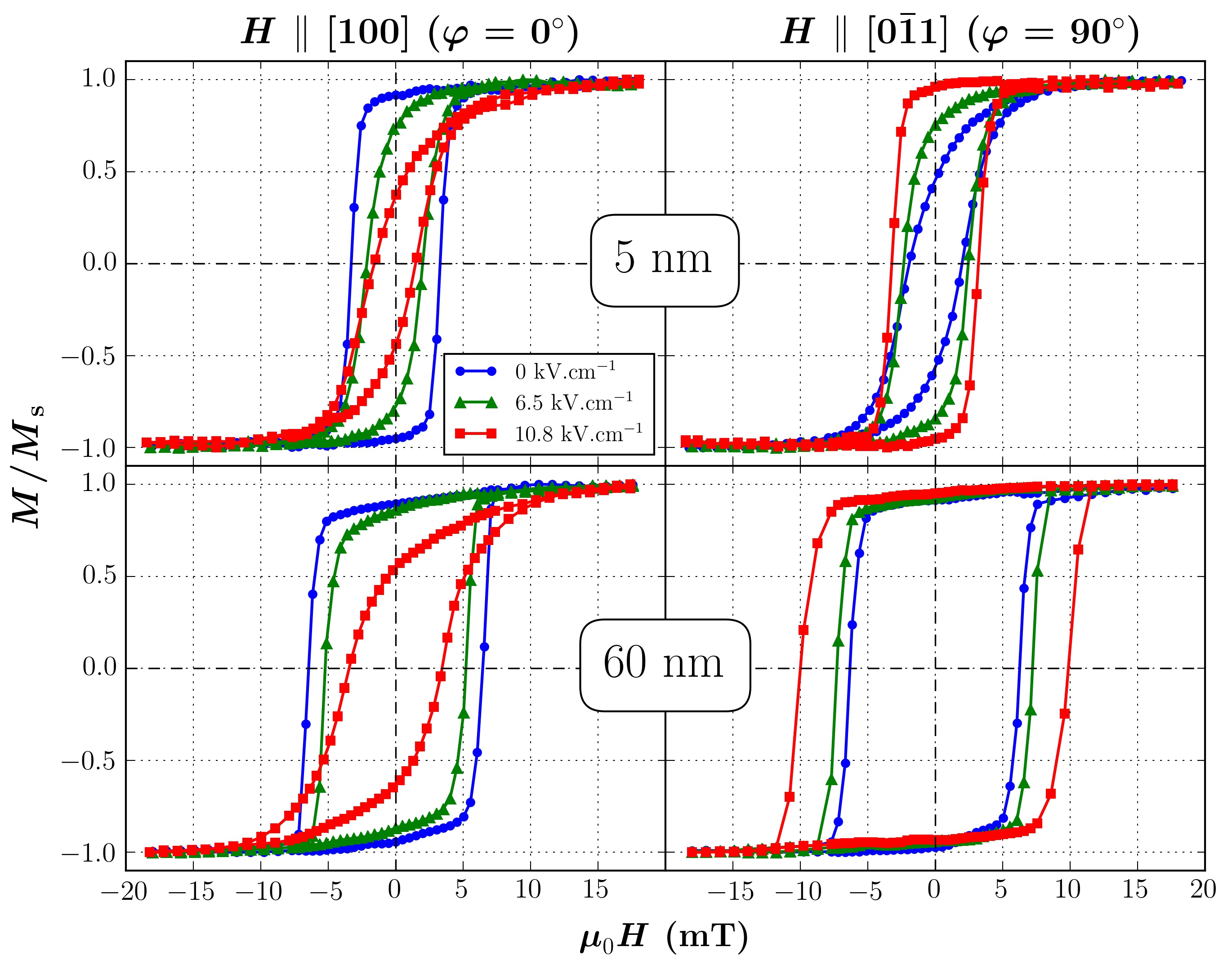}
    \caption{Hysteresis loops of the normalized magnetization reversal of the Ta(10 nm)/FeGa($t\textsubscript{FM}=5$; 60 nm)/PMN-PZT(0.3 mm)/Cu(200 nm), measured in-plane with the magnetic field $\mu_0H$ respectively parallel to [100] ($\varphi$ = 0\degree, i.e. along the deposition field axis $\mu_0H$\textsubscript{dep}) and [0$\bar{1}$1] ($\varphi$ = 90\degree, under three values of electric field $E$ = 0, 6.5 and 10.8 kV.cm$^{-1}$.}
    \label{fig:Fig2}
\end{figure}

Electric-field control of magnetization was carried out using MOKE that supported an additional \textit{in situ} set-up to enable the application of an electric field across the sample thickness. The $E$ pointing from the PMN-PZT to the FeGa film was defined as the positive $E$ (Fig.\ref{fig:fig1}). Routinely, the magnetization hysteresis loops were measured along the [100] direction ($\varphi$ = 0\degree) and the [0$\overline{1}$1] direction ($\varphi$ = 90\degree) under different values of applied electric field $E$: 0 kV.cm$^{-1}$, 6.5 kV.cm$^{-1}$ and 10 kV.cm$^{-1}$, as presented in Fig.\ref{fig:Fig2}. We will characterize these loops using the quantities $\mu_0H$\textsubscript{c} as the coercive field, and $M\textsubscript{r}^n$ as the remanent magnetization normalized to the saturation magnetization, which is equivalent to the squareness (i.e. $M\textsubscript{r}^n=M\textsubscript{r}/M\textsubscript{s}$).

Under no applied $E$, the 5 nm-thick FeGa sample's $M(\mu_0H)$ loops show an angular dependency when comparing $\mu_0H$\textsubscript{c} and $M\textsubscript{r}^n$ for both presented angles $\varphi=0\degree$ and 90$\degree$. As for the 60 nm-thick sample, the $M(\mu_0H)$ loops show a much less significant angular dependency; $\mu_0H$\textsubscript{c} and $M\textsubscript{r}^n$ show very close values for both angles. The measured values of $\mu_0H$\textsubscript{c} and $M\textsubscript{r}^n$ are typical of those observed for FeGa thin films \cite{jay2019}.

For both $t\textsubscript{FM}=5$ and 60 nm, $\mu_0H$\textsubscript{c} and $M\textsubscript{r}^n$ values along [100] decrease when $E$ increases, making the cycles more slanted than square. An easy magnetic anisotropy axis along [100] clearly becomes a harder axis under the application of $E$. The situation changes when looking at the [0$\overline{1}$1] direction, where $\mu_0H$\textsubscript{c} and $M\textsubscript{r}^n$ values increase when $E$ increases, and the cycles become more square than slanted. In this case a hard anisotropy axis along [0$\overline{1}$1], especially for the 5 nm-thick sample, becomes an easier axis. These converse reversal behaviors are tightly related to the (011)-PMN-PZT anisotropic strain induced by $E$: simultaneously a strong in-plane compressive strain along [100] ($d_{32}$) weakens the anisotropy's easy character and a tensile strain along [0$\overline{1}$1] ($d_{31}$) weakens its hard character.

The $E$-field dependency of $M(\mu_0H)$ loops along [100] and [0$\overline{1}$1], thus, suggests a first sign of a switching of a magnetic anisotropy easy axis. Such an axis, when aligned along $\mu_0H\textsubscript{dep}$, will tend to align under $E$ in the direction of tensile stress for positive magnetostriction $\lambda$, i.e. the [0$\overline{1}$1] direction. Hence, the total energy is minimzed, including the magnetoelastic term $E\textsubscript{me} = {-}(3/2) \lambda \sigma \cos^2\theta$ \cite{Lacheisserie1993}, where $\sigma$ is the applied stress and $\theta$ the angle between magnetization and stress.

In Section C we will provide azimuthal measurement in order to  provide more insight into the anisotropy behaviour of the samples.

\subsection{Bipolar $E$-field measurements}

The strain-mediated electric control of magnetization can be extended to a bipolar measurement cycle in which the electric field is swept through positive and negative values. We have performed such cycling as follows: $E = +10.8$ kV.cm$^{-1}$ $\rightarrow$ 0 kV.cm$^{-1}$ $\rightarrow$ $-10.8$ kV.cm$^{-1}$ $\rightarrow$ 0 kV.cm$^{-1}$ and finally back to the initial $+10.8$ kV.cm$^{-1}$, with a 1.6 kV.cm$^{-1}$ step. In the strain-mediated FM/FE two-phase system, the bipolar-$E$-field controlled magnetization generally exhibits a butterfly-shaped behavior and the understanding of it has been well established in terms of the piezostrain of the FE substrate transferred to the FM layer \cite{zhao2011orwu2011(2),Huang2015, zhang2012, Yang2014,Wu2011}.

In CME reports, it is common to probe the bipolar-$E$-field tuning of the magnetic moment as $M(E)$ loops, under a constant static magnetic bias field $\mu_0H$ and quantify the relative change of $M$ \cite{Zhang2014,Staruch2016,wei2016,Wang2019,Yang2009,Huang2015}. Another way of probing the electrically-induced change in magnetization would be extracting $M\textsubscript{r}^n$ from $M(\mu_0H)$ as function of $E$ loops \cite{Jiang2015}, as the ones in Fig.\ref{fig:Fig2}. In the following, we will quantify the relative change of $M\textsubscript{r}^n$ extracted from $M(\mu_0H)$ loops under incremented $E$ values as: $\Delta{M}\textsubscript{r}^n/M\textsubscript{r}^n(0)=(M\textsubscript{r}^n(E\textsubscript{max})-M\textsubscript{r}^n(0))/M\textsubscript{r}^n(0)$. 

\begin{figure}[h!]
    \centering
    \includegraphics[width=1\textwidth]{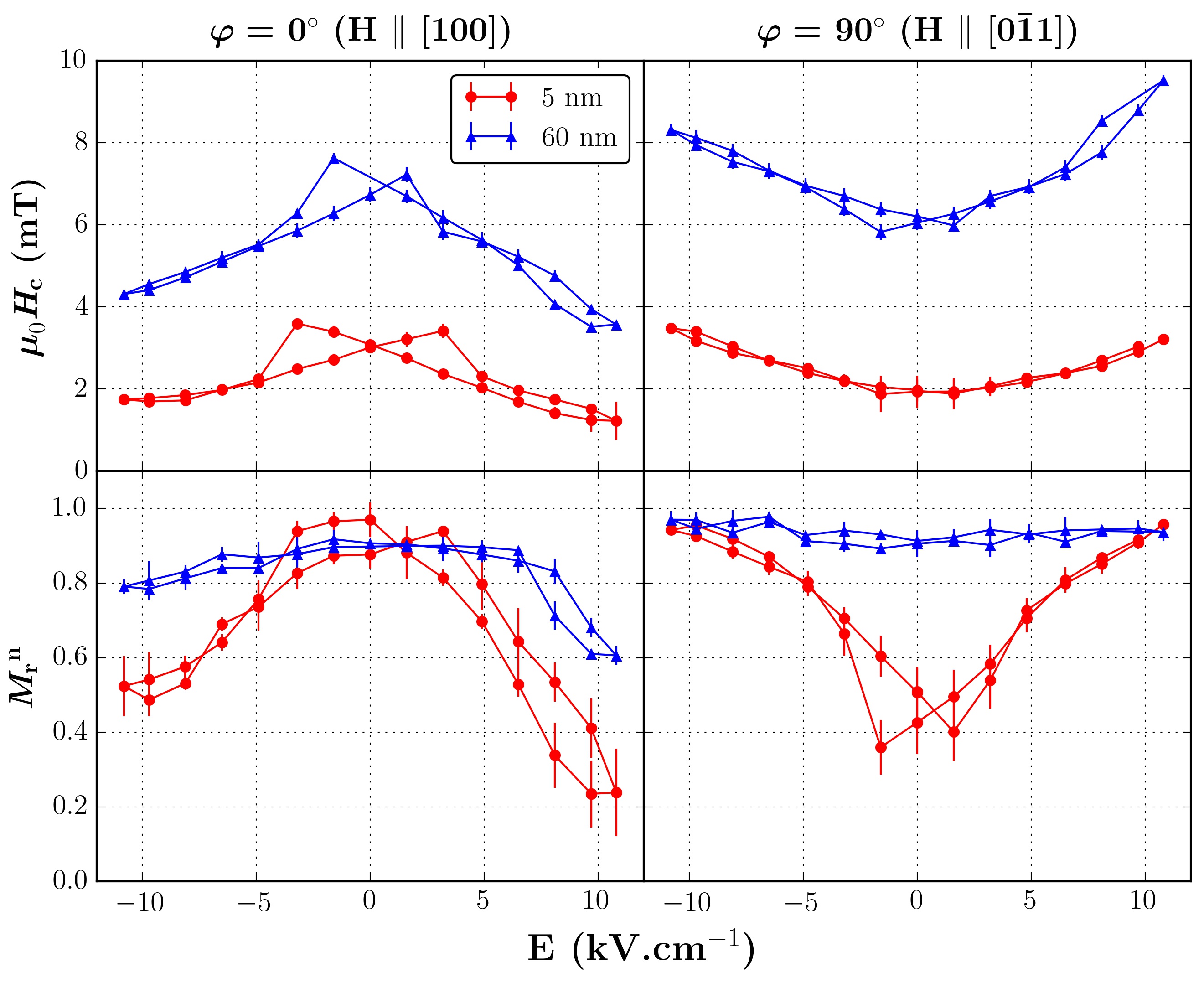}
    \caption{Electric-field control of the coercive field $\mu_0H$\textsubscript{c} (top row) and the remanent magnetization normalized to the saturation magnetization $M\textsubscript{r}^n$ (bottom row) along both directions [100] (left column) and [0$\bar{1}$1] (right column) for the Ta(10 nm)/FeGa($t\textsubscript{FM}=5$; 60 nm)/PMN-PZT(0.3 mm)/Cu(200 nm) samples. The lines are guides to the eyes.}
    \label{fig:Fig3}
\end{figure}

We can see in Fig.\ref{fig:Fig3} the nonlinear cycles representing the evolution of $\mu_0H$\textsubscript{c} and $M\textsubscript{r}^n$ as a function of $E$ along both [100] and [0$\overline{1}1$] directions. As mentioned earlier in section II, the magnetic bipolar loops in Fig.\ref{fig:Fig3} agree well with a low-hysteresis behavior of the (011)-PMN-PZT piezostrain, an almost reversible tuning of magnetization with positive and negative $E$, and most importantly a remarkable modification of the magnetic properties. For instance, by looking at the 5 nm-thick sample, $M\textsubscript{r}^n(E)$ along [100] undergoes a $\Delta{M\textsubscript{r}^n}/M\textsubscript{r}^n(0) \approx 72 \%$ relative decrease upon applying $E\textsubscript{max} = 10.8$ kV.cm$^{-1}$. Such relative decrease of magnetization has been previously observed in Co$_{40}$Fe$_{40}$B$_{20}$(20nm)/(011)-PMN-PT \cite{Zhang2014} and in FeAl(10 nm)/PIN-PMN-PT \cite{wei2016}


In the case of the 60 nm-thick sample, $\Delta{M\textsubscript{r}^n}/M\textsubscript{r}^n(0)$ is reduced to $\approx 33 \%$. This decrease with thickness can be associated with the reduced magnetoelastic coefficient in FeGa films with increasing thickness, as we have shown in our previous study \cite{jay2019}. Although the (011)-PMN-PZT piezostrain is strong enough to influence the thin FeGa films, and the direct bonding of the two materials (using the sputtering technique) is among the best techniques for maximum strain transfer, the magnetostriction of the FM phase is an important factor that drives the strain-mediated control of magnetization.

The relative change $\Delta{M\textsubscript{r}^n}/M\textsubscript{r}^n(0)$ along [0$\overline{1}$1] follows the opposite trend. $M\textsubscript{r}^n(E)$ increases by $\Delta{M\textsubscript{r}^n}/M\textsubscript{r}^n(0)\approx 80 \%$ when $E = 10.8$ kV.cm$^{-1}$ is applied for the 5 nm-thick sample. $M\textsubscript{r}^n(E)$ is practically unchanged for the 60 nm-thick sample, which is in agreement with the $M(\mu_0H)$ results in Fig.\ref{fig:Fig2}.

\begin{figure*}
    \centering
    \includegraphics[width=0.7\textwidth]{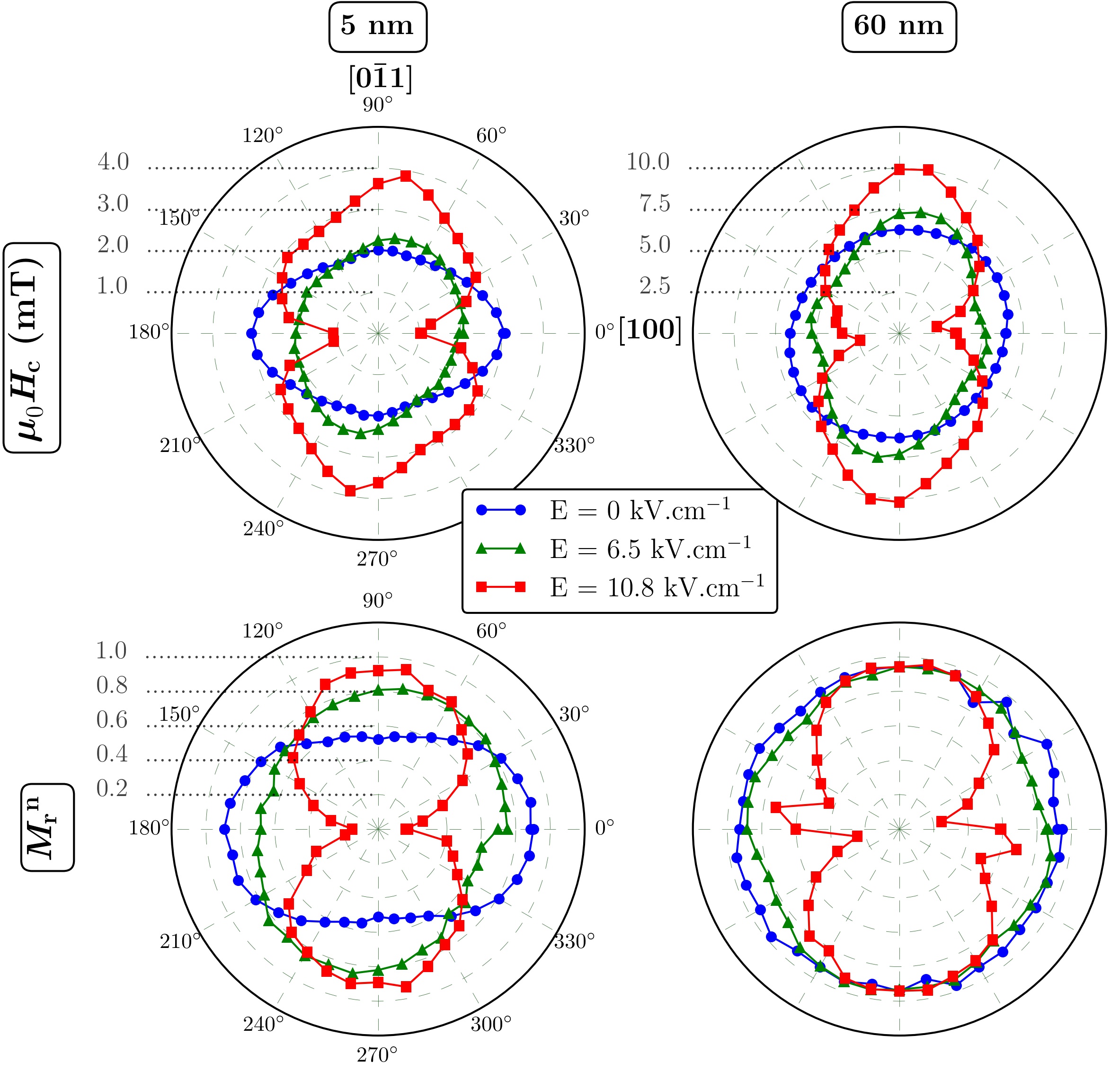}
    \caption{Azimuthal evolutions of the coercive field $\mu_0H$\textsubscript{c} (first row) and the remanent magnetization normalized to the saturation magnetization $M\textsubscript{r}^n$ (second row) extracted from $M(\mu_0H)-\varphi$ loops. The external magnetic field $\mu_0H$ is applied at the varying angle $\varphi$ on the Ta(10 nm)/FeGa($t\textsubscript{FM}=5$; 60 nm)/PMN-PZT(0.3 mm)/Cu(200 nm) samples and under the three considered values of electric field $E$ = 0, 6.5 and 10.8 kV.cm$^{-1}$. $\mu_0H$\textsubscript{dep} is applied along [100] ($\varphi$ = 0\degree).}
    \label{fig:Fig4}
\end{figure*}

We also note that the electric fields corresponding to the maximum (respectively minimum) $M\textsubscript{r}^n(E)$ along [100] (respectively [0$\overline{1}$1]) in Fig.\ref{fig:Fig3} are $\approx \pm$ 2 kV.cm$^{-1}$, which is smaller than the coercive field of PMN-PZT $E\textsubscript{c}= 4$ kV.cm$^{-1}$. This has been previously observed in similar ME composites \cite{Yang2009}. In an ideal case, a ferroelectric $P-E$ or $S-E$ loop is symmetrical, so the positive and negative $E\textsubscript{c}$ are equal and correspond to FE domains switching. Experimentally, the $E\textsubscript{c}$ value is not an absolute threshold for FE domains switching, which may start for $E$ values smaller than $E\textsubscript{c}$. These FE properties may be affected by many factors that shape a ferroelectric $P-E$ or $S-E$ loop including the thickness of the sample, presence of charged defects, mechanical stresses, preparation conditions, thermal treatment \cite{Damjanovic2006-book}, and relaxation effects of switching FE domains \cite{Zhang2014}.


In addition, Fig.\ref{fig:Fig3} shows an asymmetry of the $M\textsubscript{r}^n(E)$ and $\mu_0H\textsubscript{c}(E)$ curves by looking at the highest $E=\pm10.8$ kV.cm$^{-1}$, most noticeably the $M\textsubscript{r}^n(E)$ curves along [100]. This asymmetry may arise from the aforementioned factors, and has been observed previously in magnetoelectric composites \cite{Wang2019, Jiang2015, Yang2009, wei2016, Staruch2016, Nan2014}. Small remanent strain states may be responsible for such asymmetry \cite{Jiang2015}. It is also presumed that an internal field in the PMN-PZT may be generated, originating from different kinds of defects (e.g. structure, fatigue, relaxation) in the FE substrate \cite{Guo2016, chen2009, Wu2011, Lupascu2004, Yang2014, gopalan1996, noguchi2000}.

\subsection{Magnetic azimuthal evolutions under $E$}

The previous electric-field control of magnetization results have so far been presented for the two characteristic directions [100] ($\varphi=0\degree$) and [0$\overline{1}1$] ($\varphi=90\degree$) of the (011)-PMN-PZT. Indeed depending on thickness, FeGa exhibits a considerable angular dependency of magnetic properties  as shown in Fig.\ref{fig:Fig2}. It would be of interest to study this azimuthal dependency to further understand the anisotropy configuration by rotating the sample in its plane by the angle $\varphi$ with respect to $\mu_0H$\textsubscript{dep}. In-plane magnetization reversal measurements were performed each 10 degrees. The azimuthal evolutions of $\mu_0H$\textsubscript{c} and $M\textsubscript{r}^n$ as a function of $E$ are reported for all samples in Fig.\ref{fig:Fig4}.

Under $E=0$ kV.cm$^{-1}$, the 5 nm-thick sample shows two maxima of $\mu_0H$\textsubscript{c} and $M\textsubscript{r}^n$ lying along [100] or $\varphi=0\degree$. The axis carrying these maxima will be referred to as the ``maxima axis", and in this case lies along the $\mu_0H$\textsubscript{dep} direction. Two smaller local maxima of $\mu_0H$\textsubscript{c} also appear at $\varphi=90\degree$. The angular dependency of $\mu_0H$\textsubscript{c} indicates a cubic component of the magnetic anisotropy. This has been already observed in our previous work which revealed a magnetic anisotropy with a predominant in-plane cubic component in thin FeGa films (5 nm) deposited onto glass substrate \cite{jay2019}.

Applying an electric field $E=6.5$ kV.cm$^{-1}$ rotates the maxima axis by almost 70$\degree$, and a stronger field $E=10.8$ kV.cm$^{-1}$ rotates it further towards the [0$\overline{1}1$] direction by 80$\degree$. Indeed applying $E=10.8$ kV.cm$^{-1}$ also changes the azimuthal shape of $\mu_0H$\textsubscript{c} and $M\textsubscript{r}^n$ which is typical of a uniaxial anisotropy (i.e. two $\mu_0H$\textsubscript{c} maxima at $\varphi=80\degree$, two $M\textsubscript{r}^n$ maxima at $\varphi=90\degree$, and two minima at $\varphi=0\degree$). This amount of maxima axis rotation is higher than the reported 55$\degree$ value in FeAl(10 nm)/PIN-PMN-PT under $E=10-12$ kV.cm$^{-1}$ \cite{wei2016}. It is worth noting that the $\mu_0H$\textsubscript{c} maxima have higher values $\mu_0H\textsubscript{c,max}=4$ mT under the highest $E=10.8$ kV.cm$^{-1}$ compared to the maxima $\mu_0H\textsubscript{c,max}=3$ mT under no applied $E$. The same remark can be done for the 60 nm-thick sample \cite{Zhang2014}.

Under no applied $E$, the situation is different for the 60 nm-thick sample: the angular dependency of $\mu_0H\textsubscript{c}$ is quasi-circular, revealing a random magnetic anisotropy dispersion \cite{Cullen2007,Begue2019}. This has also been observed in our previous paper \cite{jay2019}, in which the cubic component of the 5 nm-thick films faded away, but did not vanish, to a more random anisotropy dispersion with increasing thicknesses ($\geq$ 20 nm). Such behavior was attributed to a predominant texture in thinner films which, in thicker films, is replaced by a non-preferential polycrystalline arrangement. \citet{Cullen2007} suggests that a competition between coherent and randomly oriented local anisotropies leads to zero or very small net anisotropy in FeGa thin films with almost the same composition ($\sim$ 20 \% Ga content). In a recent study \cite{Begue2019}, magnetic domain structure observations in FeGa/MgO thin films evidenced this competition between the coherent cubic anisotropy and the random anisotropy contribution.
 
Under the highest field $E=10.8$ kV.cm$^{-1}$, two maxima of $\mu_0H$\textsubscript{c} and $M\textsubscript{r}^n$ appear at $\varphi=80\degree$ and are accompanied by the development of two local maxima of $M\textsubscript{r}^n$ close to $\varphi\sim0\degree$. It is clear that the random anisotropy character vanishes, leaving out a coexistence of the coherent cubic and uniaxial components under the strain. Under the field $E=6.5$ kV.cm$^{-1}$, an intermediate behavior is observed. 

\subsection{Magnetoelectric coefficient $\alpha\textsubscript{CME}$}

To quantify the electric-field-induced variation of magnetic properties, it is convenient to introduce the converse magnetoelectric coupling coefficient defined by $\alpha\textsubscript{CME}$ (expressed in s.m$^{-1}$), which represents the variation of the magnetization under an applied electric field. 

Several methods may be used to calculate $\alpha\textsubscript{CME}$. The first one (method \textit{a}) is to consider $\alpha\textsubscript{CME}$ as the slope (first derivative) of a $M(E)$ loop, i.e. directly measuring the magnetization change as a function of a changing electric field either under a constant static magnetic bias field $\mu_0H$ \cite{Thiele2007,Lian2018-2,Eerenstein2007,Zhang2014,Yang2009,wei2016,cherifi2014}, or by saturating the FM film with $\mu_0H$ and then removing it \cite{Wang2019}. In another method (method \textit{b}), $\alpha\textsubscript{CME}$ is determined by computing the magnetization from magnetoresistance loops measurements in a spin-valve device \cite{Heron2014}. In our case, an alternative method (method \textit{c}) consists of calculating  $\alpha\textsubscript{CME}$ using the following equation \cite{Staruch2016,alberca2015}:

\begin{equation*}
\begin{aligned}
\alpha\textsubscript{CME}(\mu_0H) 
&= \mu_0\frac{\Delta{M_{E_0}(\mu_0H)}}{\Delta{E}}\\
&= \mu_0\frac{M_{E=E_0}(\mu_0H)-M_{E=0}(\mu_0H)}{E_0},
\end{aligned}
\end{equation*}
with $E_0=6.5$ or 10.8 kV.cm$^{-1}$. This relative change in magnetization under an applied electric field $E$ is directly computed from the $M(\mu_0H)$ loops presented in Fig.\ref{fig:Fig2}.

\begin{figure}[h!]
    \centering
    \includegraphics[width=1\textwidth]{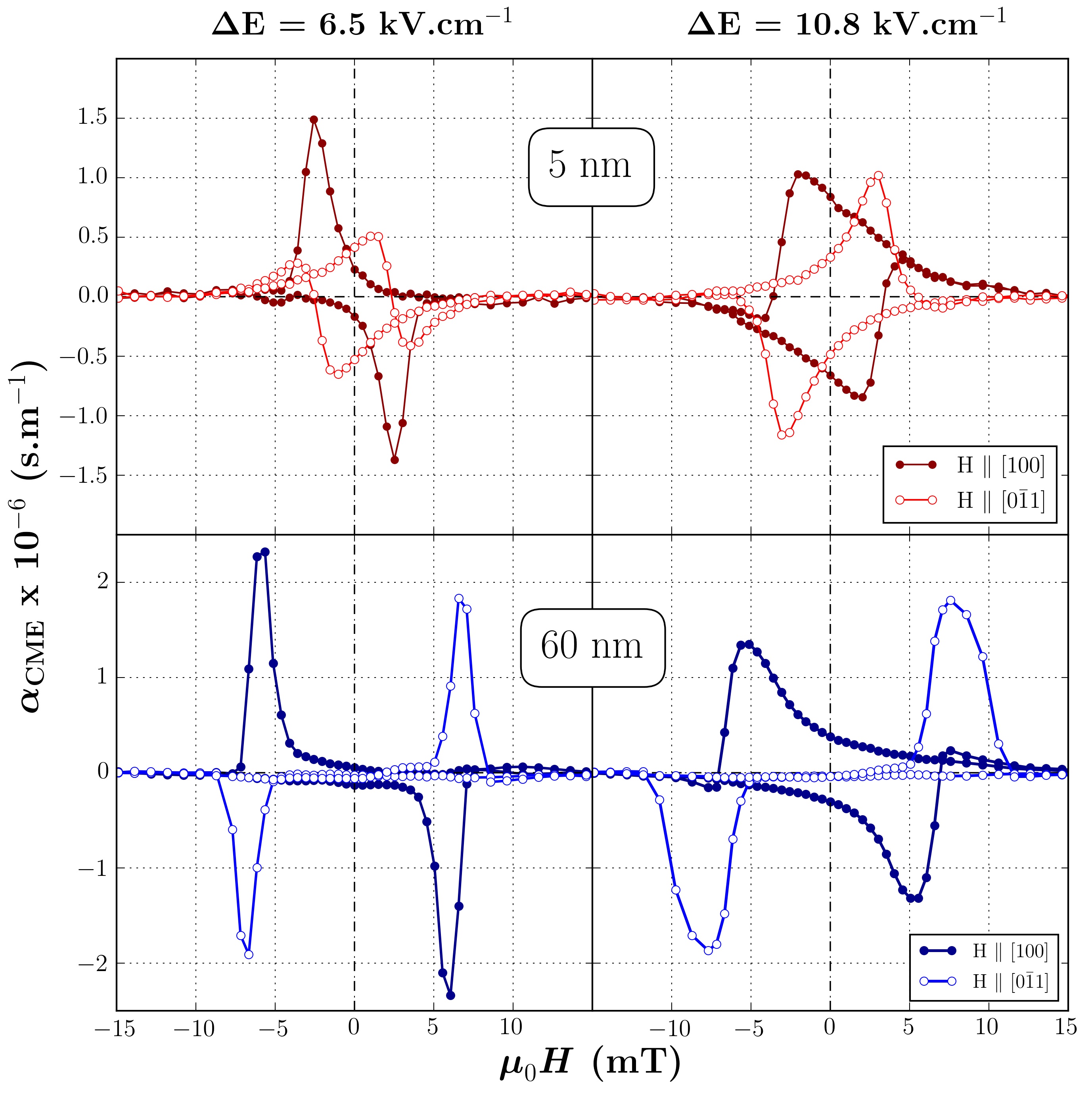}
    \caption{The converse magnetoelectric coupling coefficient $\alpha$\textsubscript{CME} as calculated from $M(\mu_0H)$ data in Fig.\ref{fig:Fig2} for the Ta(10 nm)/FeGa($t\textsubscript{FM}=5$; 60 nm)/PMN-PZT(0.3 mm)/Cu(200 nm) samples, under $\Delta{E}$ = 6.5 and 10.8 kV.cm$^{-1}$.}
    \label{fig:Fig5}
\end{figure}

The FeGa thin films in this study have a saturation magnetization $\mu_0 M\textsubscript{s}=1.15$ T \cite{jay2019}; we can thus deduce the $\alpha$\textsubscript{CME} values in s.m$^{-1}$ as shown in Fig.\ref{fig:Fig5} along both characteristic directions [100] and [0$\overline{1}1$] of the (011)-PMN-PZT. 

The results reveal a maximum $\alpha\textsubscript{CME}=2.4\times{10^{-6}}$ s.m$^{-1}$ for the 60 nm-thick sample along [100]. This value is indeed obtained for lower $\Delta{E}=6.5$ kV.cm$^{-1}$, and for magnetic bias field values near $\mu_0H$\textsubscript{c}. Also, for the 5 nm-thick sample, a higher $\alpha\textsubscript{CME}=1.5\times{10^{-6}}$ s.m$^{-1}$ is obtained along [100] than along [0$\overline{1}1$]. Indeed, these results are correlated with the fact that the magnetization reversal loops in Fig.\ref{fig:Fig2} are more significantly modified along [100] than along [0$\overline{1}1$] under $E=6.5$ kV.cm$^{-1}$. 

Besides, it is interesting to see non-zero $\alpha\textsubscript{CME}$ values at zero bias field $\mu_0H=0$, especially for the 5 nm-thick sample: $\alpha\textsubscript{CME}\sim0.7\times{10^{-6}}$ s.m$^{-1}$ for H//[100] and $\Delta{E}=10.8$ kV.cm$^{-1}$. These values correspond to a remanent magnetoelectric coupling at zero bias field $\mu_0H$, which is related to the hysteretic magnetic behavior and the strong remanent magnetization \cite{zhou2016, Lian2018-2}. This is truly encouraging for the dynamic CME self-biased potential of FeGa/PMN-PZT in applications, which is revealed in measurements that assess the magnetization change under an alternating AC electric field superimposed to a DC magnetic bias field $\mu_0H$ (yielding a non-zero remanent CME($\mu_0H$) at $\mu_0H=0$) \cite{Yang2015,Fitchorov2011a,Zhang2014-selfbiasCME,ChulYang2011-selfbiasCME}.


\begin{figure}[h!]
    \centering
    \includegraphics[width=1\textwidth]{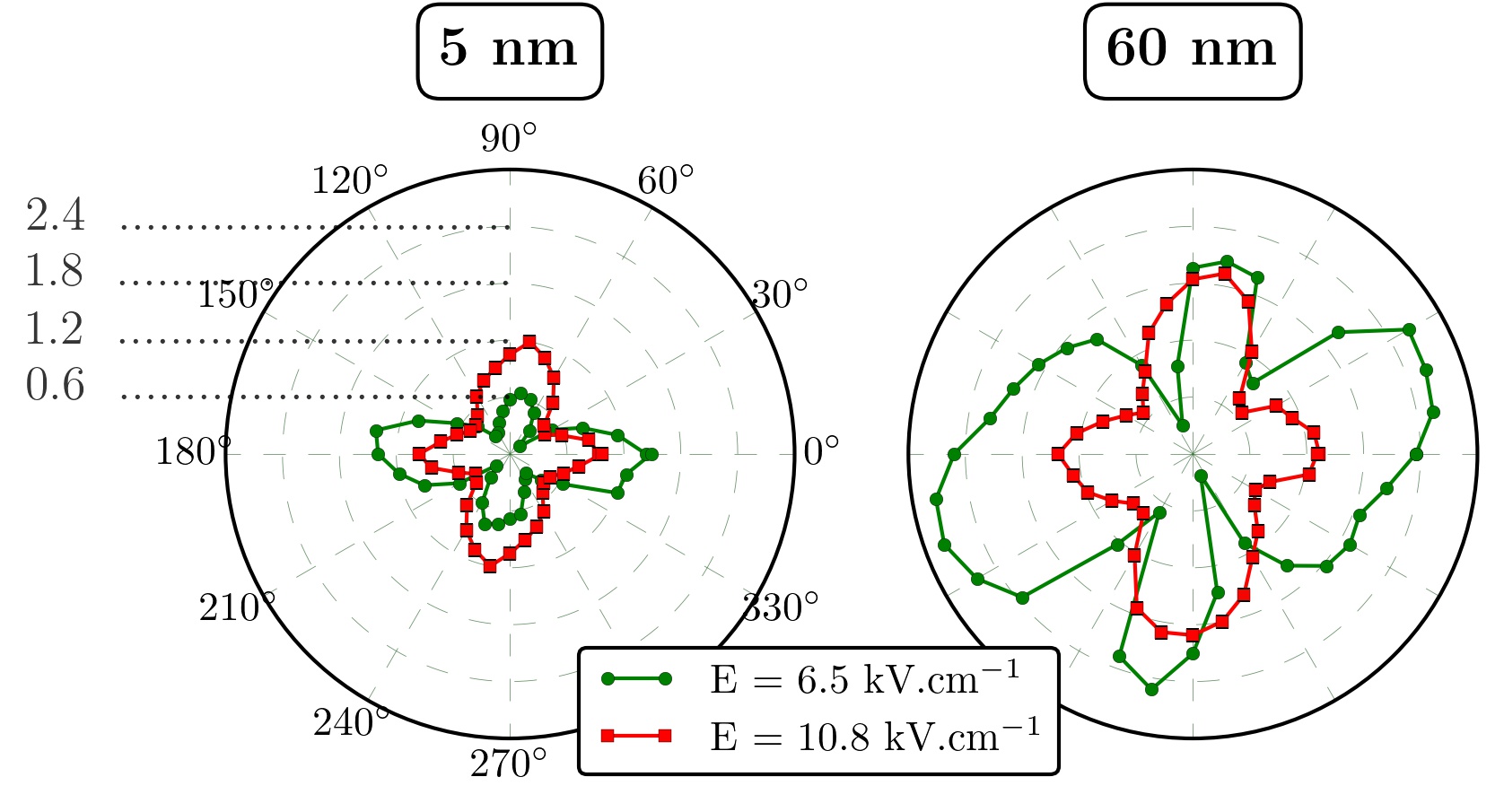}
    \caption{Azimuthal evolution of the maximum of the converse magnetoelectric coupling coefficient $\alpha$\textsubscript{CME,max} as calculated in Fig.\ref{fig:Fig5}, and extracted from $M(\mu_0H)-\varphi$ azimuthal data of the Ta(10 nm)/FeGa($t\textsubscript{FM}=5$; 60 nm)/PMN-PZT(0.3 mm)/Cu(200 nm), under $\Delta{E}$ = 6.5 and 10.8 kV.cm$^{-1}$.}
    \label{fig:Fig6}
\end{figure}

Furthermore, as we have performed azimuthal measurements shown in Fig.\ref{fig:Fig4} to assess the angular dependencies of $\mu_0H$\textsubscript{c} and $M\textsubscript{r}^n$, this led us to believe that the CME behavior may rely specifically on certain non trivial orientations $\varphi$ rather than only the crystallographic [100] and [0$\overline{1}$1] directions. We have, thus, calculated the angular dependency of $\alpha$\textsubscript{CME} by applying the same method used in Fig.\ref{fig:Fig5} to the $M(\mu_0H)-\varphi$ loops used in Fig.\ref{fig:Fig4}. The maximum value of $\alpha$\textsubscript{CME} is subsequently plotted as $\alpha$\textsubscript{CME,max} for each angle $\varphi$ in Fig.\ref{fig:Fig6}. 

The angular behavior of $\alpha$\textsubscript{CME,max} shows peculiar symmetry shapes especially under $\Delta{E} = 6.5$ kV.cm$^{-1}$. This confirms that the as-considered $\alpha$\textsubscript{CME} method (\textit{c}) depends exclusively on the measured $M(\mu_0H)-\varphi$ loops' characteristics related to both $\mu_0H$\textsubscript{c} and $M\textsubscript{r}^n$. We remind the reader that these $\alpha$\textsubscript{CME,max} correspond to peak values of $\alpha$\textsubscript{CME} for bias field values $\mu_0H$ close to $\mu_0H$\textsubscript{c} (Fig.\ref{fig:Fig5}), and not $\alpha$\textsubscript{CME} taken at zero bias field $\mu_0H$. The zero bias field $\mu_0H$ corresponds exclusively to the $M\textsubscript{r}^n(E)$ case that was explored in Fig.\ref{fig:Fig3}. 

By looking at the 60 nm-thick sample, we can deduce an even higher value than the ones obtained at $\varphi=0\degree$ and 90$\degree$ of $\alpha\textsubscript{CME,max}\approx2.7\times{10^{-6}}$ s.m$^{-1}$ at $\varphi=30\degree$. Such results are reported for the first time and indeed confirm the anisotropic nature of CME in anisotropic FeGa thin films. A stronger electric field $E=10.8$ kV.cm$^{-1}$ brings about a more uniform four-fold symmtery of the $\alpha$\textsubscript{CME,max} for all samples. Two maxima values $\alpha\textsubscript{CME,max}\approx1.8\times{10^{-6}}$ s.m$^{-1}$ are found around $\varphi=90\degree$ and two lower maxima around $\varphi=0\degree$. This behavior is similar to the stronger $\mu_0H$\textsubscript{c} modification by $E$ along [0$\overline{1}1$].\\

To our knowledge, this value of $\alpha$\textsubscript{CME} obtained at room temperature is several orders of magnitude higher than the reported values for single-phase multiferroics in the CME literature, as well as higher than or comparable to reported values in other composite multiferroics. In Table \ref{table:table} a comprehensive list is given of the reported CME values. 

It is important to note that such a high achievable $\alpha$\textsubscript{CME} is obtained within a unipolar measurement of $\lvert\Delta{E}\rvert=\lvert0-6.5\rvert$ kV.cm$^{-1}$ with a non-180$\degree$ polarization switching of the (011)-PMN-PZT ferroelectric domains (as explored in the bipolar measurements in Fig.\ref{fig:Fig3}), which is expected to fatigue the FE single crystal, induce relaxation effects, and alter the performance of ME devices \cite{zhao2011orwu2011(2), chen2009}. Therefore, electric-field control of magnetization in the unipolar case should be preferred for example for the high-speed applications of SME-RAMs (\textit{Strain-mediated MagnetoElectric Random Access Memory}) in similar structures \cite{Hu2011}.  Another key point is that our value of $\alpha$\textsubscript{CME} coupling is obtained in a static non-resonant mode of measurement. This is  not only promising for realizing non-resonant ME devices, but also appealing to perform dynamic measurements which are expected to boost this value of $\alpha$\textsubscript{CME}, but are beyond the scope of this study. 

Thus, the tunable converse ME effect reported here is particularly significant in terms of strong magnetization reversal variation.
Finally, even if a higher $\alpha\textsubscript{CME}$ is achieved with the thicker 60 nm FeGa film, it is noteworthy that $\alpha\textsubscript{CME}$ of the thinner 5 nm film is still high enough compared to the other reported values for thicker films in Table \ref{table:table}. The 5 nm film offers also the advantage of a non-zero remanent $\alpha\textsubscript{CME}$, i.e. at $\mu_0H=0$.

\begin{table}[h!]
\centering
\footnotesize
\setlength{\tabcolsep}{0mm}
\begin{tabularx}{\linewidth}{cccc}
\toprule
\rowcolor{gray!25}
\textbf{Multiferroic system} & \textbf{\boldsymbol{$\alpha$}$_{\textbf{CME}}$ (s.m$^{\textbf{-1}}$)} & \textbf{T(K)} & \textbf{ref.} \\
\midrule
CoFe$_2$O$_4$(200 nm)/PMN-PT & 3.2 $\times$ 10$^{-8}$ * & 300  & \cite{Yang2009} \\
\hline 
\makecell{La$_{0.7}$Sr$_{0.3}$MnO$_3$(20–50nm)/\\PMN-PT} & 6.0 $\times$ 10$^{-8}$ * & 330  & \cite{Thiele2007} \\
\hline
\makecell{Co$_{0.9}$Fe$_{0.1}$(2.3nm)/Cu/ \\ Co$_{0.9}$Fe$_{0.1}$(2.5nm)/BiFeO$_3$} & 1.0 $\times$ 10$^{-7}$ ** & 300  & \cite{Heron2014} \\ 
\hline
\makecell{La$_{0.67}$Sr$_{0.33}$MnO$_3$(40nm)/\\BaTiO$_3$} & 2.3 $\times$ 10$^{-7}$ * & 157  & \cite{Eerenstein2007} \\
\hline
YIG(600nm)/PMN-PZT
 & 3.1 $\times$ 10$^{-7}$ * & 300  & \cite{Lian2018-2} \\ 
\hline
\makecell{La$_{0.7}$Ca$_{0.3}$MnO$_3$(10 nm)/\\BaTiO$_3$} & 5 $\times$ 10$^{-7}$ *** & 20  & \cite{alberca2015} \\ 
\hline
Terfenol-D/PZT & 7.8 $\times$ 10$^{-7}$ & 300  & \cite{wu2016}\\
\hline
FeRh(22nm)/BaTiO$_3$ & 1.4 $\times$ 10$^{-6}$ * & 385  & \cite{cherifi2014} \\
\hline
Fe$_{81}$Al$_{19}$(10 nm)/PIN-PMN-PT & 1.6 $\times$ 10$^{-6}$ * & 300  & \cite{wei2016} \\
\hline
Co$_{40}$Fe$_{40}$B$_{20}$(20nm)/PMN-PT & 2.0 $\times$ 10$^{-6}$ * & 300  & \cite{Zhang2014} \\ 
\hline
\textbf{Fe$_{81}$Ga$_{19}$(60m)/PMN-PZT} & \textbf{ 2.7 $\times$ 10$^{-6}$} *** & \textbf{300} & \textbf{\makecell{this\\\textcolor{white}{.}work\textcolor{white}{.}}} \\ 
\hline
\makecell{Fe$_{50}$Co$_{50}$(80nm)/Ag/\\PIN-PMN-PT} & 3.5 $\times$ 10$^{-6}$ *** & 300  & \cite{Staruch2016} \\
\hline
Co$_{40}$Fe$_{40}$B$_{20}$(50nm)/PMN-PT & 8.0 $\times$ 10$^{-6}$ * & 300  & \cite{Wang2019} \\
\bottomrule
\end{tabularx}
\caption{\label{table:table}Literature recap of converse magnetoelectric coupling coefficient $\alpha$\textsubscript{CME} values (in s.m$^{-1}$) for different magnetoelectric composite materials. *: method \textit{a} of $\alpha$\textsubscript{CME} computing, **: method \textit{b} and ***: method \textit{c}.} 
\end{table}

\subsection{Thermo-mechanical effects }

In the previous sections the magnetoelectric effect has been investigated in the FeGa/PMN-PZT composite. The strain driven by an applied electric-field on the PMN-PZT substrate indeed manipulates the FeGa magnetization state and anisotropy. 

Another parameter that may modify the internal strain and stress -- and, thus, the magnetic anisotropy through inverse magnetostriction -- is temperature through thermal expansion.
Temperature is an important factor for the stability of ME devices operating in complex environments. ME effect has only rarely been reported at low temperatures for power generation (DME) \cite{Han2018-temperature,Burdin2012-temperature}. However measuring the temperature-dependent ME effect in our samples is beyond the scope of this study.  
Instead we have aimed to examine the potential influence of the PMN-PZT substrate's thermal expansion on the FeGa magnetic properties as a function of temperature, in order to gain more insight on how thermal strain can act on a magnetostrictive material deposited on ferroelectric single crystals.

\begin{figure}[h!]
    \centering
    \includegraphics[width=1\textwidth]{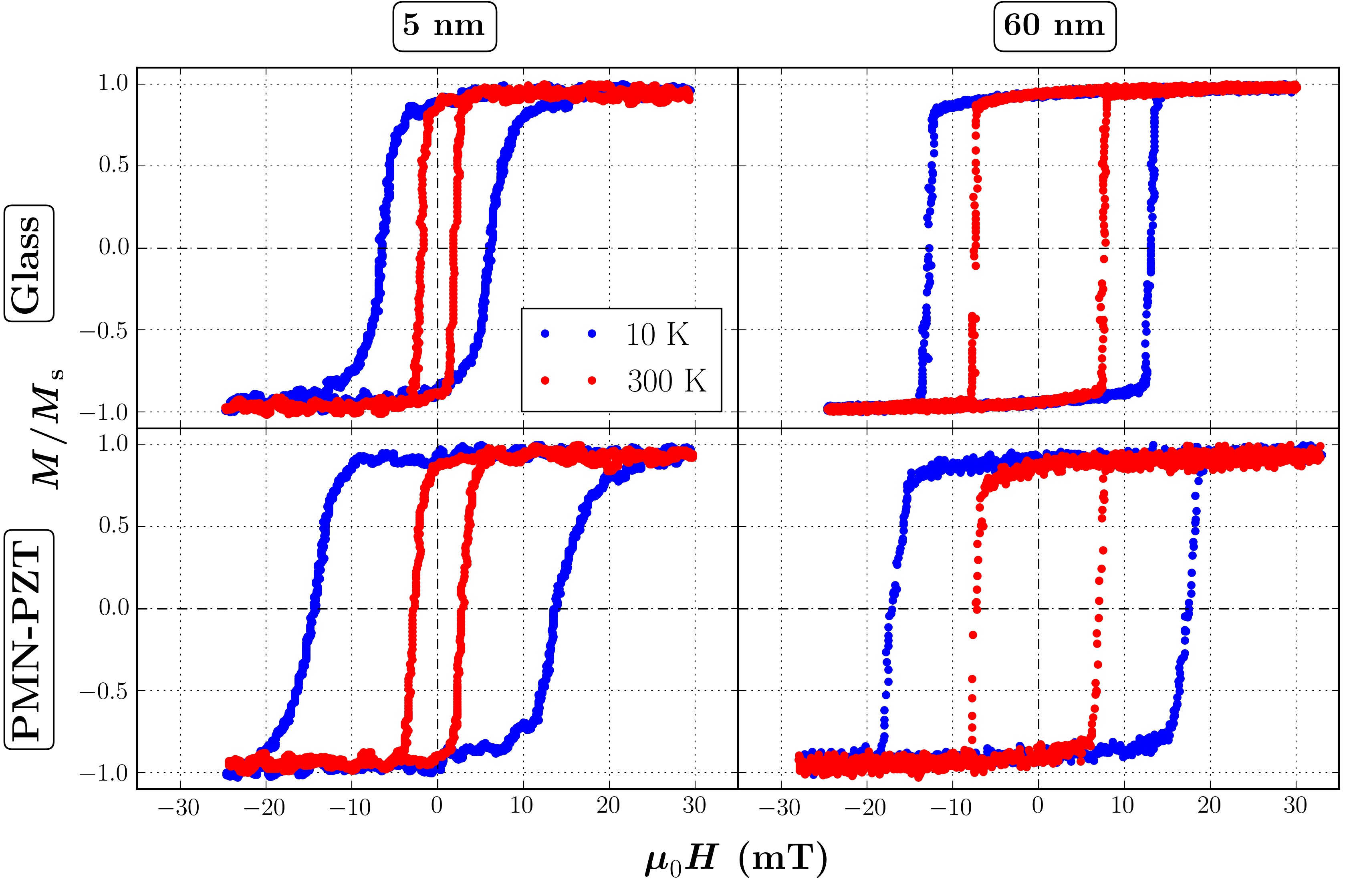}
    \caption{Hysteresis loops of the normalized magnetization reversal at 300 K and 10 K of the Ta(10 nm)/FeGa($t\textsubscript{FM}=5$; 60 nm)/PMN-PZT(0.3 mm) and FeGa($t\textsubscript{FM}=5$; 60 nm/Glass)/Ta(10 nm), measured in-plane with the magnetic field $\mu_0H$ parallel to [100] ($\varphi$ = 0\degree, i.e. along the deposition field axis $\mu_0H$\textsubscript{dep}).}
    \label{fig:Fig7}
\end{figure}

We have performed magnetic temperature measurements,  using a \textit{Cryogenic} free physical properties measurement platform with a VSM insert. The sample is zero-field cooled (ZFC) to the desired temperature (from 300 K down to 10 K as an example), then $M(\mu_0H)$ is measured, from which we can extract the coercive field $\mu_0H$\textsubscript{c}. We note that the results were the same in both ZFC and field-cooled (FC) modes. We used two different substrates for the purpose of discriminating between the FM thermal expansion and that of the substrate: amorphous glass and the single-crystalline ferroelectric PMN-PZT. Samples, thus, consist of Ta(10 nm)/FeGa($t\textsubscript{FM}=5$; 60 nm)/PMN-PZT(0.3 mm) and the reference bilayer Ta(10 nm)/FeGa($t\textsubscript{FM}=5$; 60 nm)/Glass(0.5 mm). The  corresponding $M(\mu_0H)$ loops are presented in Fig.\ref{fig:Fig7} for both temperatures 300 K and 10 K. 

$M(\mu_0H)$ loops in Fig. \ref{fig:Fig7} show a drastic influence of temperature on $\mu_0H$\textsubscript{c} values. The remanence however is not affected. For both thicknesses, $\mu_0H$\textsubscript{c} values are similar at room temperature 300 K on both substrates. At 10 K, these values diverge depending on the substrate.

\begin{figure}[h!]
    \centering
    \includegraphics[width=1\textwidth]{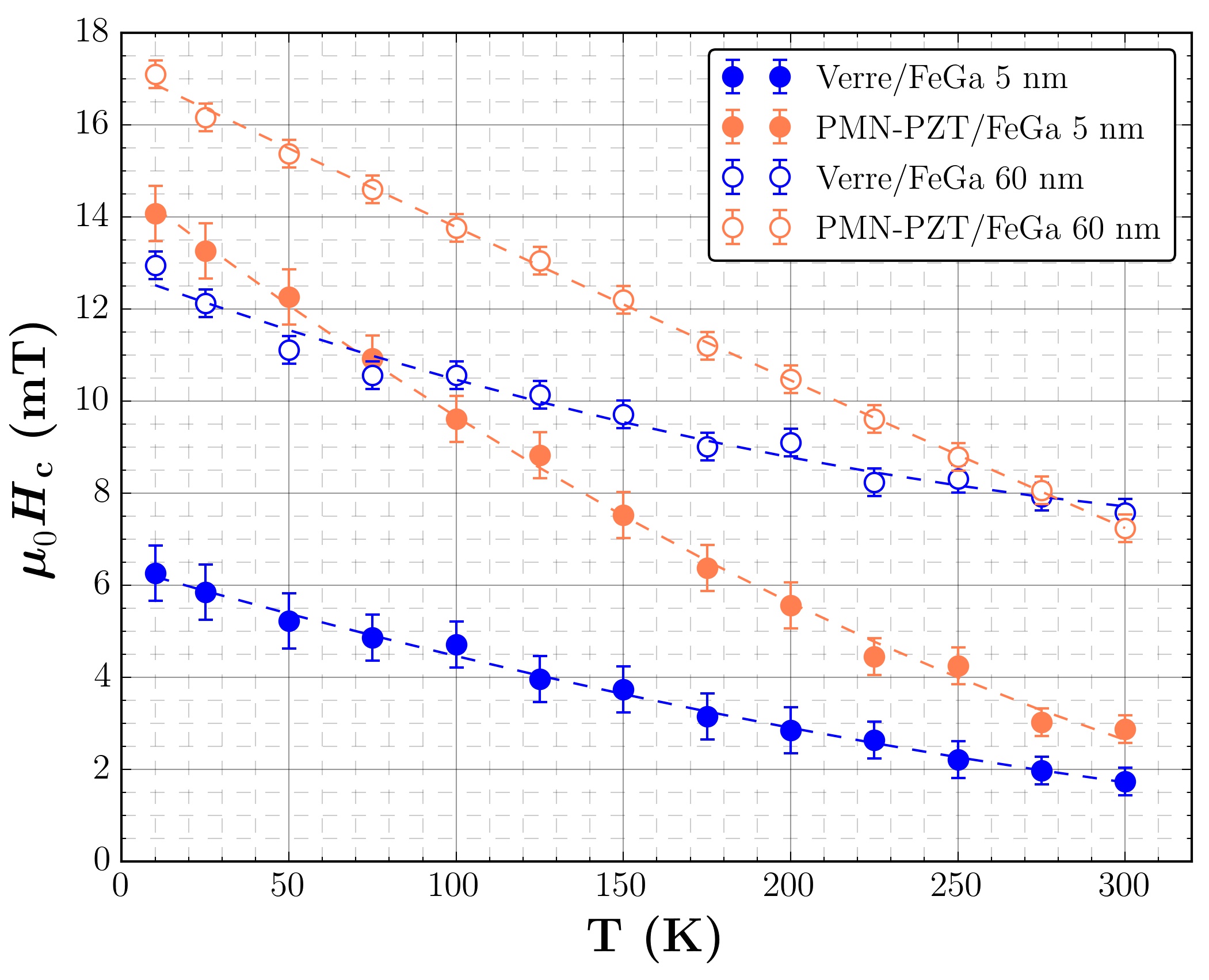}
    \caption{Temperature-dependent evolution of the coercive field $\mu_0H$\textsubscript{c} of 5 and 60 nm-thick FeGa samples (respectively filled and empty circles) deposited onto two different substrates: Glass (blue circles) and (011)-PMN-PZT (orange circles). The dashed lines are guides to the eyes.}
    \label{fig:Fig8}
\end{figure}

The evolution of $\mu_0H$\textsubscript{c} as a function of the measuring temperature for both substrates and $t$\textsubscript{FM} are presented in Fig.\ref{fig:Fig8}. $\mu_0H$\textsubscript{c} typically decreases with increasing temperature, as a result of thermal agitation \cite{Richy2018, jay2019}. Nevertheless, comparing one $t$\textsubscript{FM} of FeGa on both substrates (blue and orange circles) reveals a clear difference in the curve slope $\Delta(\mu_0H\textsubscript{c}$)/$\Delta$T. Therefore below 300 K, $\mu_0H$\textsubscript{c} values of one FeGa $t$\textsubscript{FM} gradually diverge when comparing both substrates. In particular, $\mu_0H$\textsubscript{c} of the 5 nm-thick sample increases between 300 K and 10 K approximately sixfold on PMN-PZT, whereas on glass it increases threefold. It is also noteworthy that the $\mu_0H$\textsubscript{c}(T) curve slope $\Delta (\mu_0H\textsubscript{c}$)/$\Delta$T is independent of the film thickness, by revealing almost parallel curves (color-filled or open circles) of both $t$\textsubscript{FM} on a single substrate. This has also been shown in Fig.7 of our previous work \cite{jay2019}. 

A similar substrate-dependent trend of $\mu_0H$\textsubscript{c} evolution at low temperatures was also observed when comparing exchange biased FeGa/IrMn bilayers deposited onto piezoelectric PVDF for electrical manipulation, and on Si substrates \cite{zhang2016_thermal_substrate_EB}.

Consequently, at low temperatures, when FeGa is grown on the piezoelectric PMN-PZT, its coercivity responds differently to thermal strain compared to the growth on glass. This indeed reveals a combined thermo-magneto-mechanical effect, which has been previously observed in other systems \cite{liu2014-thermal,Liu2015_thermal,zhang2016_thermal_substrate_EB}. Such phenomena may be attributed to the difference in the elastic properties of both glass and PMN-PZT substrates \cite{Burdin2012-temperature,wang2009}.



\section{Conclusion}

In summary, we demonstrate an electrical modulation at room temperature of magnetic properties in an extrinsic multiferroic fabricated by sputtering polycrystalline Fe$_{81}$Ga$_{19}$ thin films onto (011)-cut and poled PMN-PZT ferroelectric single crystals. 

In-plane azimuthal magnetization reversal measurements enabled to assess, under an applied electric field, the modification of the magnetic anisotropy character : the cubic component in thinner films (5 nm) shifts towards a predominant uniaxial component. This shift is accompanied by a 90$\degree$ rotation of the as-deposited preferential anisotropy axis, and a relative remanent magnetization change of about 80\%. In thicker films (60 nm), the random anisotropy character vanishes under $E$ leaving out a coexisting cubic and uniaxial character.

With the assistance of small magnetic fields near $\mu_0H$\textsubscript{c} values, an electric field is capable of switching the magnetization state, and achieve a converse magnetoelectric coupling coefficient $\alpha\textsubscript{CME}=2.7\times{10^{-6}}$ s.m$^{-1}$ at a given angle $\varphi=30\degree$ between $\mu_0H$ and the deposition field direction $\mu_0H\textsubscript{dep}$. This sizable ME coupling is attributed to many important factors: the high in-plane piezoelectric coefficients of (011)-PMN-PZT, the reasonable magnetic properties of FeGa such as the high saturation magnetization and the high magnetoelastic coefficients, as well as the angular dependency of FeGa magnetic properties. This reported value of $\alpha$\textsubscript{CME} is among the highest previously reported values in single-phase multiferroics or similar composites. We have also shed light for the first time on an azimuthal behavior of the $\alpha$\textsubscript{CME} driven by, not only the anisotropic ferroelectric substrate, but also the anisotropic FeGa.

This ME multiferroic composite promises good results regarding the self-bias effect in a dynamic measurement to achieve remanent CME, at zero magnetic bias field. Such highly performant combined materials may increase the pace towards novel multifunctional devices such as microwave devices, where anisotropy is required, and electrically tunable magnetoelectric memories.

We also find that the FeGa anisotropy characters manifest the same at room temperature on two different types of substrates, i.e. amorphous glass and single-crystalline ferroelectric PMN-PZT. However at low temperatures the mechanical nature of the substrate strongly influences the magnetic behavior. This provides a useful insight into the bonding nature of magnetostrictive and piezoelectric materials, and ultimately into the performance of CME composites in complex environments involving cryogenic temperatures.

\begin{acknowledgments}
We wish to acknowledge the support of Region Bretagne (ARED), France, in co-funding the PhD of W.J. This work was also supported by South African National Research Foundation (Grant No 80880) and the URC and FRC of the University of Johannesburg, South Africa.
\end{acknowledgments}

\bibliography{apssamp}

\begin{thebibliography}{100}%
\makeatletter
\providecommand \@ifxundefined [1]{%
 \@ifx{#1\undefined}
}%
\providecommand \@ifnum [1]{%
 \ifnum #1\expandafter \@firstoftwo
 \else \expandafter \@secondoftwo
 \fi
}%
\providecommand \@ifx [1]{%
 \ifx #1\expandafter \@firstoftwo
 \else \expandafter \@secondoftwo
 \fi
}%
\providecommand \natexlab [1]{#1}%
\providecommand \enquote  [1]{``#1''}%
\providecommand \bibnamefont  [1]{#1}%
\providecommand \bibfnamefont [1]{#1}%
\providecommand \citenamefont [1]{#1}%
\providecommand \href@noop [0]{\@secondoftwo}%
\providecommand \href [0]{\begingroup \@sanitize@url \@href}%
\providecommand \@href[1]{\@@startlink{#1}\@@href}%
\providecommand \@@href[1]{\endgroup#1\@@endlink}%
\providecommand \@sanitize@url [0]{\catcode `\\12\catcode `\$12\catcode
  `\&12\catcode `\#12\catcode `\^12\catcode `\_12\catcode `\%12\relax}%
\providecommand \@@startlink[1]{}%
\providecommand \@@endlink[0]{}%
\providecommand \url  [0]{\begingroup\@sanitize@url \@url }%
\providecommand \@url [1]{\endgroup\@href {#1}{\urlprefix }}%
\providecommand \urlprefix  [0]{URL }%
\providecommand \Eprint [0]{\href }%
\providecommand \doibase [0]{https://doi.org/}%
\providecommand \selectlanguage [0]{\@gobble}%
\providecommand \bibinfo  [0]{\@secondoftwo}%
\providecommand \bibfield  [0]{\@secondoftwo}%
\providecommand \translation [1]{[#1]}%
\providecommand \BibitemOpen [0]{}%
\providecommand \bibitemStop [0]{}%
\providecommand \bibitemNoStop [0]{.\EOS\space}%
\providecommand \EOS [0]{\spacefactor3000\relax}%
\providecommand \BibitemShut  [1]{\csname bibitem#1\endcsname}%
\let\auto@bib@innerbib\@empty
\bibitem [{\citenamefont {Spaldin}\ and\ \citenamefont
  {Fiebig}(2005)}]{Spaldin05}%
  \BibitemOpen
  \bibfield  {author} {\bibinfo {author} {\bibfnamefont {N.~A.}\ \bibnamefont
  {Spaldin}}\ and\ \bibinfo {author} {\bibfnamefont {M.}~\bibnamefont
  {Fiebig}},\ }\bibfield  {title} {\bibinfo {title} {The renaissance of
  magnetoelectric multiferroics},\ }\href@noop {} {\bibfield  {journal}
  {\bibinfo  {journal} {Science}\ }\textbf {\bibinfo {volume} {309}},\ \bibinfo
  {pages} {391} (\bibinfo {year} {2005})}\BibitemShut {NoStop}%
\bibitem [{\citenamefont {Fusil}\ \emph {et~al.}(2014)\citenamefont {Fusil},
  \citenamefont {Garcia}, \citenamefont {Barth\'{e}l\'{e}my},\ and\
  \citenamefont {Bibes}}]{Fusil2014}%
  \BibitemOpen
  \bibfield  {author} {\bibinfo {author} {\bibfnamefont {S.}~\bibnamefont
  {Fusil}}, \bibinfo {author} {\bibfnamefont {V.}~\bibnamefont {Garcia}},
  \bibinfo {author} {\bibfnamefont {A.}~\bibnamefont {Barth\'{e}l\'{e}my}},\
  and\ \bibinfo {author} {\bibfnamefont {M.}~\bibnamefont {Bibes}},\ }\bibfield
   {title} {\bibinfo {title} {Magnetoelectric devices for spintronics},\ }\href
  {https://doi.org/10.1146/annurev-matsci-070813-113315} {\bibfield  {journal}
  {\bibinfo  {journal} {Annual Review of Materials Research}\ }\textbf
  {\bibinfo {volume} {44}},\ \bibinfo {pages} {91} (\bibinfo {year} {2014})},\
  \Eprint
  {https://arxiv.org/abs/https://doi.org/10.1146/annurev-matsci-070813-113315}
  {https://doi.org/10.1146/annurev-matsci-070813-113315} \BibitemShut {NoStop}%
\bibitem [{\citenamefont {Eerenstein}\ \emph {et~al.}(2006)\citenamefont
  {Eerenstein}, \citenamefont {Mathur},\ and\ \citenamefont
  {Scott}}]{eerenstein06}%
  \BibitemOpen
  \bibfield  {author} {\bibinfo {author} {\bibfnamefont {W.}~\bibnamefont
  {Eerenstein}}, \bibinfo {author} {\bibfnamefont {N.~D.}\ \bibnamefont
  {Mathur}},\ and\ \bibinfo {author} {\bibfnamefont {J.~F.}\ \bibnamefont
  {Scott}},\ }\bibfield  {title} {\bibinfo {title} {Multiferroic and
  magnetoelectric materials},\ }\href {http://dx.doi.org/10.1038/nature05023}
  {\bibfield  {journal} {\bibinfo  {journal} {Nature}\ }\textbf {\bibinfo
  {volume} {442}},\ \bibinfo {pages} {759} (\bibinfo {year}
  {2006})}\BibitemShut {NoStop}%
\bibitem [{\citenamefont {Matsukura}\ \emph {et~al.}(2015)\citenamefont
  {Matsukura}, \citenamefont {Tokura},\ and\ \citenamefont
  {Ohno}}]{Matsukura15}%
  \BibitemOpen
  \bibfield  {author} {\bibinfo {author} {\bibfnamefont {F.}~\bibnamefont
  {Matsukura}}, \bibinfo {author} {\bibfnamefont {Y.}~\bibnamefont {Tokura}},\
  and\ \bibinfo {author} {\bibfnamefont {H.}~\bibnamefont {Ohno}},\ }\bibfield
  {title} {\bibinfo {title} {Control of magnetism by electric fields},\ }\href
  {http://dx.doi.org/10.1038/nnano.2015.22} {\bibfield  {journal} {\bibinfo
  {journal} {Nature Nanotechnology}\ }\textbf {\bibinfo {volume} {10}},\
  \bibinfo {pages} {209} (\bibinfo {year} {2015})}\BibitemShut {NoStop}%
\bibitem [{\citenamefont {Hu}\ \emph {et~al.}(2011)\citenamefont {Hu},
  \citenamefont {Li}, \citenamefont {Chen},\ and\ \citenamefont
  {Nan}}]{Hu2011}%
  \BibitemOpen
  \bibfield  {author} {\bibinfo {author} {\bibfnamefont {J.-M.}\ \bibnamefont
  {Hu}}, \bibinfo {author} {\bibfnamefont {Z.}~\bibnamefont {Li}}, \bibinfo
  {author} {\bibfnamefont {L.-Q.}\ \bibnamefont {Chen}},\ and\ \bibinfo
  {author} {\bibfnamefont {C.-W.}\ \bibnamefont {Nan}},\ }\bibfield  {title}
  {\bibinfo {title} {High-density magnetoresistive random access memory
  operating at ultralow voltage at room temperature},\ }\bibfield  {journal}
  {\bibinfo  {journal} {Nature Communications}\ }\textbf {\bibinfo {volume}
  {2}},\ \href {https://doi.org/10.1038/ncomms1564} {10.1038/ncomms1564}
  (\bibinfo {year} {2011})\BibitemShut {NoStop}%
\bibitem [{\citenamefont {Chiba}\ \emph {et~al.}(2008)\citenamefont {Chiba},
  \citenamefont {Sawicki}, \citenamefont {Nishitani}, \citenamefont {Nakatani},
  \citenamefont {Matsukura},\ and\ \citenamefont {Ohno}}]{Chiba2008}%
  \BibitemOpen
  \bibfield  {author} {\bibinfo {author} {\bibfnamefont {D.}~\bibnamefont
  {Chiba}}, \bibinfo {author} {\bibfnamefont {M.}~\bibnamefont {Sawicki}},
  \bibinfo {author} {\bibfnamefont {Y.}~\bibnamefont {Nishitani}}, \bibinfo
  {author} {\bibfnamefont {Y.}~\bibnamefont {Nakatani}}, \bibinfo {author}
  {\bibfnamefont {F.}~\bibnamefont {Matsukura}},\ and\ \bibinfo {author}
  {\bibfnamefont {H.}~\bibnamefont {Ohno}},\ }\bibfield  {title} {\bibinfo
  {title} {Magnetization vector manipulation by electric fields},\ }\href@noop
  {} {\bibfield  {journal} {\bibinfo  {journal} {Nature}\ }\textbf {\bibinfo
  {volume} {455}},\ \bibinfo {pages} {515} (\bibinfo {year}
  {2008})}\BibitemShut {NoStop}%
\bibitem [{\citenamefont {Heron}\ \emph {et~al.}(2011)\citenamefont {Heron},
  \citenamefont {Trassin}, \citenamefont {Ashraf}, \citenamefont {Gajek},
  \citenamefont {He}, \citenamefont {Yang}, \citenamefont {Nikonov},
  \citenamefont {Chu}, \citenamefont {Salahuddin},\ and\ \citenamefont
  {Ramesh}}]{Heron2011}%
  \BibitemOpen
  \bibfield  {author} {\bibinfo {author} {\bibfnamefont {J.~T.}\ \bibnamefont
  {Heron}}, \bibinfo {author} {\bibfnamefont {M.}~\bibnamefont {Trassin}},
  \bibinfo {author} {\bibfnamefont {K.}~\bibnamefont {Ashraf}}, \bibinfo
  {author} {\bibfnamefont {M.}~\bibnamefont {Gajek}}, \bibinfo {author}
  {\bibfnamefont {Q.}~\bibnamefont {He}}, \bibinfo {author} {\bibfnamefont
  {S.~Y.}\ \bibnamefont {Yang}}, \bibinfo {author} {\bibfnamefont {D.~E.}\
  \bibnamefont {Nikonov}}, \bibinfo {author} {\bibfnamefont {Y.-H.}\
  \bibnamefont {Chu}}, \bibinfo {author} {\bibfnamefont {S.}~\bibnamefont
  {Salahuddin}},\ and\ \bibinfo {author} {\bibfnamefont {R.}~\bibnamefont
  {Ramesh}},\ }\bibfield  {title} {\bibinfo {title}
  {Electric-{{Field}}-{{Induced Magnetization Reversal}} in a
  {{Ferromagnet}}-{{Multiferroic Heterostructure}}},\ }\href
  {10.1103/PhysRevLett.107.217202} {\bibfield  {journal} {\bibinfo  {journal}
  {Phys. Rev. Lett.}\ }\textbf {\bibinfo {volume} {107}} (\bibinfo {year}
  {2011})}\BibitemShut {NoStop}%
\bibitem [{\citenamefont {Eerenstein}\ \emph {et~al.}(2007)\citenamefont
  {Eerenstein}, \citenamefont {Wiora}, \citenamefont {Prieto}, \citenamefont
  {Scott},\ and\ \citenamefont {Mathur}}]{Eerenstein2007}%
  \BibitemOpen
  \bibfield  {author} {\bibinfo {author} {\bibfnamefont {W.}~\bibnamefont
  {Eerenstein}}, \bibinfo {author} {\bibfnamefont {M.}~\bibnamefont {Wiora}},
  \bibinfo {author} {\bibfnamefont {J.~L.}\ \bibnamefont {Prieto}}, \bibinfo
  {author} {\bibfnamefont {J.~F.}\ \bibnamefont {Scott}},\ and\ \bibinfo
  {author} {\bibfnamefont {N.~D.}\ \bibnamefont {Mathur}},\ }\bibfield  {title}
  {\bibinfo {title} {Giant sharp and persistent converse magnetoelectric
  effects in multiferroic epitaxial heterostructures},\ }\href
  {https://doi.org/10.1038/nmat1886} {\bibfield  {journal} {\bibinfo  {journal}
  {Nature Materials}\ }\textbf {\bibinfo {volume} {6}},\ \bibinfo {pages} {348}
  (\bibinfo {year} {2007})}\BibitemShut {NoStop}%
\bibitem [{\citenamefont {Zhao}\ \emph {et~al.}(2006)\citenamefont {Zhao},
  \citenamefont {Scholl}, \citenamefont {Zavaliche}, \citenamefont {Lee},
  \citenamefont {Barry}, \citenamefont {Doran}, \citenamefont {Cruz},
  \citenamefont {Chu}, \citenamefont {Ederer}, \citenamefont {Spaldin},
  \citenamefont {Das}, \citenamefont {Kim}, \citenamefont {Baek}, \citenamefont
  {Eom},\ and\ \citenamefont {Ramesh}}]{Zhao2006}%
  \BibitemOpen
  \bibfield  {author} {\bibinfo {author} {\bibfnamefont {T.}~\bibnamefont
  {Zhao}}, \bibinfo {author} {\bibfnamefont {A.}~\bibnamefont {Scholl}},
  \bibinfo {author} {\bibfnamefont {F.}~\bibnamefont {Zavaliche}}, \bibinfo
  {author} {\bibfnamefont {K.}~\bibnamefont {Lee}}, \bibinfo {author}
  {\bibfnamefont {M.}~\bibnamefont {Barry}}, \bibinfo {author} {\bibfnamefont
  {A.}~\bibnamefont {Doran}}, \bibinfo {author} {\bibfnamefont {M.~P.}\
  \bibnamefont {Cruz}}, \bibinfo {author} {\bibfnamefont {Y.~H.}\ \bibnamefont
  {Chu}}, \bibinfo {author} {\bibfnamefont {C.}~\bibnamefont {Ederer}},
  \bibinfo {author} {\bibfnamefont {N.~A.}\ \bibnamefont {Spaldin}}, \bibinfo
  {author} {\bibfnamefont {R.~R.}\ \bibnamefont {Das}}, \bibinfo {author}
  {\bibfnamefont {D.~M.}\ \bibnamefont {Kim}}, \bibinfo {author} {\bibfnamefont
  {S.~H.}\ \bibnamefont {Baek}}, \bibinfo {author} {\bibfnamefont {C.~B.}\
  \bibnamefont {Eom}},\ and\ \bibinfo {author} {\bibfnamefont {R.}~\bibnamefont
  {Ramesh}},\ }\bibfield  {title} {\bibinfo {title} {Electrical control of
  antiferromagnetic domains in multiferroic {{BiFeO3}} films at room
  temperature},\ }\href {http://dx.doi.org/10.1038/nmat1731} {\bibfield
  {journal} {\bibinfo  {journal} {Nature Materials}\ }\textbf {\bibinfo
  {volume} {5}},\ \bibinfo {pages} {823} (\bibinfo {year} {2006})}\BibitemShut
  {NoStop}%
\bibitem [{\citenamefont {Hur}\ \emph {et~al.}(2004)\citenamefont {Hur},
  \citenamefont {Park}, \citenamefont {Sharma}, \citenamefont {Ahn},
  \citenamefont {Guha},\ and\ \citenamefont {Cheong}}]{hur2004}%
  \BibitemOpen
  \bibfield  {author} {\bibinfo {author} {\bibfnamefont {N.}~\bibnamefont
  {Hur}}, \bibinfo {author} {\bibfnamefont {S.}~\bibnamefont {Park}}, \bibinfo
  {author} {\bibfnamefont {P.~A.}\ \bibnamefont {Sharma}}, \bibinfo {author}
  {\bibfnamefont {J.~S.}\ \bibnamefont {Ahn}}, \bibinfo {author} {\bibfnamefont
  {S.}~\bibnamefont {Guha}},\ and\ \bibinfo {author} {\bibfnamefont {S.-W.}\
  \bibnamefont {Cheong}},\ }\bibfield  {title} {\bibinfo {title} {Electric
  polarization reversal and memory in a multiferroic material induced by
  magnetic fields},\ }\href {https://doi.org/10.1038/nature02572} {\bibfield
  {journal} {\bibinfo  {journal} {Nature}\ }\textbf {\bibinfo {volume} {429}},\
  \bibinfo {pages} {392} (\bibinfo {year} {2004})}\BibitemShut {NoStop}%
\bibitem [{\citenamefont {Catalan}\ and\ \citenamefont
  {Scott}(2009)}]{Catalan09}%
  \BibitemOpen
  \bibfield  {author} {\bibinfo {author} {\bibfnamefont {G.}~\bibnamefont
  {Catalan}}\ and\ \bibinfo {author} {\bibfnamefont {J.~F.}\ \bibnamefont
  {Scott}},\ }\bibfield  {title} {\bibinfo {title} {Physics and applications of
  bismuth ferrite},\ }\href {10.1002/adma.200802849} {\bibfield  {journal}
  {\bibinfo  {journal} {Advanced Materials}\ }\textbf {\bibinfo {volume}
  {21}},\ \bibinfo {pages} {2463} (\bibinfo {year} {2009})}\BibitemShut
  {NoStop}%
\bibitem [{\citenamefont {Hauguel}\ \emph {et~al.}(2012)\citenamefont
  {Hauguel}, \citenamefont {Pogossian}, \citenamefont {Dekadjevi},
  \citenamefont {Spenato}, \citenamefont {Jay},\ and\ \citenamefont
  {Ben~Youssef}}]{tony12}%
  \BibitemOpen
  \bibfield  {author} {\bibinfo {author} {\bibfnamefont {T.}~\bibnamefont
  {Hauguel}}, \bibinfo {author} {\bibfnamefont {S.~P.}\ \bibnamefont
  {Pogossian}}, \bibinfo {author} {\bibfnamefont {D.~T.}\ \bibnamefont
  {Dekadjevi}}, \bibinfo {author} {\bibfnamefont {D.}~\bibnamefont {Spenato}},
  \bibinfo {author} {\bibfnamefont {J.-P.}\ \bibnamefont {Jay}},\ and\ \bibinfo
  {author} {\bibfnamefont {J.}~\bibnamefont {Ben~Youssef}},\ }\bibfield
  {title} {\bibinfo {title} {Driving mechanism of exchange bias and magnetic
  anisotropy in multiferroic polycrystalline {BiFeO}3/permalloy bilayers},\
  }\href {10.1063/1.4763480} {\bibfield  {journal} {\bibinfo  {journal}
  {Journal of Applied Physics}\ }\textbf {\bibinfo {volume} {112}},\ \bibinfo
  {pages} {093904} (\bibinfo {year} {2012})}\BibitemShut {NoStop}%
\bibitem [{\citenamefont {Richy}\ \emph {et~al.}(2018)\citenamefont {Richy},
  \citenamefont {Hauguel}, \citenamefont {Jay}, \citenamefont {Pogossian},
  \citenamefont {{Warot-Fonrose}}, \citenamefont {Sheppard}, \citenamefont
  {Snyman}, \citenamefont {Strydom}, \citenamefont {Youssef}, \citenamefont
  {Prinsloo}, \citenamefont {Spenato},\ and\ \citenamefont
  {Dekadjevi}}]{Richy2018}%
  \BibitemOpen
  \bibfield  {author} {\bibinfo {author} {\bibfnamefont {J.}~\bibnamefont
  {Richy}}, \bibinfo {author} {\bibfnamefont {T.}~\bibnamefont {Hauguel}},
  \bibinfo {author} {\bibfnamefont {J.-P.}\ \bibnamefont {Jay}}, \bibinfo
  {author} {\bibfnamefont {S.~P.}\ \bibnamefont {Pogossian}}, \bibinfo {author}
  {\bibfnamefont {B.}~\bibnamefont {{Warot-Fonrose}}}, \bibinfo {author}
  {\bibfnamefont {C.~J.}\ \bibnamefont {Sheppard}}, \bibinfo {author}
  {\bibfnamefont {J.~L.}\ \bibnamefont {Snyman}}, \bibinfo {author}
  {\bibfnamefont {A.~M.}\ \bibnamefont {Strydom}}, \bibinfo {author}
  {\bibfnamefont {J.~B.}\ \bibnamefont {Youssef}}, \bibinfo {author}
  {\bibfnamefont {A.~R.~E.}\ \bibnamefont {Prinsloo}}, \bibinfo {author}
  {\bibfnamefont {D.}~\bibnamefont {Spenato}},\ and\ \bibinfo {author}
  {\bibfnamefont {D.~T.}\ \bibnamefont {Dekadjevi}},\ }\bibfield  {title}
  {\bibinfo {title} {Temperature dependence of exchange biased multiferroic
  {{BiFeO}} {\textsubscript{3}} /{{Ni}} {\textsubscript{81}} {{Fe}}
  {\textsubscript{19}} polycrystalline bilayer"},\ }\href
  {10.1088/1361-6463/aab023} {\bibfield  {journal} {\bibinfo  {journal} {J.
  Phys. Appl. Phys.}\ }\textbf {\bibinfo {volume} {51}},\ \bibinfo {pages}
  {125308} (\bibinfo {year} {2018})}\BibitemShut {NoStop}%
\bibitem [{\citenamefont {Jahjah}\ \emph {et~al.}(2018)\citenamefont {Jahjah},
  \citenamefont {Jay}, \citenamefont {Le~Grand}, \citenamefont {Fessant},
  \citenamefont {Richy}, \citenamefont {Marcelot}, \citenamefont
  {{Warot-Fonrose}}, \citenamefont {Prinsloo}, \citenamefont {Sheppard},
  \citenamefont {Dekadjevi},\ and\ \citenamefont {Spenato}}]{Jahjah2018}%
  \BibitemOpen
  \bibfield  {author} {\bibinfo {author} {\bibfnamefont {W.}~\bibnamefont
  {Jahjah}}, \bibinfo {author} {\bibfnamefont {J.-P.}\ \bibnamefont {Jay}},
  \bibinfo {author} {\bibfnamefont {Y.}~\bibnamefont {Le~Grand}}, \bibinfo
  {author} {\bibfnamefont {A.}~\bibnamefont {Fessant}}, \bibinfo {author}
  {\bibfnamefont {J.}~\bibnamefont {Richy}}, \bibinfo {author} {\bibfnamefont
  {C.}~\bibnamefont {Marcelot}}, \bibinfo {author} {\bibfnamefont
  {B.}~\bibnamefont {{Warot-Fonrose}}}, \bibinfo {author} {\bibfnamefont
  {A.~R.~E.}\ \bibnamefont {Prinsloo}}, \bibinfo {author} {\bibfnamefont
  {C.~J.}\ \bibnamefont {Sheppard}}, \bibinfo {author} {\bibfnamefont {D.~T.}\
  \bibnamefont {Dekadjevi}},\ and\ \bibinfo {author} {\bibfnamefont
  {D.}~\bibnamefont {Spenato}},\ }\bibfield  {title} {\bibinfo {title}
  {Influence of mesoporous or parasitic {{BiFeO}} 3 structural state on the
  magnetization reversal in multiferroic {{BiFeO}} 3/{{Ni 81Fe}} 19
  polycrystalline bilayers},\ }\href {https://doi.org/10.1063/1.5049546}
  {\bibfield  {journal} {\bibinfo  {journal} {Journal of Applied Physics}\
  }\textbf {\bibinfo {volume} {124}},\ \bibinfo {pages} {235309} (\bibinfo
  {year} {2018})}\BibitemShut {NoStop}%
\bibitem [{\citenamefont {Palneedi}\ \emph {et~al.}(2016)\citenamefont
  {Palneedi}, \citenamefont {Annapureddy}, \citenamefont {Priya},\ and\
  \citenamefont {Ryu}}]{Palneedi2016}%
  \BibitemOpen
  \bibfield  {author} {\bibinfo {author} {\bibfnamefont {H.}~\bibnamefont
  {Palneedi}}, \bibinfo {author} {\bibfnamefont {V.}~\bibnamefont
  {Annapureddy}}, \bibinfo {author} {\bibfnamefont {S.}~\bibnamefont {Priya}},\
  and\ \bibinfo {author} {\bibfnamefont {J.}~\bibnamefont {Ryu}},\ }\bibfield
  {title} {\bibinfo {title} {Status and {{Perspectives}} of {{Multiferroic
  Magnetoelectric Composite Materials}} and {{Applications}}},\ }\href
  {https://doi.org/10.3390/act5010009} {\bibfield  {journal} {\bibinfo
  {journal} {Actuators}\ }\textbf {\bibinfo {volume} {5}},\ \bibinfo {pages}
  {9} (\bibinfo {year} {2016})}\BibitemShut {NoStop}%
\bibitem [{\citenamefont {Wang}\ \emph {et~al.}(2010)\citenamefont {Wang},
  \citenamefont {Hu}, \citenamefont {Lin},\ and\ \citenamefont
  {Nan}}]{Wang2010}%
  \BibitemOpen
  \bibfield  {author} {\bibinfo {author} {\bibfnamefont {Y.}~\bibnamefont
  {Wang}}, \bibinfo {author} {\bibfnamefont {J.}~\bibnamefont {Hu}}, \bibinfo
  {author} {\bibfnamefont {Y.}~\bibnamefont {Lin}},\ and\ \bibinfo {author}
  {\bibfnamefont {C.-W.}\ \bibnamefont {Nan}},\ }\bibfield  {title} {\bibinfo
  {title} {Multiferroic magnetoelectric composite nanostructures},\ }\href
  {https://doi.org/10.1038/asiamat.2010.32} {\bibfield  {journal} {\bibinfo
  {journal} {NPG Asia Materials}\ }\textbf {\bibinfo {volume} {2}},\ \bibinfo
  {pages} {61} (\bibinfo {year} {2010})}\BibitemShut {NoStop}%
\bibitem [{\citenamefont {Srinivasan}(2010)}]{Srinivasan2010}%
  \BibitemOpen
  \bibfield  {author} {\bibinfo {author} {\bibfnamefont {G.}~\bibnamefont
  {Srinivasan}},\ }\bibfield  {title} {\bibinfo {title} {Magnetoelectric
  {{Composites}}},\ }\href
  {https://doi.org/10.1146/annurev-matsci-070909-104459} {\bibfield  {journal}
  {\bibinfo  {journal} {Annual Review of Materials Research}\ }\textbf
  {\bibinfo {volume} {40}},\ \bibinfo {pages} {153} (\bibinfo {year}
  {2010})}\BibitemShut {NoStop}%
\bibitem [{\citenamefont {Nan}\ \emph {et~al.}(2008)\citenamefont {Nan},
  \citenamefont {Bichurin}, \citenamefont {Dong}, \citenamefont {Viehland},\
  and\ \citenamefont {Srinivasan}}]{Nan2008}%
  \BibitemOpen
  \bibfield  {author} {\bibinfo {author} {\bibfnamefont {C.-W.}\ \bibnamefont
  {Nan}}, \bibinfo {author} {\bibfnamefont {M.~I.}\ \bibnamefont {Bichurin}},
  \bibinfo {author} {\bibfnamefont {S.}~\bibnamefont {Dong}}, \bibinfo {author}
  {\bibfnamefont {D.}~\bibnamefont {Viehland}},\ and\ \bibinfo {author}
  {\bibfnamefont {G.}~\bibnamefont {Srinivasan}},\ }\bibfield  {title}
  {\bibinfo {title} {Multiferroic magnetoelectric composites: {{Historical}}
  perspective, status, and future directions},\ }\href
  {https://doi.org/10.1063/1.2836410} {\bibfield  {journal} {\bibinfo
  {journal} {Journal of Applied Physics}\ }\textbf {\bibinfo {volume} {103}},\
  \bibinfo {pages} {031101} (\bibinfo {year} {2008})}\BibitemShut {NoStop}%
\bibitem [{\citenamefont {Vaz}(2012)}]{vaz2012}%
  \BibitemOpen
  \bibfield  {author} {\bibinfo {author} {\bibfnamefont {C.~A.~F.}\
  \bibnamefont {Vaz}},\ }\bibfield  {title} {\bibinfo {title} {Electric field
  control of magnetism in multiferroic heterostructures},\ }\href
  {http://stacks.iop.org/0953-8984/24/i=33/a=333201} {\bibfield  {journal}
  {\bibinfo  {journal} {Journal of Physics: Condensed Matter}\ }\textbf
  {\bibinfo {volume} {24}},\ \bibinfo {pages} {333201} (\bibinfo {year}
  {2012})}\BibitemShut {NoStop}%
\bibitem [{\citenamefont {Nan}\ \emph {et~al.}(2014)\citenamefont {Nan},
  \citenamefont {Zhou}, \citenamefont {Liu}, \citenamefont {Yang},
  \citenamefont {Gao}, \citenamefont {Assaf}, \citenamefont {Lin},
  \citenamefont {Velu}, \citenamefont {Wang}, \citenamefont {Luo},
  \citenamefont {Chen}, \citenamefont {Akhtar}, \citenamefont {Hu},
  \citenamefont {Rajiv}, \citenamefont {Krishnan}, \citenamefont {Sreedhar},
  \citenamefont {Heiman}, \citenamefont {Howe}, \citenamefont {Brown},\ and\
  \citenamefont {Sun}}]{Nan2014}%
  \BibitemOpen
  \bibfield  {author} {\bibinfo {author} {\bibfnamefont {T.}~\bibnamefont
  {Nan}}, \bibinfo {author} {\bibfnamefont {Z.}~\bibnamefont {Zhou}}, \bibinfo
  {author} {\bibfnamefont {M.}~\bibnamefont {Liu}}, \bibinfo {author}
  {\bibfnamefont {X.}~\bibnamefont {Yang}}, \bibinfo {author} {\bibfnamefont
  {Y.}~\bibnamefont {Gao}}, \bibinfo {author} {\bibfnamefont {B.~A.}\
  \bibnamefont {Assaf}}, \bibinfo {author} {\bibfnamefont {H.}~\bibnamefont
  {Lin}}, \bibinfo {author} {\bibfnamefont {S.}~\bibnamefont {Velu}}, \bibinfo
  {author} {\bibfnamefont {X.}~\bibnamefont {Wang}}, \bibinfo {author}
  {\bibfnamefont {H.}~\bibnamefont {Luo}}, \bibinfo {author} {\bibfnamefont
  {J.}~\bibnamefont {Chen}}, \bibinfo {author} {\bibfnamefont {S.}~\bibnamefont
  {Akhtar}}, \bibinfo {author} {\bibfnamefont {E.}~\bibnamefont {Hu}}, \bibinfo
  {author} {\bibfnamefont {R.}~\bibnamefont {Rajiv}}, \bibinfo {author}
  {\bibfnamefont {K.}~\bibnamefont {Krishnan}}, \bibinfo {author}
  {\bibfnamefont {S.}~\bibnamefont {Sreedhar}}, \bibinfo {author}
  {\bibfnamefont {D.}~\bibnamefont {Heiman}}, \bibinfo {author} {\bibfnamefont
  {B.~M.}\ \bibnamefont {Howe}}, \bibinfo {author} {\bibfnamefont {G.~J.}\
  \bibnamefont {Brown}},\ and\ \bibinfo {author} {\bibfnamefont {N.~X.}\
  \bibnamefont {Sun}},\ }\bibfield  {title} {\bibinfo {title} {Quantification
  of strain and charge co-mediated magnetoelectric coupling on ultra-thin
  {{Permalloy}}/{{PMN}}-{{PT}} interface},\ }\href
  {https://doi.org/10.1038/srep03688} {\bibfield  {journal} {\bibinfo
  {journal} {Scientific Reports}\ }\textbf {\bibinfo {volume} {4}},\ \bibinfo
  {pages} {3688} (\bibinfo {year} {2014})}\BibitemShut {NoStop}%
\bibitem [{\citenamefont {Zhou}\ \emph {et~al.}(2015)\citenamefont {Zhou},
  \citenamefont {Howe}, \citenamefont {Liu}, \citenamefont {Nan}, \citenamefont
  {Chen}, \citenamefont {Mahalingam}, \citenamefont {Sun},\ and\ \citenamefont
  {Brown}}]{zhou2015}%
  \BibitemOpen
  \bibfield  {author} {\bibinfo {author} {\bibfnamefont {Z.}~\bibnamefont
  {Zhou}}, \bibinfo {author} {\bibfnamefont {B.~M.}\ \bibnamefont {Howe}},
  \bibinfo {author} {\bibfnamefont {M.}~\bibnamefont {Liu}}, \bibinfo {author}
  {\bibfnamefont {T.}~\bibnamefont {Nan}}, \bibinfo {author} {\bibfnamefont
  {X.}~\bibnamefont {Chen}}, \bibinfo {author} {\bibfnamefont {K.}~\bibnamefont
  {Mahalingam}}, \bibinfo {author} {\bibfnamefont {N.~X.}\ \bibnamefont
  {Sun}},\ and\ \bibinfo {author} {\bibfnamefont {G.~J.}\ \bibnamefont
  {Brown}},\ }\bibfield  {title} {\bibinfo {title} {Interfacial charge-mediated
  non-volatile magnetoelectric coupling in
  {{Co0}}.{{3Fe0}}.7/{{Ba0}}.{{6Sr0}}.{{4TiO3}}/{{Nb}}:{{SrTiO3}} multiferroic
  heterostructures},\ }\href@noop {} {\bibfield  {journal} {\bibinfo  {journal}
  {Scientific Reports}\ }\textbf {\bibinfo {volume} {5}},\ \bibinfo {pages}
  {7740} (\bibinfo {year} {2015})}\BibitemShut {NoStop}%
\bibitem [{\citenamefont {Laukhin}\ \emph {et~al.}(2006)\citenamefont
  {Laukhin}, \citenamefont {Skumryev}, \citenamefont {Mart\'{\i}},
  \citenamefont {Hrabovsky}, \citenamefont {S\'anchez}, \citenamefont
  {Garc\'{\i}a-Cuenca}, \citenamefont {Ferrater}, \citenamefont {Varela},
  \citenamefont {L\"uders}, \citenamefont {Bobo},\ and\ \citenamefont
  {Fontcuberta}}]{Laukhin2006}%
  \BibitemOpen
  \bibfield  {author} {\bibinfo {author} {\bibfnamefont {V.}~\bibnamefont
  {Laukhin}}, \bibinfo {author} {\bibfnamefont {V.}~\bibnamefont {Skumryev}},
  \bibinfo {author} {\bibfnamefont {X.}~\bibnamefont {Mart\'{\i}}}, \bibinfo
  {author} {\bibfnamefont {D.}~\bibnamefont {Hrabovsky}}, \bibinfo {author}
  {\bibfnamefont {F.}~\bibnamefont {S\'anchez}}, \bibinfo {author}
  {\bibfnamefont {M.~V.}\ \bibnamefont {Garc\'{\i}a-Cuenca}}, \bibinfo {author}
  {\bibfnamefont {C.}~\bibnamefont {Ferrater}}, \bibinfo {author}
  {\bibfnamefont {M.}~\bibnamefont {Varela}}, \bibinfo {author} {\bibfnamefont
  {U.}~\bibnamefont {L\"uders}}, \bibinfo {author} {\bibfnamefont {J.~F.}\
  \bibnamefont {Bobo}},\ and\ \bibinfo {author} {\bibfnamefont
  {J.}~\bibnamefont {Fontcuberta}},\ }\bibfield  {title} {\bibinfo {title}
  {Electric-field control of exchange bias in multiferroic epitaxial
  heterostructures},\ }\href {https://doi.org/10.1103/PhysRevLett.97.227201}
  {\bibfield  {journal} {\bibinfo  {journal} {Phys. Rev. Lett.}\ }\textbf
  {\bibinfo {volume} {97}},\ \bibinfo {pages} {227201} (\bibinfo {year}
  {2006})}\BibitemShut {NoStop}%
\bibitem [{\citenamefont {Yang}\ \emph
  {et~al.}(2015{\natexlab{a}})\citenamefont {Yang}, \citenamefont {Gong},
  \citenamefont {Ma}, \citenamefont {Shen}, \citenamefont {Wang}, \citenamefont
  {Cao}, \citenamefont {Zhong},\ and\ \citenamefont
  {Du}}]{yang2015_EBcontrolV}%
  \BibitemOpen
  \bibfield  {author} {\bibinfo {author} {\bibfnamefont {Y.}~\bibnamefont
  {Yang}}, \bibinfo {author} {\bibfnamefont {Y.}~\bibnamefont {Gong}}, \bibinfo
  {author} {\bibfnamefont {S.}~\bibnamefont {Ma}}, \bibinfo {author}
  {\bibfnamefont {C.}~\bibnamefont {Shen}}, \bibinfo {author} {\bibfnamefont
  {D.}~\bibnamefont {Wang}}, \bibinfo {author} {\bibfnamefont {Q.}~\bibnamefont
  {Cao}}, \bibinfo {author} {\bibfnamefont {Z.}~\bibnamefont {Zhong}},\ and\
  \bibinfo {author} {\bibfnamefont {Y.}~\bibnamefont {Du}},\ }\bibfield
  {title} {\bibinfo {title} {Electric-field control of exchange bias field in a
  {{Mn}} 50.1 {{Ni}} 39.3 {{Sn}} 10.6 /piezoelectric laminate},\ }\href
  {https://doi.org/10.1016/j.jallcom.2014.08.244} {\bibfield  {journal}
  {\bibinfo  {journal} {Journal of Alloys and Compounds}\ }\textbf {\bibinfo
  {volume} {619}},\ \bibinfo {pages} {1} (\bibinfo {year}
  {2015}{\natexlab{a}})}\BibitemShut {NoStop}%
\bibitem [{\citenamefont {Liu}\ \emph {et~al.}(2011)\citenamefont {Liu},
  \citenamefont {Lou}, \citenamefont {Li},\ and\ \citenamefont
  {Sun}}]{liu2011_EB_FeGa}%
  \BibitemOpen
  \bibfield  {author} {\bibinfo {author} {\bibfnamefont {M.}~\bibnamefont
  {Liu}}, \bibinfo {author} {\bibfnamefont {J.}~\bibnamefont {Lou}}, \bibinfo
  {author} {\bibfnamefont {S.}~\bibnamefont {Li}},\ and\ \bibinfo {author}
  {\bibfnamefont {N.~X.}\ \bibnamefont {Sun}},\ }\bibfield  {title} {\bibinfo
  {title} {E-{{Field Control}} of {{Exchange Bias}} and {{Deterministic
  Magnetization Switching}} in {{AFM}}/{{FM}}/{{FE Multiferroic
  Heterostructures}}},\ }\href {https://doi.org/10.1002/adfm.201002485}
  {\bibfield  {journal} {\bibinfo  {journal} {Advanced Functional Materials}\
  }\textbf {\bibinfo {volume} {21}},\ \bibinfo {pages} {2593} (\bibinfo {year}
  {2011})}\BibitemShut {NoStop}%
\bibitem [{\citenamefont {Thiele}\ \emph {et~al.}(2007)\citenamefont {Thiele},
  \citenamefont {D\"orr}, \citenamefont {Bilani}, \citenamefont {R\"odel},\
  and\ \citenamefont {Schultz}}]{Thiele2007}%
  \BibitemOpen
  \bibfield  {author} {\bibinfo {author} {\bibfnamefont {C.}~\bibnamefont
  {Thiele}}, \bibinfo {author} {\bibfnamefont {K.}~\bibnamefont {D\"orr}},
  \bibinfo {author} {\bibfnamefont {O.}~\bibnamefont {Bilani}}, \bibinfo
  {author} {\bibfnamefont {J.}~\bibnamefont {R\"odel}},\ and\ \bibinfo {author}
  {\bibfnamefont {L.}~\bibnamefont {Schultz}},\ }\bibfield  {title} {\bibinfo
  {title} {Influence of strain on the magnetization and magnetoelectric effect
  in {{La}} 0.7 {{A}} 0.3 {{Mn O}} 3 $\slash$ {{PMN}} - {{PT}} ( 001 ) ( {{A}}
  = {{Sr}} , {{Ca}} )},\ }\bibfield  {journal} {\bibinfo  {journal} {Physical
  Review B}\ }\textbf {\bibinfo {volume} {75}},\ \href
  {https://doi.org/10.1103/PhysRevB.75.054408} {10.1103/PhysRevB.75.054408}
  (\bibinfo {year} {2007})\BibitemShut {NoStop}%
\bibitem [{\citenamefont {Yang}\ \emph {et~al.}(2009)\citenamefont {Yang},
  \citenamefont {Zhao}, \citenamefont {Tian}, \citenamefont {Luo},
  \citenamefont {Zhang}, \citenamefont {He},\ and\ \citenamefont
  {Luo}}]{Yang2009}%
  \BibitemOpen
  \bibfield  {author} {\bibinfo {author} {\bibfnamefont {J.~J.}\ \bibnamefont
  {Yang}}, \bibinfo {author} {\bibfnamefont {Y.~G.}\ \bibnamefont {Zhao}},
  \bibinfo {author} {\bibfnamefont {H.~F.}\ \bibnamefont {Tian}}, \bibinfo
  {author} {\bibfnamefont {L.~B.}\ \bibnamefont {Luo}}, \bibinfo {author}
  {\bibfnamefont {H.~Y.}\ \bibnamefont {Zhang}}, \bibinfo {author}
  {\bibfnamefont {Y.~J.}\ \bibnamefont {He}},\ and\ \bibinfo {author}
  {\bibfnamefont {H.~S.}\ \bibnamefont {Luo}},\ }\bibfield  {title} {\bibinfo
  {title} {Electric field manipulation of magnetization at room temperature in
  multiferroic {{CoFe2O4}}/{{Pb}}({{Mg1}}/{{3Nb2}}/3)0.{{7Ti0}}.{{3O3}}
  heterostructures},\ }\href {https://doi.org/10.1063/1.3143622} {\bibfield
  {journal} {\bibinfo  {journal} {Applied Physics Letters}\ }\textbf {\bibinfo
  {volume} {94}},\ \bibinfo {pages} {212504} (\bibinfo {year}
  {2009})}\BibitemShut {NoStop}%
\bibitem [{\citenamefont {Alberca}\ \emph {et~al.}(2015)\citenamefont
  {Alberca}, \citenamefont {Munuera}, \citenamefont {Azpeitia}, \citenamefont
  {Kirby}, \citenamefont {Nemes}, \citenamefont {{Perez-Mu\~noz}},
  \citenamefont {Tornos}, \citenamefont {Mompean}, \citenamefont {Leon},
  \citenamefont {Santamaria},\ and\ \citenamefont
  {{Garcia-Hernandez}}}]{alberca2015}%
  \BibitemOpen
  \bibfield  {author} {\bibinfo {author} {\bibfnamefont {A.}~\bibnamefont
  {Alberca}}, \bibinfo {author} {\bibfnamefont {C.}~\bibnamefont {Munuera}},
  \bibinfo {author} {\bibfnamefont {J.}~\bibnamefont {Azpeitia}}, \bibinfo
  {author} {\bibfnamefont {B.}~\bibnamefont {Kirby}}, \bibinfo {author}
  {\bibfnamefont {N.~M.}\ \bibnamefont {Nemes}}, \bibinfo {author}
  {\bibfnamefont {A.~M.}\ \bibnamefont {{Perez-Mu\~noz}}}, \bibinfo {author}
  {\bibfnamefont {J.}~\bibnamefont {Tornos}}, \bibinfo {author} {\bibfnamefont
  {F.~J.}\ \bibnamefont {Mompean}}, \bibinfo {author} {\bibfnamefont
  {C.}~\bibnamefont {Leon}}, \bibinfo {author} {\bibfnamefont {J.}~\bibnamefont
  {Santamaria}},\ and\ \bibinfo {author} {\bibfnamefont {M.}~\bibnamefont
  {{Garcia-Hernandez}}},\ }\bibfield  {title} {\bibinfo {title} {Phase
  separation enhanced magneto-electric coupling in
  {{La0}}.{{7Ca0}}.{{3MnO3}}/{{BaTiO3}} ultra-thin films},\ }\href
  {https://doi.org/10.1038/srep17926} {\bibfield  {journal} {\bibinfo
  {journal} {Scientific Reports}\ }\textbf {\bibinfo {volume} {5}},\ \bibinfo
  {pages} {17926} (\bibinfo {year} {2015})}\BibitemShut {NoStop}%
\bibitem [{\citenamefont {Staruch}\ \emph {et~al.}(2016)\citenamefont
  {Staruch}, \citenamefont {Gopman}, \citenamefont {Iunin}, \citenamefont
  {Shull}, \citenamefont {Cheng}, \citenamefont {Bussmann},\ and\ \citenamefont
  {Finkel}}]{Staruch2016}%
  \BibitemOpen
  \bibfield  {author} {\bibinfo {author} {\bibfnamefont {M.}~\bibnamefont
  {Staruch}}, \bibinfo {author} {\bibfnamefont {D.~B.}\ \bibnamefont {Gopman}},
  \bibinfo {author} {\bibfnamefont {Y.~L.}\ \bibnamefont {Iunin}}, \bibinfo
  {author} {\bibfnamefont {R.~D.}\ \bibnamefont {Shull}}, \bibinfo {author}
  {\bibfnamefont {S.~F.}\ \bibnamefont {Cheng}}, \bibinfo {author}
  {\bibfnamefont {K.}~\bibnamefont {Bussmann}},\ and\ \bibinfo {author}
  {\bibfnamefont {P.}~\bibnamefont {Finkel}},\ }\bibfield  {title} {\bibinfo
  {title} {Reversible strain control of magnetic anisotropy in magnetoelectric
  heterostructures at room temperature},\ }\bibfield  {journal} {\bibinfo
  {journal} {Scientific Reports}\ }\textbf {\bibinfo {volume} {6}},\ \href
  {https://doi.org/10.1038/srep37429} {10.1038/srep37429} (\bibinfo {year}
  {2016})\BibitemShut {NoStop}%
\bibitem [{\citenamefont {Wu}\ \emph {et~al.}(2011)\citenamefont {Wu},
  \citenamefont {Bur}, \citenamefont {Mohanchandra}, \citenamefont {Wong},
  \citenamefont {Wang}, \citenamefont {Lynch},\ and\ \citenamefont
  {Carman}}]{Wu2011}%
  \BibitemOpen
  \bibfield  {author} {\bibinfo {author} {\bibfnamefont {T.}~\bibnamefont
  {Wu}}, \bibinfo {author} {\bibfnamefont {A.}~\bibnamefont {Bur}}, \bibinfo
  {author} {\bibfnamefont {K.~P.}\ \bibnamefont {Mohanchandra}}, \bibinfo
  {author} {\bibfnamefont {K.}~\bibnamefont {Wong}}, \bibinfo {author}
  {\bibfnamefont {K.~L.}\ \bibnamefont {Wang}}, \bibinfo {author}
  {\bibfnamefont {C.~S.}\ \bibnamefont {Lynch}},\ and\ \bibinfo {author}
  {\bibfnamefont {G.~P.}\ \bibnamefont {Carman}},\ }\bibfield  {title}
  {\bibinfo {title} {Giant electric-field-induced reversible and permanent
  magnetization reorientation on magnetoelectric {{Ni}}/(011)
  [{{Pb}}({{Mg1}}/{{3Nb2}}/3){{O3}}](1-x)\textendash{}[{{PbTiO3}}]x
  heterostructure},\ }\href {https://doi.org/10.1063/1.3534788} {\bibfield
  {journal} {\bibinfo  {journal} {Applied Physics Letters}\ }\textbf {\bibinfo
  {volume} {98}},\ \bibinfo {pages} {012504} (\bibinfo {year}
  {2011})}\BibitemShut {NoStop}%
\bibitem [{\citenamefont {Zhang}\ \emph {et~al.}(2015)\citenamefont {Zhang},
  \citenamefont {Zhao}, \citenamefont {Xiao}, \citenamefont {Wu}, \citenamefont
  {Rizwan}, \citenamefont {Yang}, \citenamefont {Li}, \citenamefont {Wang},
  \citenamefont {Zhu}, \citenamefont {Zhang}, \citenamefont {Jin},\ and\
  \citenamefont {Han}}]{Zhang2014}%
  \BibitemOpen
  \bibfield  {author} {\bibinfo {author} {\bibfnamefont {S.}~\bibnamefont
  {Zhang}}, \bibinfo {author} {\bibfnamefont {Y.}~\bibnamefont {Zhao}},
  \bibinfo {author} {\bibfnamefont {X.}~\bibnamefont {Xiao}}, \bibinfo {author}
  {\bibfnamefont {Y.}~\bibnamefont {Wu}}, \bibinfo {author} {\bibfnamefont
  {S.}~\bibnamefont {Rizwan}}, \bibinfo {author} {\bibfnamefont
  {L.}~\bibnamefont {Yang}}, \bibinfo {author} {\bibfnamefont {P.}~\bibnamefont
  {Li}}, \bibinfo {author} {\bibfnamefont {J.}~\bibnamefont {Wang}}, \bibinfo
  {author} {\bibfnamefont {M.}~\bibnamefont {Zhu}}, \bibinfo {author}
  {\bibfnamefont {H.}~\bibnamefont {Zhang}}, \bibinfo {author} {\bibfnamefont
  {X.}~\bibnamefont {Jin}},\ and\ \bibinfo {author} {\bibfnamefont
  {X.}~\bibnamefont {Han}},\ }\bibfield  {title} {\bibinfo {title} {Giant
  electrical modulation of magnetization in
  {{Co40Fe40B20}}/{{Pb}}({{Mg1}}/{{3Nb2}}/3)0.{{7Ti0}}.{{3O3}}(011)
  heterostructure},\ }\bibfield  {journal} {\bibinfo  {journal} {Scientific
  Reports}\ }\textbf {\bibinfo {volume} {4}},\ \href
  {https://doi.org/10.1038/srep03727} {10.1038/srep03727} (\bibinfo {year}
  {2015})\BibitemShut {NoStop}%
\bibitem [{\citenamefont {Yang}\ \emph
  {et~al.}(2015{\natexlab{b}})\citenamefont {Yang}, \citenamefont {Li},
  \citenamefont {Wen}, \citenamefont {Yang}, \citenamefont {Wang},
  \citenamefont {Zhang},\ and\ \citenamefont {Zhang}}]{Yang2015}%
  \BibitemOpen
  \bibfield  {author} {\bibinfo {author} {\bibfnamefont {C.}~\bibnamefont
  {Yang}}, \bibinfo {author} {\bibfnamefont {P.}~\bibnamefont {Li}}, \bibinfo
  {author} {\bibfnamefont {Y.}~\bibnamefont {Wen}}, \bibinfo {author}
  {\bibfnamefont {A.}~\bibnamefont {Yang}}, \bibinfo {author} {\bibfnamefont
  {D.}~\bibnamefont {Wang}}, \bibinfo {author} {\bibfnamefont {F.}~\bibnamefont
  {Zhang}},\ and\ \bibinfo {author} {\bibfnamefont {J.}~\bibnamefont {Zhang}},\
  }\bibfield  {title} {\bibinfo {title} {Giant {{Converse Magnetoelectric
  Effect}} in {{PZT}}/{{FeCuNbSiB}}/{{FeGa}}/{{FeCuNbSiB}}/{{PZT Laminates
  Without Magnetic Bias Field}}},\ }\href
  {https://doi.org/10.1109/TMAG.2015.2435010} {\bibfield  {journal} {\bibinfo
  {journal} {IEEE Transactions on Magnetics}\ }\textbf {\bibinfo {volume}
  {51}},\ \bibinfo {pages} {1} (\bibinfo {year}
  {2015}{\natexlab{b}})}\BibitemShut {NoStop}%
\bibitem [{\citenamefont {Biswas}\ \emph {et~al.}(2017)\citenamefont {Biswas},
  \citenamefont {Ahmad}, \citenamefont {Atulasimha},\ and\ \citenamefont
  {Bandyopadhyay}}]{Biswas2017}%
  \BibitemOpen
  \bibfield  {author} {\bibinfo {author} {\bibfnamefont {A.~K.}\ \bibnamefont
  {Biswas}}, \bibinfo {author} {\bibfnamefont {H.}~\bibnamefont {Ahmad}},
  \bibinfo {author} {\bibfnamefont {J.}~\bibnamefont {Atulasimha}},\ and\
  \bibinfo {author} {\bibfnamefont {S.}~\bibnamefont {Bandyopadhyay}},\
  }\bibfield  {title} {\bibinfo {title} {Experimental {{Demonstration}} of
  {{Complete}} 180$^\circ$ {{Reversal}} of {{Magnetization}} in {{Isolated Co
  Nanomagnets}} on a {{PMN}}\textendash{{PT Substrate}} with {{Voltage
  Generated Strain}}},\ }\href@noop {} {\bibfield  {journal} {\bibinfo
  {journal} {Nano letters}\ }\textbf {\bibinfo {volume} {17}},\ \bibinfo
  {pages} {3478} (\bibinfo {year} {2017})}\BibitemShut {NoStop}%
\bibitem [{\citenamefont {Cheng}\ \emph {et~al.}(2018)\citenamefont {Cheng},
  \citenamefont {Peng}, \citenamefont {Hu}, \citenamefont {Zhou},\ and\
  \citenamefont {Liu}}]{cheng2018}%
  \BibitemOpen
  \bibfield  {author} {\bibinfo {author} {\bibfnamefont {Y.}~\bibnamefont
  {Cheng}}, \bibinfo {author} {\bibfnamefont {B.}~\bibnamefont {Peng}},
  \bibinfo {author} {\bibfnamefont {Z.}~\bibnamefont {Hu}}, \bibinfo {author}
  {\bibfnamefont {Z.}~\bibnamefont {Zhou}},\ and\ \bibinfo {author}
  {\bibfnamefont {M.}~\bibnamefont {Liu}},\ }\bibfield  {title} {\bibinfo
  {title} {Recent development and status of magnetoelectric materials and
  devices},\ }\href {https://doi.org/10.1016/j.physleta.2018.07.014} {\bibfield
   {journal} {\bibinfo  {journal} {Physics Letters A}\ }\textbf {\bibinfo
  {volume} {382}},\ \bibinfo {pages} {3018} (\bibinfo {year}
  {2018})}\BibitemShut {NoStop}%
\bibitem [{\citenamefont {Du~Tr\'{e}molet~de
  Lacheisserie}(1993)}]{Lacheisserie1993}%
  \BibitemOpen
  \bibfield  {author} {\bibinfo {author} {\bibfnamefont {E.}~\bibnamefont
  {Du~Tr\'{e}molet~de Lacheisserie}},\ }\href@noop {} {\emph {\bibinfo {title}
  {Magnetostriction: theory and applications of magnetoelasticity}}}\ (\bibinfo
   {publisher} {Boca Raton: CRC Press},\ \bibinfo {year} {1993})\BibitemShut
  {NoStop}%
\bibitem [{\citenamefont {Zhang}\ \emph {et~al.}(2007)\citenamefont {Zhang},
  \citenamefont {Lee}, \citenamefont {Kim}, \citenamefont {Lee},\ and\
  \citenamefont {Shrout}}]{Zhang2007}%
  \BibitemOpen
  \bibfield  {author} {\bibinfo {author} {\bibfnamefont {S.}~\bibnamefont
  {Zhang}}, \bibinfo {author} {\bibfnamefont {S.-M.}\ \bibnamefont {Lee}},
  \bibinfo {author} {\bibfnamefont {D.-H.}\ \bibnamefont {Kim}}, \bibinfo
  {author} {\bibfnamefont {H.-Y.}\ \bibnamefont {Lee}},\ and\ \bibinfo {author}
  {\bibfnamefont {T.~R.}\ \bibnamefont {Shrout}},\ }\bibfield  {title}
  {\bibinfo {title} {Electromechanical properties of pmn–pzt piezoelectric
  single crystals near morphotropic phase boundary compositions},\ }\href
  {https://doi.org/10.1111/j.1551-2916.2007.02004.x} {\bibfield  {journal}
  {\bibinfo  {journal} {Journal of the American Ceramic Society}\ }\textbf
  {\bibinfo {volume} {90}},\ \bibinfo {pages} {3859} (\bibinfo {year}
  {2007})}\BibitemShut {NoStop}%
\bibitem [{\citenamefont {Zhang}\ and\ \citenamefont
  {Shrout}(2010)}]{Zhang2010}%
  \BibitemOpen
  \bibfield  {author} {\bibinfo {author} {\bibfnamefont {S.}~\bibnamefont
  {Zhang}}\ and\ \bibinfo {author} {\bibfnamefont {T.~R.}\ \bibnamefont
  {Shrout}},\ }\bibfield  {title} {\bibinfo {title} {Relaxor-{{PT}} single
  crystals: Observations and developments},\ }\href@noop {} {\bibfield
  {journal} {\bibinfo  {journal} {IEEE transactions on ultrasonics,
  ferroelectrics, and frequency control}\ }\textbf {\bibinfo {volume} {57}}
  (\bibinfo {year} {2010})}\BibitemShut {NoStop}%
\bibitem [{\citenamefont {Richter}\ \emph {et~al.}(2008)\citenamefont
  {Richter}, \citenamefont {Denneler}, \citenamefont {Schuh}, \citenamefont
  {Suvaci},\ and\ \citenamefont {Moos}}]{Richter2008}%
  \BibitemOpen
  \bibfield  {author} {\bibinfo {author} {\bibfnamefont {T.}~\bibnamefont
  {Richter}}, \bibinfo {author} {\bibfnamefont {S.}~\bibnamefont {Denneler}},
  \bibinfo {author} {\bibfnamefont {C.}~\bibnamefont {Schuh}}, \bibinfo
  {author} {\bibfnamefont {E.}~\bibnamefont {Suvaci}},\ and\ \bibinfo {author}
  {\bibfnamefont {R.}~\bibnamefont {Moos}},\ }\bibfield  {title} {\bibinfo
  {title} {Textured {{PMN}}\textendash{{PT}} and {{PMN}}\textendash{{PZT}}},\
  }\href {https://doi.org/10.1111/j.1551-2916.2007.02216.x} {\bibfield
  {journal} {\bibinfo  {journal} {Journal of the American Ceramic Society}\
  }\textbf {\bibinfo {volume} {91}},\ \bibinfo {pages} {929} (\bibinfo {year}
  {2008})}\BibitemShut {NoStop}%
\bibitem [{IEE()}]{IEEEStandardRelaxorPiezo}%
  \BibitemOpen
  \bibfield  {title} {\bibinfo {title} {{{IEEE Standard}} for
  {{Relaxor}}-{{Based Single Crystals}} for {{Transducer}} and {{Actuator
  Applications}}}\ }\href {https://doi.org/10.1109/IEEESTD.2017.8241013}
  {10.1109/IEEESTD.2017.8241013}\BibitemShut {NoStop}%
\bibitem [{\citenamefont {Palneedi}\ \emph {et~al.}(2018)\citenamefont
  {Palneedi}, \citenamefont {Na}, \citenamefont {Hwang}, \citenamefont
  {Peddigari}, \citenamefont {Shin}, \citenamefont {Kim},\ and\ \citenamefont
  {Ryu}}]{Palneedi2018}%
  \BibitemOpen
  \bibfield  {author} {\bibinfo {author} {\bibfnamefont {H.}~\bibnamefont
  {Palneedi}}, \bibinfo {author} {\bibfnamefont {S.-M.}\ \bibnamefont {Na}},
  \bibinfo {author} {\bibfnamefont {G.-T.}\ \bibnamefont {Hwang}}, \bibinfo
  {author} {\bibfnamefont {M.}~\bibnamefont {Peddigari}}, \bibinfo {author}
  {\bibfnamefont {K.~W.}\ \bibnamefont {Shin}}, \bibinfo {author}
  {\bibfnamefont {K.~H.}\ \bibnamefont {Kim}},\ and\ \bibinfo {author}
  {\bibfnamefont {J.}~\bibnamefont {Ryu}},\ }\bibfield  {title} {\bibinfo
  {title} {Highly tunable magnetoelectric response in dimensional gradient
  laminate composites of {{Fe}}-{{Ga}} alloy and
  {{Pb}}({{Mg1}}/{{3Nb2}}/3){{O3}}-{{Pb}}({{Zr}},{{Ti}}){{O3}} single
  crystal},\ }\href {https://doi.org/10.1016/j.jallcom.2018.05.122} {\bibfield
  {journal} {\bibinfo  {journal} {Journal of Alloys and Compounds}\ }\textbf
  {\bibinfo {volume} {765}},\ \bibinfo {pages} {764} (\bibinfo {year}
  {2018})}\BibitemShut {NoStop}%
\bibitem [{\citenamefont {Wang}\ \emph {et~al.}(2009)\citenamefont {Wang},
  \citenamefont {Or}, \citenamefont {Zhao},\ and\ \citenamefont
  {Luo}}]{wang2009}%
  \BibitemOpen
  \bibfield  {author} {\bibinfo {author} {\bibfnamefont {F.}~\bibnamefont
  {Wang}}, \bibinfo {author} {\bibfnamefont {S.~W.}\ \bibnamefont {Or}},
  \bibinfo {author} {\bibfnamefont {X.}~\bibnamefont {Zhao}},\ and\ \bibinfo
  {author} {\bibfnamefont {H.}~\bibnamefont {Luo}},\ }\bibfield  {title}
  {\bibinfo {title} {Cryogenic dielectric and piezoelectric activities in
  rhombohedral (1 - {\emph{x}} ){{Pb}}({{Mg}} {\textsubscript{1/3}} {{Nb}}
  {\textsubscript{2/3}} ){{O}} {\textsubscript{3}} \textendash{} {\emph{x}}
  {{PbTiO}} {\textsubscript{3}} single crystals with different crystallographic
  orientations},\ }\href {https://doi.org/10.1088/0022-3727/42/18/182001}
  {\bibfield  {journal} {\bibinfo  {journal} {Journal of Physics D: Applied
  Physics}\ }\textbf {\bibinfo {volume} {42}},\ \bibinfo {pages} {182001}
  (\bibinfo {year} {2009})}\BibitemShut {NoStop}%
\bibitem [{\citenamefont {Hwang}\ \emph {et~al.}(2018)\citenamefont {Hwang},
  \citenamefont {Palneedi}, \citenamefont {Jung}, \citenamefont {Kwon},
  \citenamefont {Peddigari}, \citenamefont {Min}, \citenamefont {Kim},
  \citenamefont {Ahn}, \citenamefont {Choi}, \citenamefont {Hahn},
  \citenamefont {Choi}, \citenamefont {Yoon}, \citenamefont {Park},
  \citenamefont {Lee}, \citenamefont {Choe}, \citenamefont {Kim},\ and\
  \citenamefont {Ryu}}]{Hwang2018}%
  \BibitemOpen
  \bibfield  {author} {\bibinfo {author} {\bibfnamefont {G.-T.}\ \bibnamefont
  {Hwang}}, \bibinfo {author} {\bibfnamefont {H.}~\bibnamefont {Palneedi}},
  \bibinfo {author} {\bibfnamefont {B.~M.}\ \bibnamefont {Jung}}, \bibinfo
  {author} {\bibfnamefont {S.~J.}\ \bibnamefont {Kwon}}, \bibinfo {author}
  {\bibfnamefont {M.}~\bibnamefont {Peddigari}}, \bibinfo {author}
  {\bibfnamefont {Y.}~\bibnamefont {Min}}, \bibinfo {author} {\bibfnamefont
  {J.-W.}\ \bibnamefont {Kim}}, \bibinfo {author} {\bibfnamefont {C.-W.}\
  \bibnamefont {Ahn}}, \bibinfo {author} {\bibfnamefont {J.-J.}\ \bibnamefont
  {Choi}}, \bibinfo {author} {\bibfnamefont {B.-D.}\ \bibnamefont {Hahn}},
  \bibinfo {author} {\bibfnamefont {J.-H.}\ \bibnamefont {Choi}}, \bibinfo
  {author} {\bibfnamefont {W.-H.}\ \bibnamefont {Yoon}}, \bibinfo {author}
  {\bibfnamefont {D.-S.}\ \bibnamefont {Park}}, \bibinfo {author}
  {\bibfnamefont {S.-B.}\ \bibnamefont {Lee}}, \bibinfo {author} {\bibfnamefont
  {Y.}~\bibnamefont {Choe}}, \bibinfo {author} {\bibfnamefont {K.-H.}\
  \bibnamefont {Kim}},\ and\ \bibinfo {author} {\bibfnamefont {J.}~\bibnamefont
  {Ryu}},\ }\bibfield  {title} {\bibinfo {title} {Enhancement of
  {{Magnetoelectric Conversion Achieved}} by {{Optimization}} of {{Interfacial
  Adhesion Layer}} in {{Laminate Composites}}},\ }\href
  {https://doi.org/10.1021/acsami.8b09848} {\bibfield  {journal} {\bibinfo
  {journal} {ACS Applied Materials \& Interfaces}\ }\textbf {\bibinfo {volume}
  {10}},\ \bibinfo {pages} {32323} (\bibinfo {year} {2018})}\BibitemShut
  {NoStop}%
\bibitem [{\citenamefont {Luo}\ \emph {et~al.}(2006)\citenamefont {Luo},
  \citenamefont {Wang}, \citenamefont {Tang}, \citenamefont {Zhao},
  \citenamefont {Feng}, \citenamefont {Lin},\ and\ \citenamefont
  {Luo}}]{luo2006}%
  \BibitemOpen
  \bibfield  {author} {\bibinfo {author} {\bibfnamefont {L.}~\bibnamefont
  {Luo}}, \bibinfo {author} {\bibfnamefont {H.}~\bibnamefont {Wang}}, \bibinfo
  {author} {\bibfnamefont {Y.}~\bibnamefont {Tang}}, \bibinfo {author}
  {\bibfnamefont {X.}~\bibnamefont {Zhao}}, \bibinfo {author} {\bibfnamefont
  {Z.}~\bibnamefont {Feng}}, \bibinfo {author} {\bibfnamefont {D.}~\bibnamefont
  {Lin}},\ and\ \bibinfo {author} {\bibfnamefont {H.}~\bibnamefont {Luo}},\
  }\bibfield  {title} {\bibinfo {title} {Ultrahigh transverse strain and
  piezoelectric behavior in
  (1-x){{Pb}}({{Mg1}}{$\slash$}{{3Nb2}}{$\slash$}3){{O3}}\textendash{{xPbTiO3}}
  crystals},\ }\href {https://doi.org/10.1063/1.2161947} {\bibfield  {journal}
  {\bibinfo  {journal} {Journal of Applied Physics}\ }\textbf {\bibinfo
  {volume} {99}},\ \bibinfo {pages} {024104} (\bibinfo {year}
  {2006})}\BibitemShut {NoStop}%
\bibitem [{\citenamefont {Luo}\ and\ \citenamefont
  {Zhang}(2014)}]{LuoZhang2014Piezo}%
  \BibitemOpen
  \bibfield  {author} {\bibinfo {author} {\bibfnamefont {J.}~\bibnamefont
  {Luo}}\ and\ \bibinfo {author} {\bibfnamefont {S.}~\bibnamefont {Zhang}},\
  }\bibfield  {title} {\bibinfo {title} {Advances in the {{Growth}} and
  {{Characterization}} of {{Relaxor}}-{{PT}}-{{Based Ferroelectric Single
  Crystals}}},\ }\href {https://doi.org/10.3390/cryst4030306} {\bibfield
  {journal} {\bibinfo  {journal} {Crystals}\ }\textbf {\bibinfo {volume} {4}},\
  \bibinfo {pages} {306} (\bibinfo {year} {2014})}\BibitemShut {NoStop}%
\bibitem [{\citenamefont {Rajaram~Patil}\ \emph {et~al.}(2013)\citenamefont
  {Rajaram~Patil}, \citenamefont {Kambale}, \citenamefont {Chai}, \citenamefont
  {Yoon}, \citenamefont {Jeong}, \citenamefont {Park}, \citenamefont {Kim},
  \citenamefont {Choi}, \citenamefont {Ahn}, \citenamefont {Hahn},
  \citenamefont {Zhang}, \citenamefont {Hoon~Kim},\ and\ \citenamefont
  {Ryu}}]{Patil2013}%
  \BibitemOpen
  \bibfield  {author} {\bibinfo {author} {\bibfnamefont {D.}~\bibnamefont
  {Rajaram~Patil}}, \bibinfo {author} {\bibfnamefont {R.~C.}\ \bibnamefont
  {Kambale}}, \bibinfo {author} {\bibfnamefont {Y.}~\bibnamefont {Chai}},
  \bibinfo {author} {\bibfnamefont {W.-H.}\ \bibnamefont {Yoon}}, \bibinfo
  {author} {\bibfnamefont {D.-Y.}\ \bibnamefont {Jeong}}, \bibinfo {author}
  {\bibfnamefont {D.-S.}\ \bibnamefont {Park}}, \bibinfo {author}
  {\bibfnamefont {J.-W.}\ \bibnamefont {Kim}}, \bibinfo {author} {\bibfnamefont
  {J.-J.}\ \bibnamefont {Choi}}, \bibinfo {author} {\bibfnamefont {C.-W.}\
  \bibnamefont {Ahn}}, \bibinfo {author} {\bibfnamefont {B.-D.}\ \bibnamefont
  {Hahn}}, \bibinfo {author} {\bibfnamefont {S.}~\bibnamefont {Zhang}},
  \bibinfo {author} {\bibfnamefont {K.}~\bibnamefont {Hoon~Kim}},\ and\
  \bibinfo {author} {\bibfnamefont {J.}~\bibnamefont {Ryu}},\ }\bibfield
  {title} {\bibinfo {title} {Multiple broadband magnetoelectric response in
  thickness-controlled {{Ni}}/[011] {{Pb}}({{Mg}} {\textsubscript{1/3}} {{Nb}}
  {\textsubscript{2/3}} ){{O}} {\textsubscript{3}} -{{Pb}}({{Zr}},{{Ti}}){{O}}
  {\textsubscript{3}} single crystal/{{Ni}} laminates},\ }\href
  {https://doi.org/10.1063/1.4817383} {\bibfield  {journal} {\bibinfo
  {journal} {Applied Physics Letters}\ }\textbf {\bibinfo {volume} {103}},\
  \bibinfo {pages} {052907} (\bibinfo {year} {2013})}\BibitemShut {NoStop}%
\bibitem [{\citenamefont {Kambale}\ \emph {et~al.}(2013)\citenamefont
  {Kambale}, \citenamefont {Yoon}, \citenamefont {Park}, \citenamefont {Choi},
  \citenamefont {Ahn}, \citenamefont {Kim}, \citenamefont {Hahn}, \citenamefont
  {Jeong}, \citenamefont {Chul~Lee}, \citenamefont {Chung},\ and\ \citenamefont
  {Ryu}}]{Kambale2013}%
  \BibitemOpen
  \bibfield  {author} {\bibinfo {author} {\bibfnamefont {R.~C.}\ \bibnamefont
  {Kambale}}, \bibinfo {author} {\bibfnamefont {W.-H.}\ \bibnamefont {Yoon}},
  \bibinfo {author} {\bibfnamefont {D.-S.}\ \bibnamefont {Park}}, \bibinfo
  {author} {\bibfnamefont {J.-J.}\ \bibnamefont {Choi}}, \bibinfo {author}
  {\bibfnamefont {C.-W.}\ \bibnamefont {Ahn}}, \bibinfo {author} {\bibfnamefont
  {J.-W.}\ \bibnamefont {Kim}}, \bibinfo {author} {\bibfnamefont {B.-D.}\
  \bibnamefont {Hahn}}, \bibinfo {author} {\bibfnamefont {D.-Y.}\ \bibnamefont
  {Jeong}}, \bibinfo {author} {\bibfnamefont {B.}~\bibnamefont {Chul~Lee}},
  \bibinfo {author} {\bibfnamefont {G.-S.}\ \bibnamefont {Chung}},\ and\
  \bibinfo {author} {\bibfnamefont {J.}~\bibnamefont {Ryu}},\ }\bibfield
  {title} {\bibinfo {title} {Magnetoelectric properties and magnetomechanical
  energy harvesting from stray vibration and electromagnetic wave by
  {{Pb}}({{Mg}} {\textsubscript{1/3}} {{Nb}} {\textsubscript{2/3}} ){{O}}
  {\textsubscript{3}} -{{Pb}}({{Zr}},{{Ti}}){{O}} {\textsubscript{3}} single
  crystal/{{Ni}} cantilever},\ }\href {https://doi.org/10.1063/1.4804959}
  {\bibfield  {journal} {\bibinfo  {journal} {Journal of Applied Physics}\
  }\textbf {\bibinfo {volume} {113}},\ \bibinfo {pages} {204108} (\bibinfo
  {year} {2013})}\BibitemShut {NoStop}%
\bibitem [{\citenamefont {Ryu}\ \emph {et~al.}(2015)\citenamefont {Ryu},
  \citenamefont {Kang}, \citenamefont {Zhou}, \citenamefont {Choi},
  \citenamefont {Yoon}, \citenamefont {Park}, \citenamefont {Choi},
  \citenamefont {Hahn}, \citenamefont {Ahn}, \citenamefont {Kim}, \citenamefont
  {Kim}, \citenamefont {Priya}, \citenamefont {Lee}, \citenamefont {Jeong},\
  and\ \citenamefont {Jeong}}]{Ryu2015}%
  \BibitemOpen
  \bibfield  {author} {\bibinfo {author} {\bibfnamefont {J.}~\bibnamefont
  {Ryu}}, \bibinfo {author} {\bibfnamefont {J.-E.}\ \bibnamefont {Kang}},
  \bibinfo {author} {\bibfnamefont {Y.}~\bibnamefont {Zhou}}, \bibinfo {author}
  {\bibfnamefont {S.-Y.}\ \bibnamefont {Choi}}, \bibinfo {author}
  {\bibfnamefont {W.-H.}\ \bibnamefont {Yoon}}, \bibinfo {author}
  {\bibfnamefont {D.-S.}\ \bibnamefont {Park}}, \bibinfo {author}
  {\bibfnamefont {J.-J.}\ \bibnamefont {Choi}}, \bibinfo {author}
  {\bibfnamefont {B.-D.}\ \bibnamefont {Hahn}}, \bibinfo {author}
  {\bibfnamefont {C.-W.}\ \bibnamefont {Ahn}}, \bibinfo {author} {\bibfnamefont
  {J.-W.}\ \bibnamefont {Kim}}, \bibinfo {author} {\bibfnamefont {Y.-D.}\
  \bibnamefont {Kim}}, \bibinfo {author} {\bibfnamefont {S.}~\bibnamefont
  {Priya}}, \bibinfo {author} {\bibfnamefont {S.~Y.}\ \bibnamefont {Lee}},
  \bibinfo {author} {\bibfnamefont {S.}~\bibnamefont {Jeong}},\ and\ \bibinfo
  {author} {\bibfnamefont {D.-Y.}\ \bibnamefont {Jeong}},\ }\bibfield  {title}
  {\bibinfo {title} {Ubiquitous magneto-mechano-electric generator},\ }\href
  {https://doi.org/10.1039/C5EE00414D} {\bibfield  {journal} {\bibinfo
  {journal} {Energy \& Environmental Science}\ }\textbf {\bibinfo {volume}
  {8}},\ \bibinfo {pages} {2402} (\bibinfo {year} {2015})}\BibitemShut
  {NoStop}%
\bibitem [{\citenamefont {Palneedi}\ \emph {et~al.}(2017)\citenamefont
  {Palneedi}, \citenamefont {Annapureddy}, \citenamefont {Lee}, \citenamefont
  {Choi}, \citenamefont {Choi}, \citenamefont {Chung}, \citenamefont {Kang},\
  and\ \citenamefont {Ryu}}]{Palneedi2017}%
  \BibitemOpen
  \bibfield  {author} {\bibinfo {author} {\bibfnamefont {H.}~\bibnamefont
  {Palneedi}}, \bibinfo {author} {\bibfnamefont {V.}~\bibnamefont
  {Annapureddy}}, \bibinfo {author} {\bibfnamefont {H.-Y.}\ \bibnamefont
  {Lee}}, \bibinfo {author} {\bibfnamefont {J.-J.}\ \bibnamefont {Choi}},
  \bibinfo {author} {\bibfnamefont {S.-Y.}\ \bibnamefont {Choi}}, \bibinfo
  {author} {\bibfnamefont {S.-Y.}\ \bibnamefont {Chung}}, \bibinfo {author}
  {\bibfnamefont {S.-J.~L.}\ \bibnamefont {Kang}},\ and\ \bibinfo {author}
  {\bibfnamefont {J.}~\bibnamefont {Ryu}},\ }\bibfield  {title} {\bibinfo
  {title} {Strong and anisotropic magnetoelectricity in composites of
  magnetostrictive {{Ni}} and solid-state grown lead-free piezoelectric
  {{BZT}}\textendash{{BCT}} single crystals},\ }\href
  {https://doi.org/10.1016/j.jascer.2016.12.005} {\bibfield  {journal}
  {\bibinfo  {journal} {Journal of Asian Ceramic Societies}\ }\textbf {\bibinfo
  {volume} {5}},\ \bibinfo {pages} {36} (\bibinfo {year} {2017})}\BibitemShut
  {NoStop}%
\bibitem [{\citenamefont {Bilgen}\ \emph {et~al.}(2011)\citenamefont {Bilgen},
  \citenamefont {Amin~Karami}, \citenamefont {Inman},\ and\ \citenamefont
  {Friswell}}]{Bilgen2011}%
  \BibitemOpen
  \bibfield  {author} {\bibinfo {author} {\bibfnamefont {O.}~\bibnamefont
  {Bilgen}}, \bibinfo {author} {\bibfnamefont {M.}~\bibnamefont {Amin~Karami}},
  \bibinfo {author} {\bibfnamefont {D.~J.}\ \bibnamefont {Inman}},\ and\
  \bibinfo {author} {\bibfnamefont {M.~I.}\ \bibnamefont {Friswell}},\
  }\bibfield  {title} {\bibinfo {title} {The actuation characterization of
  cantilevered unimorph beams with single crystal piezoelectric materials},\
  }\href {https://doi.org/10.1088/0964-1726/20/5/055024} {\bibfield  {journal}
  {\bibinfo  {journal} {Smart Materials and Structures}\ }\textbf {\bibinfo
  {volume} {20}},\ \bibinfo {pages} {055024} (\bibinfo {year}
  {2011})}\BibitemShut {NoStop}%
\bibitem [{\citenamefont {Lian}\ \emph {et~al.}(2018)\citenamefont {Lian},
  \citenamefont {Ponchel}, \citenamefont {Tiercelin}, \citenamefont {Chen},
  \citenamefont {R\'emiens}, \citenamefont {Lasri}, \citenamefont {Wang},
  \citenamefont {Pernod}, \citenamefont {Zhang},\ and\ \citenamefont
  {Dong}}]{Lian2018-2}%
  \BibitemOpen
  \bibfield  {author} {\bibinfo {author} {\bibfnamefont {J.}~\bibnamefont
  {Lian}}, \bibinfo {author} {\bibfnamefont {F.}~\bibnamefont {Ponchel}},
  \bibinfo {author} {\bibfnamefont {N.}~\bibnamefont {Tiercelin}}, \bibinfo
  {author} {\bibfnamefont {Y.}~\bibnamefont {Chen}}, \bibinfo {author}
  {\bibfnamefont {D.}~\bibnamefont {R\'emiens}}, \bibinfo {author}
  {\bibfnamefont {T.}~\bibnamefont {Lasri}}, \bibinfo {author} {\bibfnamefont
  {G.}~\bibnamefont {Wang}}, \bibinfo {author} {\bibfnamefont {P.}~\bibnamefont
  {Pernod}}, \bibinfo {author} {\bibfnamefont {W.}~\bibnamefont {Zhang}},\ and\
  \bibinfo {author} {\bibfnamefont {X.}~\bibnamefont {Dong}},\ }\bibfield
  {title} {\bibinfo {title} {Electric field tuning of magnetism in
  heterostructure of yttrium iron garnet film/lead magnesium niobate-lead
  zirconate titanate ceramic},\ }\href {https://doi.org/10.1063/1.5023885}
  {\bibfield  {journal} {\bibinfo  {journal} {Applied Physics Letters}\
  }\textbf {\bibinfo {volume} {112}},\ \bibinfo {pages} {162904} (\bibinfo
  {year} {2018})}\BibitemShut {NoStop}%
\bibitem [{\citenamefont {Park}\ \emph {et~al.}(2010)\citenamefont {Park},
  \citenamefont {Cho}, \citenamefont {Arat}, \citenamefont {Evey},\ and\
  \citenamefont {Priya}}]{Park2010}%
  \BibitemOpen
  \bibfield  {author} {\bibinfo {author} {\bibfnamefont {C.-S.}\ \bibnamefont
  {Park}}, \bibinfo {author} {\bibfnamefont {K.-H.}\ \bibnamefont {Cho}},
  \bibinfo {author} {\bibfnamefont {M.~A.}\ \bibnamefont {Arat}}, \bibinfo
  {author} {\bibfnamefont {J.}~\bibnamefont {Evey}},\ and\ \bibinfo {author}
  {\bibfnamefont {S.}~\bibnamefont {Priya}},\ }\bibfield  {title} {\bibinfo
  {title} {High magnetic field sensitivity in
  {{Pb}}({{Zr}},{{Ti}}){{O3}}\textendash{{Pb}}({{Mg1}}/{{3Nb2}}/3){{O3}} single
  crystal/{{Terfenol}}-{{D}}/{{Metglas}} magnetoelectric laminate composites},\
  }\href {https://doi.org/10.1063/1.3406142} {\bibfield  {journal} {\bibinfo
  {journal} {Journal of Applied Physics}\ }\textbf {\bibinfo {volume} {107}},\
  \bibinfo {pages} {094109} (\bibinfo {year} {2010})}\BibitemShut {NoStop}%
\bibitem [{\citenamefont {Annapureddy}\ \emph {et~al.}(2018)\citenamefont
  {Annapureddy}, \citenamefont {Na}, \citenamefont {Hwang}, \citenamefont
  {Kang}, \citenamefont {Sriramdas}, \citenamefont {Palneedi}, \citenamefont
  {Yoon}, \citenamefont {Hahn}, \citenamefont {Kim}, \citenamefont {Ahn},
  \citenamefont {Park}, \citenamefont {Choi}, \citenamefont {Jeong},
  \citenamefont {Flatau}, \citenamefont {Peddigari}, \citenamefont {Priya},
  \citenamefont {Kim},\ and\ \citenamefont {Ryu}}]{Annapureddy2018}%
  \BibitemOpen
  \bibfield  {author} {\bibinfo {author} {\bibfnamefont {V.}~\bibnamefont
  {Annapureddy}}, \bibinfo {author} {\bibfnamefont {S.-M.}\ \bibnamefont {Na}},
  \bibinfo {author} {\bibfnamefont {G.-T.}\ \bibnamefont {Hwang}}, \bibinfo
  {author} {\bibfnamefont {M.~G.}\ \bibnamefont {Kang}}, \bibinfo {author}
  {\bibfnamefont {R.}~\bibnamefont {Sriramdas}}, \bibinfo {author}
  {\bibfnamefont {H.}~\bibnamefont {Palneedi}}, \bibinfo {author}
  {\bibfnamefont {W.-H.}\ \bibnamefont {Yoon}}, \bibinfo {author}
  {\bibfnamefont {B.-D.}\ \bibnamefont {Hahn}}, \bibinfo {author}
  {\bibfnamefont {J.-W.}\ \bibnamefont {Kim}}, \bibinfo {author} {\bibfnamefont
  {C.-W.}\ \bibnamefont {Ahn}}, \bibinfo {author} {\bibfnamefont {D.-S.}\
  \bibnamefont {Park}}, \bibinfo {author} {\bibfnamefont {J.-J.}\ \bibnamefont
  {Choi}}, \bibinfo {author} {\bibfnamefont {D.-Y.}\ \bibnamefont {Jeong}},
  \bibinfo {author} {\bibfnamefont {A.~B.}\ \bibnamefont {Flatau}}, \bibinfo
  {author} {\bibfnamefont {M.}~\bibnamefont {Peddigari}}, \bibinfo {author}
  {\bibfnamefont {S.}~\bibnamefont {Priya}}, \bibinfo {author} {\bibfnamefont
  {K.-H.}\ \bibnamefont {Kim}},\ and\ \bibinfo {author} {\bibfnamefont
  {J.}~\bibnamefont {Ryu}},\ }\bibfield  {title} {\bibinfo {title} {Exceeding
  milli-watt powering magneto-mechano-electric generator for standalone-powered
  electronics},\ }\bibfield  {journal} {\bibinfo  {journal} {Energy \&
  Environmental Science}\ }\href {https://doi.org/10.1039/C7EE03429F}
  {10.1039/C7EE03429F} (\bibinfo {year} {2018})\BibitemShut {NoStop}%
\bibitem [{\citenamefont {Chu}\ \emph {et~al.}(2018)\citenamefont {Chu},
  \citenamefont {Annapureddy}, \citenamefont {PourhosseiniAsl}, \citenamefont
  {Palneedi}, \citenamefont {Ryu},\ and\ \citenamefont {Dong}}]{Chu2018}%
  \BibitemOpen
  \bibfield  {author} {\bibinfo {author} {\bibfnamefont {Z.}~\bibnamefont
  {Chu}}, \bibinfo {author} {\bibfnamefont {V.}~\bibnamefont {Annapureddy}},
  \bibinfo {author} {\bibfnamefont {M.}~\bibnamefont {PourhosseiniAsl}},
  \bibinfo {author} {\bibfnamefont {H.}~\bibnamefont {Palneedi}}, \bibinfo
  {author} {\bibfnamefont {J.}~\bibnamefont {Ryu}},\ and\ \bibinfo {author}
  {\bibfnamefont {S.}~\bibnamefont {Dong}},\ }\bibfield  {title} {\bibinfo
  {title} {Dual-stimulus magnetoelectric energy harvesting},\ }\href
  {https://doi.org/10.1557/mrs.2018.31} {\bibfield  {journal} {\bibinfo
  {journal} {MRS Bulletin}\ }\textbf {\bibinfo {volume} {43}},\ \bibinfo
  {pages} {199} (\bibinfo {year} {2018})}\BibitemShut {NoStop}%
\bibitem [{\citenamefont {{Clark}}\ \emph {et~al.}(2000)\citenamefont
  {{Clark}}, \citenamefont {{Restorff}}, \citenamefont {{Wun-Fogle}},
  \citenamefont {{Lograsso}},\ and\ \citenamefont {{Schlagel}}}]{Clark2000}%
  \BibitemOpen
  \bibfield  {author} {\bibinfo {author} {\bibfnamefont {A.~E.}\ \bibnamefont
  {{Clark}}}, \bibinfo {author} {\bibfnamefont {J.~B.}\ \bibnamefont
  {{Restorff}}}, \bibinfo {author} {\bibfnamefont {M.}~\bibnamefont
  {{Wun-Fogle}}}, \bibinfo {author} {\bibfnamefont {T.~A.}\ \bibnamefont
  {{Lograsso}}},\ and\ \bibinfo {author} {\bibfnamefont {D.~L.}\ \bibnamefont
  {{Schlagel}}},\ }\bibfield  {title} {\bibinfo {title} {Magnetostrictive
  properties of body-centered cubic fe-ga and fe-ga-al alloys},\ }\href
  {https://doi.org/10.1109/20.908752} {\bibfield  {journal} {\bibinfo
  {journal} {IEEE Transactions on Magnetics}\ }\textbf {\bibinfo {volume}
  {36}},\ \bibinfo {pages} {3238} (\bibinfo {year} {2000})}\BibitemShut
  {NoStop}%
\bibitem [{\citenamefont {Atulasimha}\ and\ \citenamefont
  {Flatau}(2011)}]{Atulasimha2011}%
  \BibitemOpen
  \bibfield  {author} {\bibinfo {author} {\bibfnamefont {J.}~\bibnamefont
  {Atulasimha}}\ and\ \bibinfo {author} {\bibfnamefont {A.~B.}\ \bibnamefont
  {Flatau}},\ }\bibfield  {title} {\bibinfo {title} {A review of
  magnetostrictive iron{\textendash}gallium alloys},\ }\href
  {https://doi.org/10.1088/0964-1726/20/4/043001} {\bibfield  {journal}
  {\bibinfo  {journal} {Smart Materials and Structures}\ }\textbf {\bibinfo
  {volume} {20}},\ \bibinfo {pages} {043001} (\bibinfo {year}
  {2011})}\BibitemShut {NoStop}%
\bibitem [{\citenamefont {Jahjah}\ \emph {et~al.}(2019)\citenamefont {Jahjah},
  \citenamefont {Manach}, \citenamefont {Grand}, \citenamefont {Fessant},
  \citenamefont {{Warot-Fonrose}}, \citenamefont {Prinsloo}, \citenamefont
  {Sheppard}, \citenamefont {Dekadjevi}, \citenamefont {Spenato},\ and\
  \citenamefont {Jay}}]{jay2019}%
  \BibitemOpen
  \bibfield  {author} {\bibinfo {author} {\bibfnamefont {W.}~\bibnamefont
  {Jahjah}}, \bibinfo {author} {\bibfnamefont {R.}~\bibnamefont {Manach}},
  \bibinfo {author} {\bibfnamefont {Y.~L.}\ \bibnamefont {Grand}}, \bibinfo
  {author} {\bibfnamefont {A.}~\bibnamefont {Fessant}}, \bibinfo {author}
  {\bibfnamefont {B.}~\bibnamefont {{Warot-Fonrose}}}, \bibinfo {author}
  {\bibfnamefont {A.~R.~E.}\ \bibnamefont {Prinsloo}}, \bibinfo {author}
  {\bibfnamefont {C.~J.}\ \bibnamefont {Sheppard}}, \bibinfo {author}
  {\bibfnamefont {D.~T.}\ \bibnamefont {Dekadjevi}}, \bibinfo {author}
  {\bibfnamefont {D.}~\bibnamefont {Spenato}},\ and\ \bibinfo {author}
  {\bibfnamefont {J.-P.}\ \bibnamefont {Jay}},\ }\bibfield  {title} {\bibinfo
  {title} {Thickness dependence of magnetization reversal and magnetostriction
  in {{FeGa}} thin films},\ }\href@noop {} {\bibfield  {journal} {\bibinfo
  {journal} {arXiv:1903.05397 [cond-mat]}\ } (\bibinfo {year}
  {2019})}\BibitemShut {NoStop}%
\bibitem [{\citenamefont {Finkel}\ \emph {et~al.}(2015)\citenamefont {Finkel},
  \citenamefont {P\'erez~Moyet}, \citenamefont {{Wun-Fogle}}, \citenamefont
  {Restorff}, \citenamefont {Kosior}, \citenamefont {Staruch}, \citenamefont
  {Stace},\ and\ \citenamefont {Amin}}]{finkel2015}%
  \BibitemOpen
  \bibfield  {author} {\bibinfo {author} {\bibfnamefont {P.}~\bibnamefont
  {Finkel}}, \bibinfo {author} {\bibfnamefont {R.}~\bibnamefont
  {P\'erez~Moyet}}, \bibinfo {author} {\bibfnamefont {M.}~\bibnamefont
  {{Wun-Fogle}}}, \bibinfo {author} {\bibfnamefont {J.}~\bibnamefont
  {Restorff}}, \bibinfo {author} {\bibfnamefont {J.}~\bibnamefont {Kosior}},
  \bibinfo {author} {\bibfnamefont {M.}~\bibnamefont {Staruch}}, \bibinfo
  {author} {\bibfnamefont {J.}~\bibnamefont {Stace}},\ and\ \bibinfo {author}
  {\bibfnamefont {A.}~\bibnamefont {Amin}},\ }\bibfield  {title} {\bibinfo
  {title} {Non-{{Resonant Magnetoelectric Energy Harvesting Utilizing Phase
  Transformation}} in {{Relaxor Ferroelectric Single Crystals}}},\ }\href
  {https://doi.org/10.3390/act5010002} {\bibfield  {journal} {\bibinfo
  {journal} {Actuators}\ }\textbf {\bibinfo {volume} {5}},\ \bibinfo {pages}
  {2} (\bibinfo {year} {2015})}\BibitemShut {NoStop}%
\bibitem [{\citenamefont {Wang}\ \emph {et~al.}(2011)\citenamefont {Wang},
  \citenamefont {Du}, \citenamefont {Fan}, \citenamefont {Xu}, \citenamefont
  {Zhang},\ and\ \citenamefont {Zhao}}]{wang2011}%
  \BibitemOpen
  \bibfield  {author} {\bibinfo {author} {\bibfnamefont {L.}~\bibnamefont
  {Wang}}, \bibinfo {author} {\bibfnamefont {Z.}~\bibnamefont {Du}}, \bibinfo
  {author} {\bibfnamefont {C.}~\bibnamefont {Fan}}, \bibinfo {author}
  {\bibfnamefont {L.}~\bibnamefont {Xu}}, \bibinfo {author} {\bibfnamefont
  {H.}~\bibnamefont {Zhang}},\ and\ \bibinfo {author} {\bibfnamefont
  {D.}~\bibnamefont {Zhao}},\ }\bibfield  {title} {\bibinfo {title}
  {Magnetoelectric properties of {{Fe}}\textendash{{Ga}}/{{BaTiO3}} laminate
  composites},\ }\href {https://doi.org/10.1016/j.jallcom.2010.09.083}
  {\bibfield  {journal} {\bibinfo  {journal} {Journal of Alloys and Compounds}\
  }\textbf {\bibinfo {volume} {509}},\ \bibinfo {pages} {508} (\bibinfo {year}
  {2011})}\BibitemShut {NoStop}%
\bibitem [{\citenamefont {Dong}\ \emph {et~al.}(2005)\citenamefont {Dong},
  \citenamefont {Zhai}, \citenamefont {Wang}, \citenamefont {Bai},
  \citenamefont {Li}, \citenamefont {Viehland},\ and\ \citenamefont
  {Lograsso}}]{dong2005}%
  \BibitemOpen
  \bibfield  {author} {\bibinfo {author} {\bibfnamefont {S.}~\bibnamefont
  {Dong}}, \bibinfo {author} {\bibfnamefont {J.}~\bibnamefont {Zhai}}, \bibinfo
  {author} {\bibfnamefont {N.}~\bibnamefont {Wang}}, \bibinfo {author}
  {\bibfnamefont {F.}~\bibnamefont {Bai}}, \bibinfo {author} {\bibfnamefont
  {J.}~\bibnamefont {Li}}, \bibinfo {author} {\bibfnamefont {D.}~\bibnamefont
  {Viehland}},\ and\ \bibinfo {author} {\bibfnamefont {T.~A.}\ \bibnamefont
  {Lograsso}},\ }\bibfield  {title} {\bibinfo {title}
  {Fe\textendash{{Ga}}{$\slash$}{{Pb}}({{Mg1}}{$\slash$}{{3Nb2}}{$\slash$}3){{O3}}\textendash{{PbTiO3}}
  magnetoelectric laminate composites},\ }\href
  {https://doi.org/10.1063/1.2137455} {\bibfield  {journal} {\bibinfo
  {journal} {Applied Physics Letters}\ }\textbf {\bibinfo {volume} {87}},\
  \bibinfo {pages} {222504} (\bibinfo {year} {2005})}\BibitemShut {NoStop}%
\bibitem [{\citenamefont {Parkes}\ \emph {et~al.}(2012)\citenamefont {Parkes},
  \citenamefont {Cavill}, \citenamefont {Hindmarch}, \citenamefont {Wadley},
  \citenamefont {McGee}, \citenamefont {Staddon}, \citenamefont {Edmonds},
  \citenamefont {Campion}, \citenamefont {Gallagher},\ and\ \citenamefont
  {Rushforth}}]{parkes2012}%
  \BibitemOpen
  \bibfield  {author} {\bibinfo {author} {\bibfnamefont {D.~E.}\ \bibnamefont
  {Parkes}}, \bibinfo {author} {\bibfnamefont {S.~A.}\ \bibnamefont {Cavill}},
  \bibinfo {author} {\bibfnamefont {A.~T.}\ \bibnamefont {Hindmarch}}, \bibinfo
  {author} {\bibfnamefont {P.}~\bibnamefont {Wadley}}, \bibinfo {author}
  {\bibfnamefont {F.}~\bibnamefont {McGee}}, \bibinfo {author} {\bibfnamefont
  {C.~R.}\ \bibnamefont {Staddon}}, \bibinfo {author} {\bibfnamefont {K.~W.}\
  \bibnamefont {Edmonds}}, \bibinfo {author} {\bibfnamefont {R.~P.}\
  \bibnamefont {Campion}}, \bibinfo {author} {\bibfnamefont {B.~L.}\
  \bibnamefont {Gallagher}},\ and\ \bibinfo {author} {\bibfnamefont {A.~W.}\
  \bibnamefont {Rushforth}},\ }\bibfield  {title} {\bibinfo {title}
  {Non-volatile voltage control of magnetization and magnetic domain walls in
  magnetostrictive epitaxial thin films},\ }\href
  {https://doi.org/10.1063/1.4745789} {\bibfield  {journal} {\bibinfo
  {journal} {Applied Physics Letters}\ }\textbf {\bibinfo {volume} {101}},\
  \bibinfo {pages} {072402} (\bibinfo {year} {2012})}\BibitemShut {NoStop}%
\bibitem [{\citenamefont {Xie}\ \emph {et~al.}(2014)\citenamefont {Xie},
  \citenamefont {Zhan}, \citenamefont {Liu}, \citenamefont {Dai}, \citenamefont
  {Yang}, \citenamefont {Zuo}, \citenamefont {Chen}, \citenamefont {Wang},
  \citenamefont {Zhang}, \citenamefont {Rong},\ and\ \citenamefont
  {Li}}]{Xie2014}%
  \BibitemOpen
  \bibfield  {author} {\bibinfo {author} {\bibfnamefont {Y.}~\bibnamefont
  {Xie}}, \bibinfo {author} {\bibfnamefont {Q.}~\bibnamefont {Zhan}}, \bibinfo
  {author} {\bibfnamefont {Y.}~\bibnamefont {Liu}}, \bibinfo {author}
  {\bibfnamefont {G.}~\bibnamefont {Dai}}, \bibinfo {author} {\bibfnamefont
  {H.}~\bibnamefont {Yang}}, \bibinfo {author} {\bibfnamefont {Z.}~\bibnamefont
  {Zuo}}, \bibinfo {author} {\bibfnamefont {B.}~\bibnamefont {Chen}}, \bibinfo
  {author} {\bibfnamefont {B.}~\bibnamefont {Wang}}, \bibinfo {author}
  {\bibfnamefont {Y.}~\bibnamefont {Zhang}}, \bibinfo {author} {\bibfnamefont
  {X.}~\bibnamefont {Rong}},\ and\ \bibinfo {author} {\bibfnamefont {R.-W.}\
  \bibnamefont {Li}},\ }\bibfield  {title} {\bibinfo {title} {Electric-field
  control of magnetic anisotropy in {{Fe}} {\textsubscript{81}} {{Ga}}
  {\textsubscript{19}} /{{BaTiO}} {\textsubscript{3}} heterostructure films},\
  }\href {https://doi.org/10.1063/1.4901911} {\bibfield  {journal} {\bibinfo
  {journal} {AIP Advances}\ }\textbf {\bibinfo {volume} {4}},\ \bibinfo {pages}
  {117113} (\bibinfo {year} {2014})}\BibitemShut {NoStop}%
\bibitem [{\citenamefont {Liu}\ and\ \citenamefont
  {Sun}(2014)}]{liu2014_review}%
  \BibitemOpen
  \bibfield  {author} {\bibinfo {author} {\bibfnamefont {M.}~\bibnamefont
  {Liu}}\ and\ \bibinfo {author} {\bibfnamefont {N.~X.}\ \bibnamefont {Sun}},\
  }\bibfield  {title} {\bibinfo {title} {Voltage control of magnetism in
  multiferroic heterostructures},\ }\href
  {https://doi.org/10.1098/rsta.2012.0439} {\bibfield  {journal} {\bibinfo
  {journal} {Philosophical Transactions of the Royal Society A: Mathematical,
  Physical and Engineering Sciences}\ }\textbf {\bibinfo {volume} {372}},\
  \bibinfo {pages} {20120439} (\bibinfo {year} {2014})}\BibitemShut {NoStop}%
\bibitem [{\citenamefont {Ahmad}\ \emph {et~al.}(2015)\citenamefont {Ahmad},
  \citenamefont {Atulasimha},\ and\ \citenamefont {Bandyopadhyay}}]{Ahmad2015}%
  \BibitemOpen
  \bibfield  {author} {\bibinfo {author} {\bibfnamefont {H.}~\bibnamefont
  {Ahmad}}, \bibinfo {author} {\bibfnamefont {J.}~\bibnamefont {Atulasimha}},\
  and\ \bibinfo {author} {\bibfnamefont {S.}~\bibnamefont {Bandyopadhyay}},\
  }\bibfield  {title} {\bibinfo {title} {Electric field control of magnetic
  states in isolated and dipole-coupled {{FeGa}} nanomagnets delineated on a
  {{PMN}}-{{PT}} substrate},\ }\href
  {https://doi.org/10.1088/0957-4484/26/40/401001} {\bibfield  {journal}
  {\bibinfo  {journal} {Nanotechnology}\ }\textbf {\bibinfo {volume} {26}},\
  \bibinfo {pages} {401001} (\bibinfo {year} {2015})}\BibitemShut {NoStop}%
\bibitem [{\citenamefont {Phuoc}\ and\ \citenamefont {Ong}(2017)}]{Phuoc2017}%
  \BibitemOpen
  \bibfield  {author} {\bibinfo {author} {\bibfnamefont {N.~N.}\ \bibnamefont
  {Phuoc}}\ and\ \bibinfo {author} {\bibfnamefont {C.~K.}\ \bibnamefont
  {Ong}},\ }\bibfield  {title} {\bibinfo {title} {Electrical manipulation of
  electromagnetic properties of
  {{FeGa}}/[{{Pb}}({{Mg1}}/{{3Nb2}}/3){{O3}}]0.68\textendash{}[{{PbTiO3}}]0.32(011)
  multiferroic heterostructures},\ }\href
  {https://doi.org/10.1007/s10854-016-6233-3} {\bibfield  {journal} {\bibinfo
  {journal} {Journal of Materials Science: Materials in Electronics}\ }\textbf
  {\bibinfo {volume} {28}},\ \bibinfo {pages} {5628} (\bibinfo {year}
  {2017})}\BibitemShut {NoStop}%
\bibitem [{\citenamefont {Zhang}\ \emph {et~al.}(2018)\citenamefont {Zhang},
  \citenamefont {Huang}, \citenamefont {Turghun}, \citenamefont {Duan},
  \citenamefont {Wang},\ and\ \citenamefont {Shi}}]{Zhang2018}%
  \BibitemOpen
  \bibfield  {author} {\bibinfo {author} {\bibfnamefont {Y.}~\bibnamefont
  {Zhang}}, \bibinfo {author} {\bibfnamefont {C.}~\bibnamefont {Huang}},
  \bibinfo {author} {\bibfnamefont {M.}~\bibnamefont {Turghun}}, \bibinfo
  {author} {\bibfnamefont {Z.}~\bibnamefont {Duan}}, \bibinfo {author}
  {\bibfnamefont {F.}~\bibnamefont {Wang}},\ and\ \bibinfo {author}
  {\bibfnamefont {W.}~\bibnamefont {Shi}},\ }\bibfield  {title} {\bibinfo
  {title} {Electric-regulated enhanced in-plane uniaxial anisotropy in
  {{FeGa}}/{{PMN}}\textendash{{PT}} composite using oblique pulsed laser
  deposition},\ }\bibfield  {journal} {\bibinfo  {journal} {Applied Physics A}\
  }\textbf {\bibinfo {volume} {124}},\ \href
  {https://doi.org/10.1007/s00339-018-1723-1} {10.1007/s00339-018-1723-1}
  (\bibinfo {year} {2018})\BibitemShut {NoStop}%
\bibitem [{\citenamefont {Hu}\ \emph {et~al.}(2015)\citenamefont {Hu},
  \citenamefont {Nan}, \citenamefont {Wang}, \citenamefont {Staruch},
  \citenamefont {Gao}, \citenamefont {Finkel},\ and\ \citenamefont
  {Sun}}]{Hu2015}%
  \BibitemOpen
  \bibfield  {author} {\bibinfo {author} {\bibfnamefont {Z.}~\bibnamefont
  {Hu}}, \bibinfo {author} {\bibfnamefont {T.}~\bibnamefont {Nan}}, \bibinfo
  {author} {\bibfnamefont {X.}~\bibnamefont {Wang}}, \bibinfo {author}
  {\bibfnamefont {M.}~\bibnamefont {Staruch}}, \bibinfo {author} {\bibfnamefont
  {Y.}~\bibnamefont {Gao}}, \bibinfo {author} {\bibfnamefont {P.}~\bibnamefont
  {Finkel}},\ and\ \bibinfo {author} {\bibfnamefont {N.~X.}\ \bibnamefont
  {Sun}},\ }\bibfield  {title} {\bibinfo {title} {Voltage control of magnetism
  in {{FeGaB}}/{{PIN}}-{{PMN}}-{{PT}} multiferroic heterostructures for
  high-power and high-temperature applications},\ }\href
  {https://doi.org/10.1063/1.4905855} {\bibfield  {journal} {\bibinfo
  {journal} {Applied Physics Letters}\ }\textbf {\bibinfo {volume} {106}},\
  \bibinfo {pages} {022901} (\bibinfo {year} {2015})}\BibitemShut {NoStop}%
\bibitem [{\citenamefont {Lou}\ \emph {et~al.}(2009)\citenamefont {Lou},
  \citenamefont {Liu}, \citenamefont {Reed}, \citenamefont {Ren},\ and\
  \citenamefont {Sun}}]{lou2009}%
  \BibitemOpen
  \bibfield  {author} {\bibinfo {author} {\bibfnamefont {J.}~\bibnamefont
  {Lou}}, \bibinfo {author} {\bibfnamefont {M.}~\bibnamefont {Liu}}, \bibinfo
  {author} {\bibfnamefont {D.}~\bibnamefont {Reed}}, \bibinfo {author}
  {\bibfnamefont {Y.}~\bibnamefont {Ren}},\ and\ \bibinfo {author}
  {\bibfnamefont {N.~X.}\ \bibnamefont {Sun}},\ }\bibfield  {title} {\bibinfo
  {title} {Giant {{Electric Field Tuning}} of {{Magnetism}} in {{Novel
  Multiferroic FeGaB}}/{{Lead Zinc Niobate}}-{{Lead Titanate}} ({{PZN}}-{{PT}})
  {{Heterostructures}}},\ }\href {https://doi.org/10.1002/adma.200901131}
  {\bibfield  {journal} {\bibinfo  {journal} {Advanced Materials}\ }\textbf
  {\bibinfo {volume} {21}},\ \bibinfo {pages} {4711} (\bibinfo {year}
  {2009})}\BibitemShut {NoStop}%
\bibitem [{\citenamefont {Zhang}\ \emph {et~al.}(2014)\citenamefont {Zhang},
  \citenamefont {Li}, \citenamefont {Wen}, \citenamefont {He}, \citenamefont
  {Yang}, \citenamefont {Wang}, \citenamefont {Yang},\ and\ \citenamefont
  {Lu}}]{Zhang2014-selfbiasCME}%
  \BibitemOpen
  \bibfield  {author} {\bibinfo {author} {\bibfnamefont {J.}~\bibnamefont
  {Zhang}}, \bibinfo {author} {\bibfnamefont {P.}~\bibnamefont {Li}}, \bibinfo
  {author} {\bibfnamefont {Y.}~\bibnamefont {Wen}}, \bibinfo {author}
  {\bibfnamefont {W.}~\bibnamefont {He}}, \bibinfo {author} {\bibfnamefont
  {A.}~\bibnamefont {Yang}}, \bibinfo {author} {\bibfnamefont {D.}~\bibnamefont
  {Wang}}, \bibinfo {author} {\bibfnamefont {C.}~\bibnamefont {Yang}},\ and\
  \bibinfo {author} {\bibfnamefont {C.}~\bibnamefont {Lu}},\ }\bibfield
  {title} {\bibinfo {title} {Giant self-biased converse magnetoelectric effect
  in multiferroic heterostructure with single-phase magnetostrictive
  materials},\ }\href {https://doi.org/10.1063/1.4900929} {\bibfield  {journal}
  {\bibinfo  {journal} {Appl. Phys. Lett.}\ }\textbf {\bibinfo {volume}
  {105}},\ \bibinfo {pages} {172408} (\bibinfo {year} {2014})}\BibitemShut
  {NoStop}%
\bibitem [{\citenamefont {Chul~Yang}\ \emph {et~al.}(2011)\citenamefont
  {Chul~Yang}, \citenamefont {Cho}, \citenamefont {Park},\ and\ \citenamefont
  {Priya}}]{ChulYang2011-selfbiasCME}%
  \BibitemOpen
  \bibfield  {author} {\bibinfo {author} {\bibfnamefont {S.}~\bibnamefont
  {Chul~Yang}}, \bibinfo {author} {\bibfnamefont {K.-H.}\ \bibnamefont {Cho}},
  \bibinfo {author} {\bibfnamefont {C.-S.}\ \bibnamefont {Park}},\ and\
  \bibinfo {author} {\bibfnamefont {S.}~\bibnamefont {Priya}},\ }\bibfield
  {title} {\bibinfo {title} {Self-biased converse magnetoelectric effect},\
  }\href {https://doi.org/10.1063/1.3662420} {\bibfield  {journal} {\bibinfo
  {journal} {Appl. Phys. Lett.}\ }\textbf {\bibinfo {volume} {99}},\ \bibinfo
  {pages} {202904} (\bibinfo {year} {2011})}\BibitemShut {NoStop}%
\bibitem [{\citenamefont {Fitchorov}\ \emph {et~al.}(2011)\citenamefont
  {Fitchorov}, \citenamefont {Chen}, \citenamefont {Jiang}, \citenamefont
  {Zhang}, \citenamefont {Zhao}, \citenamefont {Vittoria},\ and\ \citenamefont
  {Harris}}]{Fitchorov2011a}%
  \BibitemOpen
  \bibfield  {author} {\bibinfo {author} {\bibfnamefont {T.}~\bibnamefont
  {Fitchorov}}, \bibinfo {author} {\bibfnamefont {Y.}~\bibnamefont {Chen}},
  \bibinfo {author} {\bibfnamefont {L.}~\bibnamefont {Jiang}}, \bibinfo
  {author} {\bibfnamefont {G.}~\bibnamefont {Zhang}}, \bibinfo {author}
  {\bibfnamefont {Z.}~\bibnamefont {Zhao}}, \bibinfo {author} {\bibfnamefont
  {C.}~\bibnamefont {Vittoria}},\ and\ \bibinfo {author} {\bibfnamefont
  {V.~G.}\ \bibnamefont {Harris}},\ }\bibfield  {title} {\bibinfo {title}
  {Converse {{Magnetoelectric Effect}} in a {{Fe}}-{{Ga}}/{{PMN}}-{{PT
  Laminated Multiferroic Heterostructure}} for {{Field Generator
  Applications}}},\ }\href {https://doi.org/10.1109/TMAG.2011.2151255}
  {\bibfield  {journal} {\bibinfo  {journal} {IEEE Transactions on Magnetics}\
  }\textbf {\bibinfo {volume} {47}},\ \bibinfo {pages} {4050} (\bibinfo {year}
  {2011})}\BibitemShut {NoStop}%
\bibitem [{\citenamefont {Mandal}\ \emph {et~al.}(2011)\citenamefont {Mandal},
  \citenamefont {Sreenivasulu}, \citenamefont {Petrov},\ and\ \citenamefont
  {Srinivasan}}]{Mandal2011}%
  \BibitemOpen
  \bibfield  {author} {\bibinfo {author} {\bibfnamefont {S.~K.}\ \bibnamefont
  {Mandal}}, \bibinfo {author} {\bibfnamefont {G.}~\bibnamefont
  {Sreenivasulu}}, \bibinfo {author} {\bibfnamefont {V.~M.}\ \bibnamefont
  {Petrov}},\ and\ \bibinfo {author} {\bibfnamefont {G.}~\bibnamefont
  {Srinivasan}},\ }\bibfield  {title} {\bibinfo {title} {Magnetization-graded
  multiferroic composite and magnetoelectric effects at zero bias},\ }\href
  {https://doi.org/10.1103/PhysRevB.84.014432} {\bibfield  {journal} {\bibinfo
  {journal} {Phys. Rev. B}\ }\textbf {\bibinfo {volume} {84}},\ \bibinfo
  {pages} {014432} (\bibinfo {year} {2011})}\BibitemShut {NoStop}%
\bibitem [{\citenamefont {Zhou}\ \emph {et~al.}(2016)\citenamefont {Zhou},
  \citenamefont {Maurya}, \citenamefont {Yan}, \citenamefont {Srinivasan},
  \citenamefont {Quandt},\ and\ \citenamefont {Priya}}]{zhou2016}%
  \BibitemOpen
  \bibfield  {author} {\bibinfo {author} {\bibfnamefont {Y.}~\bibnamefont
  {Zhou}}, \bibinfo {author} {\bibfnamefont {D.}~\bibnamefont {Maurya}},
  \bibinfo {author} {\bibfnamefont {Y.}~\bibnamefont {Yan}}, \bibinfo {author}
  {\bibfnamefont {G.}~\bibnamefont {Srinivasan}}, \bibinfo {author}
  {\bibfnamefont {E.}~\bibnamefont {Quandt}},\ and\ \bibinfo {author}
  {\bibfnamefont {S.}~\bibnamefont {Priya}},\ }\bibfield  {title} {\bibinfo
  {title} {Self-{{Biased Magnetoelectric Composites}}: {{An Overview}} and
  {{Future Perspectives}}},\ }\href {https://doi.org/10.1515/ehs-2015-0003}
  {\bibfield  {journal} {\bibinfo  {journal} {Energy Harvesting and Systems}\
  }\textbf {\bibinfo {volume} {3}},\ \bibinfo {pages} {1} (\bibinfo {year}
  {2016})}\BibitemShut {NoStop}%
\bibitem [{\citenamefont {Fiebig}\ \emph {et~al.}(2016)\citenamefont {Fiebig},
  \citenamefont {Lottermoser}, \citenamefont {Meier},\ and\ \citenamefont
  {Trassin}}]{fiebig16}%
  \BibitemOpen
  \bibfield  {author} {\bibinfo {author} {\bibfnamefont {M.}~\bibnamefont
  {Fiebig}}, \bibinfo {author} {\bibfnamefont {T.}~\bibnamefont {Lottermoser}},
  \bibinfo {author} {\bibfnamefont {D.}~\bibnamefont {Meier}},\ and\ \bibinfo
  {author} {\bibfnamefont {M.}~\bibnamefont {Trassin}},\ }\bibfield  {title}
  {\bibinfo {title} {The evolution of multiferroics},\ }\href
  {10.1038/natrevmats.2016.46} {\bibfield  {journal} {\bibinfo  {journal}
  {Nature Reviews Materials}\ }\textbf {\bibinfo {volume} {1}},\ \bibinfo
  {pages} {16046} (\bibinfo {year} {2016})}\BibitemShut {NoStop}%
\bibitem [{cer()}]{ceracomp}%
  \BibitemOpen
  \href@noop {} {\bibinfo {title} {{Ceracomp Co. Ltd., Korea}}},\ \bibinfo
  {howpublished} {\url{http://www.ceracomp.com}}\BibitemShut {NoStop}%
\bibitem [{\citenamefont {Shanthi}\ \emph {et~al.}(2008)\citenamefont
  {Shanthi}, \citenamefont {Lim}, \citenamefont {Rajan},\ and\ \citenamefont
  {Jin}}]{Shanthi2008}%
  \BibitemOpen
  \bibfield  {author} {\bibinfo {author} {\bibfnamefont {M.}~\bibnamefont
  {Shanthi}}, \bibinfo {author} {\bibfnamefont {L.~C.}\ \bibnamefont {Lim}},
  \bibinfo {author} {\bibfnamefont {K.~K.}\ \bibnamefont {Rajan}},\ and\
  \bibinfo {author} {\bibfnamefont {J.}~\bibnamefont {Jin}},\ }\bibfield
  {title} {\bibinfo {title} {Complete sets of elastic, dielectric, and
  piezoelectric properties of flux-grown [011]-poled
  {{Pb}}({{Mg1}}$\slash${{3Nb2}}$\slash$3){{O3}}-(28\textendash{}32)\%
  {{PbTiO3}} single crystals},\ }\href {https://doi.org/10.1063/1.2907702}
  {\bibfield  {journal} {\bibinfo  {journal} {Applied Physics Letters}\
  }\textbf {\bibinfo {volume} {92}},\ \bibinfo {pages} {142906} (\bibinfo
  {year} {2008})}\BibitemShut {NoStop}%
\bibitem [{sch()}]{schottglass}%
  \BibitemOpen
  \href@noop {} {\bibinfo {title} {{Schott D 263 TM Glass}}},\ \bibinfo
  {howpublished} {\url
  {https://www.schott.com/nexterion/english/products/uncoated-substrates/d263.html?highlighted_text=d263}}\BibitemShut
  {NoStop}%
\bibitem [{Evi()}]{Evico}%
  \BibitemOpen
  \href@noop {} {\bibinfo {title} {{Evico Magnetics, Dresden}}},\ \bibinfo
  {howpublished} {\url{http://www.evico-magnetics.de}}\BibitemShut {NoStop}%
\bibitem [{\citenamefont {{Cryogenic Ltd., London, UK}}(2016)}]{CryogenicLtd}%
  \BibitemOpen
  \bibfield  {author} {\bibinfo {author} {\bibnamefont {{Cryogenic Ltd.,
  London, UK}}},\ }\href@noop {} {\bibinfo {title} {Cryogen free magnet system
  user manual}} (\bibinfo {year} {2016})\BibitemShut {NoStop}%
\bibitem [{\citenamefont {Zhao}\ \emph {et~al.}(2011)\citenamefont {Zhao},
  \citenamefont {Bao}, \citenamefont {Bur}, \citenamefont {Hockel},
  \citenamefont {Wong}, \citenamefont {Mohanchandra}, \citenamefont {Lynch},\
  and\ \citenamefont {Carman}}]{zhao2011orwu2011(2)}%
  \BibitemOpen
  \bibfield  {author} {\bibinfo {author} {\bibfnamefont {P.}~\bibnamefont
  {Zhao}}, \bibinfo {author} {\bibfnamefont {M.}~\bibnamefont {Bao}}, \bibinfo
  {author} {\bibfnamefont {A.}~\bibnamefont {Bur}}, \bibinfo {author}
  {\bibfnamefont {J.~L.}\ \bibnamefont {Hockel}}, \bibinfo {author}
  {\bibfnamefont {K.}~\bibnamefont {Wong}}, \bibinfo {author} {\bibfnamefont
  {K.~P.}\ \bibnamefont {Mohanchandra}}, \bibinfo {author} {\bibfnamefont
  {C.~S.}\ \bibnamefont {Lynch}},\ and\ \bibinfo {author} {\bibfnamefont
  {G.~P.}\ \bibnamefont {Carman}},\ }\bibfield  {title} {\bibinfo {title}
  {Domain engineered switchable strain states in ferroelectric (011)
  [{{Pb}}({{Mg}} {\textsubscript{1/3}} {{Nb}} {\textsubscript{2/3}} ){{O}}
  {\textsubscript{3}} ] {\textsubscript{(1-x)}} -[{{PbTiO}} {\textsubscript{3}}
  ] {\textsubscript{x}} ({{PMN}}-{{PT}}, x{$\approx$}0.32) single crystals},\
  }\href {https://doi.org/10.1063/1.3595670} {\bibfield  {journal} {\bibinfo
  {journal} {Journal of Applied Physics}\ }\textbf {\bibinfo {volume} {109}},\
  \bibinfo {pages} {124101} (\bibinfo {year} {2011})}\BibitemShut {NoStop}%
\bibitem [{\citenamefont {Huang}\ \emph {et~al.}(2015)\citenamefont {Huang},
  \citenamefont {Yang},\ and\ \citenamefont {Li}}]{Huang2015}%
  \BibitemOpen
  \bibfield  {author} {\bibinfo {author} {\bibfnamefont {W.}~\bibnamefont
  {Huang}}, \bibinfo {author} {\bibfnamefont {S.}~\bibnamefont {Yang}},\ and\
  \bibinfo {author} {\bibfnamefont {X.}~\bibnamefont {Li}},\ }\bibfield
  {title} {\bibinfo {title} {Multiferroic heterostructures and tunneling
  junctions},\ }\href {https://doi.org/10.1016/j.jmat.2015.08.002} {\bibfield
  {journal} {\bibinfo  {journal} {Journal of Materiomics}\ }\textbf {\bibinfo
  {volume} {1}},\ \bibinfo {pages} {263} (\bibinfo {year} {2015})}\BibitemShut
  {NoStop}%
\bibitem [{\citenamefont {Zhang}\ \emph {et~al.}(2012)\citenamefont {Zhang},
  \citenamefont {Zhao}, \citenamefont {Li}, \citenamefont {Yang}, \citenamefont
  {Rizwan}, \citenamefont {Zhang}, \citenamefont {Seidel}, \citenamefont {Qu},
  \citenamefont {Yang}, \citenamefont {Luo}, \citenamefont {He}, \citenamefont
  {Zou}, \citenamefont {Chen}, \citenamefont {Wang}, \citenamefont {Yang},
  \citenamefont {Sun}, \citenamefont {Wu}, \citenamefont {Xiao}, \citenamefont
  {Jin}, \citenamefont {Huang}, \citenamefont {Gao}, \citenamefont {Han},\ and\
  \citenamefont {Ramesh}}]{zhang2012}%
  \BibitemOpen
  \bibfield  {author} {\bibinfo {author} {\bibfnamefont {S.}~\bibnamefont
  {Zhang}}, \bibinfo {author} {\bibfnamefont {Y.~G.}\ \bibnamefont {Zhao}},
  \bibinfo {author} {\bibfnamefont {P.~S.}\ \bibnamefont {Li}}, \bibinfo
  {author} {\bibfnamefont {J.~J.}\ \bibnamefont {Yang}}, \bibinfo {author}
  {\bibfnamefont {S.}~\bibnamefont {Rizwan}}, \bibinfo {author} {\bibfnamefont
  {J.~X.}\ \bibnamefont {Zhang}}, \bibinfo {author} {\bibfnamefont
  {J.}~\bibnamefont {Seidel}}, \bibinfo {author} {\bibfnamefont {T.~L.}\
  \bibnamefont {Qu}}, \bibinfo {author} {\bibfnamefont {Y.~J.}\ \bibnamefont
  {Yang}}, \bibinfo {author} {\bibfnamefont {Z.~L.}\ \bibnamefont {Luo}},
  \bibinfo {author} {\bibfnamefont {Q.}~\bibnamefont {He}}, \bibinfo {author}
  {\bibfnamefont {T.}~\bibnamefont {Zou}}, \bibinfo {author} {\bibfnamefont
  {Q.~P.}\ \bibnamefont {Chen}}, \bibinfo {author} {\bibfnamefont {J.~W.}\
  \bibnamefont {Wang}}, \bibinfo {author} {\bibfnamefont {L.~F.}\ \bibnamefont
  {Yang}}, \bibinfo {author} {\bibfnamefont {Y.}~\bibnamefont {Sun}}, \bibinfo
  {author} {\bibfnamefont {Y.~Z.}\ \bibnamefont {Wu}}, \bibinfo {author}
  {\bibfnamefont {X.}~\bibnamefont {Xiao}}, \bibinfo {author} {\bibfnamefont
  {X.~F.}\ \bibnamefont {Jin}}, \bibinfo {author} {\bibfnamefont
  {J.}~\bibnamefont {Huang}}, \bibinfo {author} {\bibfnamefont
  {C.}~\bibnamefont {Gao}}, \bibinfo {author} {\bibfnamefont {X.~F.}\
  \bibnamefont {Han}},\ and\ \bibinfo {author} {\bibfnamefont {R.}~\bibnamefont
  {Ramesh}},\ }\bibfield  {title} {\bibinfo {title} {Electric-{{Field Control}}
  of {{Nonvolatile Magnetization}} in {{Co}} 40 {{Fe}} 40 {{B}} 20 / {{Pb}} (
  {{Mg}} 1 / 3 {{Nb}} 2 / 3 ) 0.7 {{Ti}} 0.3 {{O}} 3 {{Structure}} at {{Room
  Temperature}}},\ }\bibfield  {journal} {\bibinfo  {journal} {Physical Review
  Letters}\ }\textbf {\bibinfo {volume} {108}},\ \href
  {https://doi.org/10.1103/PhysRevLett.108.137203}
  {10.1103/PhysRevLett.108.137203} (\bibinfo {year} {2012})\BibitemShut
  {NoStop}%
\bibitem [{\citenamefont {Yang}\ \emph
  {et~al.}(2015{\natexlab{c}})\citenamefont {Yang}, \citenamefont {Zhao},
  \citenamefont {Zhang}, \citenamefont {Li}, \citenamefont {Gao}, \citenamefont
  {Yang}, \citenamefont {Huang}, \citenamefont {Miao}, \citenamefont {Liu},
  \citenamefont {Chen}, \citenamefont {Nan},\ and\ \citenamefont
  {Gao}}]{Yang2014}%
  \BibitemOpen
  \bibfield  {author} {\bibinfo {author} {\bibfnamefont {L.}~\bibnamefont
  {Yang}}, \bibinfo {author} {\bibfnamefont {Y.}~\bibnamefont {Zhao}}, \bibinfo
  {author} {\bibfnamefont {S.}~\bibnamefont {Zhang}}, \bibinfo {author}
  {\bibfnamefont {P.}~\bibnamefont {Li}}, \bibinfo {author} {\bibfnamefont
  {Y.}~\bibnamefont {Gao}}, \bibinfo {author} {\bibfnamefont {Y.}~\bibnamefont
  {Yang}}, \bibinfo {author} {\bibfnamefont {H.}~\bibnamefont {Huang}},
  \bibinfo {author} {\bibfnamefont {P.}~\bibnamefont {Miao}}, \bibinfo {author}
  {\bibfnamefont {Y.}~\bibnamefont {Liu}}, \bibinfo {author} {\bibfnamefont
  {A.}~\bibnamefont {Chen}}, \bibinfo {author} {\bibfnamefont {C.~W.}\
  \bibnamefont {Nan}},\ and\ \bibinfo {author} {\bibfnamefont {C.}~\bibnamefont
  {Gao}},\ }\bibfield  {title} {\bibinfo {title} {Bipolar loop-like
  non-volatile strain in the (001)-oriented
  {{Pb}}({{Mg1}}/{{3Nb2}}/3){{O3}}-{{PbTiO3}} single crystals},\ }\href
  {https://doi.org/10.1038/srep04591} {\bibfield  {journal} {\bibinfo
  {journal} {Scientific Reports}\ }\textbf {\bibinfo {volume} {4}},\ \bibinfo
  {pages} {4591} (\bibinfo {year} {2015}{\natexlab{c}})}\BibitemShut {NoStop}%
\bibitem [{\citenamefont {Wei}\ \emph {et~al.}(2016)\citenamefont {Wei},
  \citenamefont {Gao}, \citenamefont {Chen}, \citenamefont {Xi}, \citenamefont
  {Shao}, \citenamefont {Zhang}, \citenamefont {Chen},\ and\ \citenamefont
  {Li}}]{wei2016}%
  \BibitemOpen
  \bibfield  {author} {\bibinfo {author} {\bibfnamefont {Y.}~\bibnamefont
  {Wei}}, \bibinfo {author} {\bibfnamefont {C.}~\bibnamefont {Gao}}, \bibinfo
  {author} {\bibfnamefont {Z.}~\bibnamefont {Chen}}, \bibinfo {author}
  {\bibfnamefont {S.}~\bibnamefont {Xi}}, \bibinfo {author} {\bibfnamefont
  {W.}~\bibnamefont {Shao}}, \bibinfo {author} {\bibfnamefont {P.}~\bibnamefont
  {Zhang}}, \bibinfo {author} {\bibfnamefont {G.}~\bibnamefont {Chen}},\ and\
  \bibinfo {author} {\bibfnamefont {J.}~\bibnamefont {Li}},\ }\bibfield
  {title} {\bibinfo {title} {Four-state memory based on a giant and
  non-volatile converse magnetoelectric effect in
  {{FeAl}}/{{PIN}}-{{PMN}}-{{PT}} structure},\ }\bibfield  {journal} {\bibinfo
  {journal} {Scientific Reports}\ }\textbf {\bibinfo {volume} {6}},\ \href
  {https://doi.org/10.1038/srep30002} {10.1038/srep30002} (\bibinfo {year}
  {2016})\BibitemShut {NoStop}%
\bibitem [{\citenamefont {Wang}\ \emph {et~al.}(2019)\citenamefont {Wang},
  \citenamefont {Pesquera}, \citenamefont {Mansell}, \citenamefont {{van
  Dijken}}, \citenamefont {Cowburn}, \citenamefont {Ghidini},\ and\
  \citenamefont {Mathur}}]{Wang2019}%
  \BibitemOpen
  \bibfield  {author} {\bibinfo {author} {\bibfnamefont {J.}~\bibnamefont
  {Wang}}, \bibinfo {author} {\bibfnamefont {D.}~\bibnamefont {Pesquera}},
  \bibinfo {author} {\bibfnamefont {R.}~\bibnamefont {Mansell}}, \bibinfo
  {author} {\bibfnamefont {S.}~\bibnamefont {{van Dijken}}}, \bibinfo {author}
  {\bibfnamefont {R.~P.}\ \bibnamefont {Cowburn}}, \bibinfo {author}
  {\bibfnamefont {M.}~\bibnamefont {Ghidini}},\ and\ \bibinfo {author}
  {\bibfnamefont {N.~D.}\ \bibnamefont {Mathur}},\ }\bibfield  {title}
  {\bibinfo {title} {Giant non-volatile magnetoelectric effects via growth
  anisotropy in {{Co}} {\textsubscript{40}} {{Fe}} {\textsubscript{40}} {{B}}
  {\textsubscript{20}} films on {{PMN}}-{{PT}} substrates},\ }\href
  {https://doi.org/10.1063/1.5078787} {\bibfield  {journal} {\bibinfo
  {journal} {Appl. Phys. Lett.}\ }\textbf {\bibinfo {volume} {114}},\ \bibinfo
  {pages} {092401} (\bibinfo {year} {2019})}\BibitemShut {NoStop}%
\bibitem [{\citenamefont {Jiang}\ \emph {et~al.}(2015)\citenamefont {Jiang},
  \citenamefont {Zhang}, \citenamefont {Dong}, \citenamefont {Guo},\ and\
  \citenamefont {Xue}}]{Jiang2015}%
  \BibitemOpen
  \bibfield  {author} {\bibinfo {author} {\bibfnamefont {C.}~\bibnamefont
  {Jiang}}, \bibinfo {author} {\bibfnamefont {C.}~\bibnamefont {Zhang}},
  \bibinfo {author} {\bibfnamefont {C.}~\bibnamefont {Dong}}, \bibinfo {author}
  {\bibfnamefont {D.}~\bibnamefont {Guo}},\ and\ \bibinfo {author}
  {\bibfnamefont {D.}~\bibnamefont {Xue}},\ }\bibfield  {title} {\bibinfo
  {title} {Electric field tuning of non-volatile three-state magnetoelectric
  memory in {{FeCo}}-{{NiFe}} {\textsubscript{2}} {{O}} {\textsubscript{4}}
  /{{Pb}}({{Mg}} {\textsubscript{1/3}} {{Nb}} {\textsubscript{2/3}} )
  {\textsubscript{0.7}} {{Ti}} {\textsubscript{0.3}} {{O}} {\textsubscript{3}}
  heterostructures},\ }\href {https://doi.org/10.1063/1.4916569} {\bibfield
  {journal} {\bibinfo  {journal} {Applied Physics Letters}\ }\textbf {\bibinfo
  {volume} {106}},\ \bibinfo {pages} {122406} (\bibinfo {year}
  {2015})}\BibitemShut {NoStop}%
\bibitem [{\citenamefont {Damjanovic}(2006)}]{Damjanovic2006-book}%
  \BibitemOpen
  \bibfield  {author} {\bibinfo {author} {\bibfnamefont {D.}~\bibnamefont
  {Damjanovic}},\ }\bibfield  {title} {\bibinfo {title} {Hysteresis in
  {{Piezoelectric}} and {{Ferroelectric Materials}}},\ }in\ \href
  {https://doi.org/10.1016/B978-012480874-4/50022-1} {\emph {\bibinfo
  {booktitle} {The {{Science}} of {{Hysteresis}}}}}\ (\bibinfo  {publisher}
  {{Elsevier}},\ \bibinfo {year} {2006})\ pp.\ \bibinfo {pages}
  {337--465}\BibitemShut {NoStop}%
\bibitem [{\citenamefont {Guo}\ \emph {et~al.}(2016)\citenamefont {Guo},
  \citenamefont {Han}, \citenamefont {Zuo}, \citenamefont {Zhang},
  \citenamefont {Li}, \citenamefont {Cui}, \citenamefont {Wu}, \citenamefont
  {Yun}, \citenamefont {Wang}, \citenamefont {Peng},\ and\ \citenamefont
  {Xi}}]{Guo2016}%
  \BibitemOpen
  \bibfield  {author} {\bibinfo {author} {\bibfnamefont {X.}~\bibnamefont
  {Guo}}, \bibinfo {author} {\bibfnamefont {X.}~\bibnamefont {Han}}, \bibinfo
  {author} {\bibfnamefont {Y.}~\bibnamefont {Zuo}}, \bibinfo {author}
  {\bibfnamefont {J.}~\bibnamefont {Zhang}}, \bibinfo {author} {\bibfnamefont
  {D.}~\bibnamefont {Li}}, \bibinfo {author} {\bibfnamefont {B.}~\bibnamefont
  {Cui}}, \bibinfo {author} {\bibfnamefont {K.}~\bibnamefont {Wu}}, \bibinfo
  {author} {\bibfnamefont {J.}~\bibnamefont {Yun}}, \bibinfo {author}
  {\bibfnamefont {T.}~\bibnamefont {Wang}}, \bibinfo {author} {\bibfnamefont
  {Y.}~\bibnamefont {Peng}},\ and\ \bibinfo {author} {\bibfnamefont
  {L.}~\bibnamefont {Xi}},\ }\bibfield  {title} {\bibinfo {title} {Electric
  field induced magnetic anisotropy transition from fourfold to twofold
  symmetry in (001) 0.{{68Pb}}({{Mg}} {\textsubscript{1/3}} {{Nb}}
  {\textsubscript{2/3}} ){{O}} {\textsubscript{3}} -0.{{32PbTiO}}
  {\textsubscript{3}} /{{Fe}} {\textsubscript{0.86}} {{Si}}
  {\textsubscript{0.14}} epitaxial heterostructures},\ }\href
  {https://doi.org/10.1063/1.4945983} {\bibfield  {journal} {\bibinfo
  {journal} {Applied Physics Letters}\ }\textbf {\bibinfo {volume} {108}},\
  \bibinfo {pages} {152401} (\bibinfo {year} {2016})}\BibitemShut {NoStop}%
\bibitem [{\citenamefont {Chen}\ \emph {et~al.}(2009)\citenamefont {Chen},
  \citenamefont {Geiler}, \citenamefont {Fitchorov}, \citenamefont {Vittoria},\
  and\ \citenamefont {Harris}}]{chen2009}%
  \BibitemOpen
  \bibfield  {author} {\bibinfo {author} {\bibfnamefont {Y.}~\bibnamefont
  {Chen}}, \bibinfo {author} {\bibfnamefont {A.~L.}\ \bibnamefont {Geiler}},
  \bibinfo {author} {\bibfnamefont {T.}~\bibnamefont {Fitchorov}}, \bibinfo
  {author} {\bibfnamefont {C.}~\bibnamefont {Vittoria}},\ and\ \bibinfo
  {author} {\bibfnamefont {V.~G.}\ \bibnamefont {Harris}},\ }\bibfield  {title}
  {\bibinfo {title} {Time domain analyses of the converse magnetoelectric
  effect in a multiferroic metallic glass-relaxor ferroelectric
  heterostructure},\ }\href {https://doi.org/10.1063/1.3258023} {\bibfield
  {journal} {\bibinfo  {journal} {Applied Physics Letters}\ }\textbf {\bibinfo
  {volume} {95}},\ \bibinfo {pages} {182501} (\bibinfo {year}
  {2009})}\BibitemShut {NoStop}%
\bibitem [{\citenamefont {Lupascu}\ and\ \citenamefont
  {Verdier}(2004)}]{Lupascu2004}%
  \BibitemOpen
  \bibfield  {author} {\bibinfo {author} {\bibfnamefont {D.~C.}\ \bibnamefont
  {Lupascu}}\ and\ \bibinfo {author} {\bibfnamefont {C.}~\bibnamefont
  {Verdier}},\ }\bibfield  {title} {\bibinfo {title} {Fatigue anisotropy in
  lead-zirconate-titanate},\ }\href
  {https://doi.org/https://doi.org/10.1016/S0955-2219(03)00572-7} {\bibfield
  {journal} {\bibinfo  {journal} {Journal of the European Ceramic Society}\
  }\textbf {\bibinfo {volume} {24}},\ \bibinfo {pages} {1663} (\bibinfo {year}
  {2004})},\ \bibinfo {note} {electroceramics VIII}\BibitemShut {NoStop}%
\bibitem [{\citenamefont {Gopalan}\ and\ \citenamefont
  {Gupta}(1996)}]{gopalan1996}%
  \BibitemOpen
  \bibfield  {author} {\bibinfo {author} {\bibfnamefont {V.}~\bibnamefont
  {Gopalan}}\ and\ \bibinfo {author} {\bibfnamefont {M.~C.}\ \bibnamefont
  {Gupta}},\ }\bibfield  {title} {\bibinfo {title} {Observation of internal
  field in {{LiTaO}} {\textsubscript{3}} single crystals: {{Its}} origin and
  time-temperature dependence},\ }\href {https://doi.org/10.1063/1.116220}
  {\bibfield  {journal} {\bibinfo  {journal} {Applied Physics Letters}\
  }\textbf {\bibinfo {volume} {68}},\ \bibinfo {pages} {888} (\bibinfo {year}
  {1996})}\BibitemShut {NoStop}%
\bibitem [{\citenamefont {Noguchi}\ \emph {et~al.}(2000)\citenamefont
  {Noguchi}, \citenamefont {Miwa}, \citenamefont {Goshima},\ and\ \citenamefont
  {Miyayama}}]{noguchi2000}%
  \BibitemOpen
  \bibfield  {author} {\bibinfo {author} {\bibfnamefont {Y.}~\bibnamefont
  {Noguchi}}, \bibinfo {author} {\bibfnamefont {I.}~\bibnamefont {Miwa}},
  \bibinfo {author} {\bibfnamefont {Y.}~\bibnamefont {Goshima}},\ and\ \bibinfo
  {author} {\bibfnamefont {M.}~\bibnamefont {Miyayama}},\ }\bibfield  {title}
  {\bibinfo {title} {Defect {{Control}} for {{Large Remanent Polarization}} in
  {{Bismuth Titanate Ferroelectrics}} -- {{Doping Effect}} of
  {{Higher}}-{{Valent Cations}} --},\ }\href
  {https://doi.org/10.1143/JJAP.39.L1259} {\bibfield  {journal} {\bibinfo
  {journal} {Japanese Journal of Applied Physics}\ }\textbf {\bibinfo {volume}
  {39}},\ \bibinfo {pages} {L1259} (\bibinfo {year} {2000})}\BibitemShut
  {NoStop}%
\bibitem [{\citenamefont {Cullen}\ \emph {et~al.}(2007)\citenamefont {Cullen},
  \citenamefont {Zhao},\ and\ \citenamefont {Wuttig}}]{Cullen2007}%
  \BibitemOpen
  \bibfield  {author} {\bibinfo {author} {\bibfnamefont {J.}~\bibnamefont
  {Cullen}}, \bibinfo {author} {\bibfnamefont {P.}~\bibnamefont {Zhao}},\ and\
  \bibinfo {author} {\bibfnamefont {M.}~\bibnamefont {Wuttig}},\ }\bibfield
  {title} {\bibinfo {title} {Anisotropy of crystalline ferromagnets with
  defects},\ }\href {https://doi.org/10.1063/1.2749471} {\bibfield  {journal}
  {\bibinfo  {journal} {Journal of Applied Physics}\ }\textbf {\bibinfo
  {volume} {101}},\ \bibinfo {pages} {123922} (\bibinfo {year}
  {2007})}\BibitemShut {NoStop}%
\bibitem [{\citenamefont {Begu{\'e}}\ \emph {et~al.}(2019)\citenamefont
  {Begu{\'e}}, \citenamefont {Proietti}, \citenamefont {Arnaudas},\ and\
  \citenamefont {Ciria}}]{Begue2019}%
  \BibitemOpen
  \bibfield  {author} {\bibinfo {author} {\bibfnamefont {A.}~\bibnamefont
  {Begu{\'e}}}, \bibinfo {author} {\bibfnamefont {M.~G.}\ \bibnamefont
  {Proietti}}, \bibinfo {author} {\bibfnamefont {J.~I.}\ \bibnamefont
  {Arnaudas}},\ and\ \bibinfo {author} {\bibfnamefont {M.}~\bibnamefont
  {Ciria}},\ }\bibfield  {title} {\bibinfo {title} {Magnetic ripple domain
  structure in {{FeGa}}/{{MgO}} thin films},\ }\href@noop {} {\bibfield
  {journal} {\bibinfo  {journal} {arXiv:1905.09180 [cond-mat]}\ } (\bibinfo
  {year} {2019})},\ \Eprint {https://arxiv.org/abs/1905.09180}
  {arXiv:1905.09180 [cond-mat]} \BibitemShut {NoStop}%
\bibitem [{\citenamefont {Cherifi}\ \emph {et~al.}(2014)\citenamefont
  {Cherifi}, \citenamefont {Ivanovskaya}, \citenamefont {Phillips},
  \citenamefont {Zobelli}, \citenamefont {Infante}, \citenamefont {Jacquet},
  \citenamefont {Garcia}, \citenamefont {Fusil}, \citenamefont {Briddon},
  \citenamefont {Guiblin}, \citenamefont {Mougin}, \citenamefont {\"Unal},
  \citenamefont {Kronast}, \citenamefont {Valencia}, \citenamefont {Dkhil},
  \citenamefont {Barth\'el\'emy},\ and\ \citenamefont {Bibes}}]{cherifi2014}%
  \BibitemOpen
  \bibfield  {author} {\bibinfo {author} {\bibfnamefont {R.~O.}\ \bibnamefont
  {Cherifi}}, \bibinfo {author} {\bibfnamefont {V.}~\bibnamefont
  {Ivanovskaya}}, \bibinfo {author} {\bibfnamefont {L.~C.}\ \bibnamefont
  {Phillips}}, \bibinfo {author} {\bibfnamefont {A.}~\bibnamefont {Zobelli}},
  \bibinfo {author} {\bibfnamefont {I.~C.}\ \bibnamefont {Infante}}, \bibinfo
  {author} {\bibfnamefont {E.}~\bibnamefont {Jacquet}}, \bibinfo {author}
  {\bibfnamefont {V.}~\bibnamefont {Garcia}}, \bibinfo {author} {\bibfnamefont
  {S.}~\bibnamefont {Fusil}}, \bibinfo {author} {\bibfnamefont {P.~R.}\
  \bibnamefont {Briddon}}, \bibinfo {author} {\bibfnamefont {N.}~\bibnamefont
  {Guiblin}}, \bibinfo {author} {\bibfnamefont {A.}~\bibnamefont {Mougin}},
  \bibinfo {author} {\bibfnamefont {A.~A.}\ \bibnamefont {\"Unal}}, \bibinfo
  {author} {\bibfnamefont {F.}~\bibnamefont {Kronast}}, \bibinfo {author}
  {\bibfnamefont {S.}~\bibnamefont {Valencia}}, \bibinfo {author}
  {\bibfnamefont {B.}~\bibnamefont {Dkhil}}, \bibinfo {author} {\bibfnamefont
  {A.}~\bibnamefont {Barth\'el\'emy}},\ and\ \bibinfo {author} {\bibfnamefont
  {M.}~\bibnamefont {Bibes}},\ }\bibfield  {title} {\bibinfo {title}
  {Electric-field control of magnetic order above room temperature},\ }\href
  {https://doi.org/10.1038/nmat3870} {\bibfield  {journal} {\bibinfo  {journal}
  {Nature Materials}\ }\textbf {\bibinfo {volume} {13}},\ \bibinfo {pages}
  {345} (\bibinfo {year} {2014})}\BibitemShut {NoStop}%
\bibitem [{\citenamefont {Heron}\ \emph {et~al.}(2014)\citenamefont {Heron},
  \citenamefont {Bosse}, \citenamefont {He}, \citenamefont {Gao}, \citenamefont
  {Trassin}, \citenamefont {Ye}, \citenamefont {Clarkson}, \citenamefont
  {Wang}, \citenamefont {Liu}, \citenamefont {Salahuddin}, \citenamefont
  {Ralph}, \citenamefont {Schlom}, \citenamefont {\'I\~niguez}, \citenamefont
  {Huey},\ and\ \citenamefont {Ramesh}}]{Heron2014}%
  \BibitemOpen
  \bibfield  {author} {\bibinfo {author} {\bibfnamefont {J.~T.}\ \bibnamefont
  {Heron}}, \bibinfo {author} {\bibfnamefont {J.~L.}\ \bibnamefont {Bosse}},
  \bibinfo {author} {\bibfnamefont {Q.}~\bibnamefont {He}}, \bibinfo {author}
  {\bibfnamefont {Y.}~\bibnamefont {Gao}}, \bibinfo {author} {\bibfnamefont
  {M.}~\bibnamefont {Trassin}}, \bibinfo {author} {\bibfnamefont
  {L.}~\bibnamefont {Ye}}, \bibinfo {author} {\bibfnamefont {J.~D.}\
  \bibnamefont {Clarkson}}, \bibinfo {author} {\bibfnamefont {C.}~\bibnamefont
  {Wang}}, \bibinfo {author} {\bibfnamefont {J.}~\bibnamefont {Liu}}, \bibinfo
  {author} {\bibfnamefont {S.}~\bibnamefont {Salahuddin}}, \bibinfo {author}
  {\bibfnamefont {D.~C.}\ \bibnamefont {Ralph}}, \bibinfo {author}
  {\bibfnamefont {D.~G.}\ \bibnamefont {Schlom}}, \bibinfo {author}
  {\bibfnamefont {J.}~\bibnamefont {\'I\~niguez}}, \bibinfo {author}
  {\bibfnamefont {B.~D.}\ \bibnamefont {Huey}},\ and\ \bibinfo {author}
  {\bibfnamefont {R.}~\bibnamefont {Ramesh}},\ }\bibfield  {title} {\bibinfo
  {title} {Deterministic switching of ferromagnetism at room temperature using
  an electric field},\ }\href {https://doi.org/10.1038/nature14004} {\bibfield
  {journal} {\bibinfo  {journal} {Nature}\ }\textbf {\bibinfo {volume} {516}},\
  \bibinfo {pages} {370} (\bibinfo {year} {2014})}\BibitemShut {NoStop}%
\bibitem [{\citenamefont {Wu}\ \emph {et~al.}(2016)\citenamefont {Wu},
  \citenamefont {Zhang},\ and\ \citenamefont {Zhang}}]{wu2016}%
  \BibitemOpen
  \bibfield  {author} {\bibinfo {author} {\bibfnamefont {G.}~\bibnamefont
  {Wu}}, \bibinfo {author} {\bibfnamefont {R.}~\bibnamefont {Zhang}},\ and\
  \bibinfo {author} {\bibfnamefont {N.}~\bibnamefont {Zhang}},\ }\bibfield
  {title} {\bibinfo {title} {Enhanced converse magnetoelectric effect in
  cylindrical piezoelectric-magnetostrictive composites},\ }\href
  {https://doi.org/10.1051/epjap/2016150607} {\bibfield  {journal} {\bibinfo
  {journal} {The European Physical Journal Applied Physics}\ }\textbf {\bibinfo
  {volume} {76}},\ \bibinfo {pages} {10602} (\bibinfo {year}
  {2016})}\BibitemShut {NoStop}%
\bibitem [{\citenamefont {Han}\ \emph {et~al.}(2018)\citenamefont {Han},
  \citenamefont {Zhang},\ and\ \citenamefont {Gao}}]{Han2018-temperature}%
  \BibitemOpen
  \bibfield  {author} {\bibinfo {author} {\bibfnamefont {J.}~\bibnamefont
  {Han}}, \bibinfo {author} {\bibfnamefont {J.}~\bibnamefont {Zhang}},\ and\
  \bibinfo {author} {\bibfnamefont {Y.}~\bibnamefont {Gao}},\ }\bibfield
  {title} {\bibinfo {title} {A nonlinear magneto-mechanical-thermal-electric
  coupling model of {{Terfenol}}-{{D}}/{{PZT}}/{{Terfenol}}-{{D}} and
  {{Ni}}/{{PZT}}/{{Ni}} laminates},\ }\href
  {https://doi.org/10.1016/j.jmmm.2018.06.079} {\ \textbf {\bibinfo {volume}
  {466}},\ \bibinfo {pages} {200} (\bibinfo {year} {2018})}\BibitemShut
  {NoStop}%
\bibitem [{\citenamefont {Burdin}\ \emph {et~al.}(2012)\citenamefont {Burdin},
  \citenamefont {Fetisov}, \citenamefont {Chashin},\ and\ \citenamefont
  {Ekonomov}}]{Burdin2012-temperature}%
  \BibitemOpen
  \bibfield  {author} {\bibinfo {author} {\bibfnamefont {D.~A.}\ \bibnamefont
  {Burdin}}, \bibinfo {author} {\bibfnamefont {Y.~K.}\ \bibnamefont {Fetisov}},
  \bibinfo {author} {\bibfnamefont {D.~V.}\ \bibnamefont {Chashin}},\ and\
  \bibinfo {author} {\bibfnamefont {N.~A.}\ \bibnamefont {Ekonomov}},\
  }\bibfield  {title} {\bibinfo {title} {Temperature dependence of the
  characteristics of the resonant magnetoelectric effect in a lead magnesium
  niobate-lead titanate/nickel structure},\ }\href
  {https://doi.org/10.1134/S1063785012070164} {\bibfield  {journal} {\bibinfo
  {journal} {Tech. Phys. Lett.}\ }\textbf {\bibinfo {volume} {38}},\ \bibinfo
  {pages} {661} (\bibinfo {year} {2012})}\BibitemShut {NoStop}%
\bibitem [{\citenamefont {Zhang}\ \emph {et~al.}(2016)\citenamefont {Zhang},
  \citenamefont {Zhan}, \citenamefont {Rong}, \citenamefont {Li}, \citenamefont
  {Zuo}, \citenamefont {Liu}, \citenamefont {Wang},\ and\ \citenamefont
  {Li}}]{zhang2016_thermal_substrate_EB}%
  \BibitemOpen
  \bibfield  {author} {\bibinfo {author} {\bibfnamefont {Y.}~\bibnamefont
  {Zhang}}, \bibinfo {author} {\bibfnamefont {Q.}~\bibnamefont {Zhan}},
  \bibinfo {author} {\bibfnamefont {X.}~\bibnamefont {Rong}}, \bibinfo {author}
  {\bibfnamefont {H.}~\bibnamefont {Li}}, \bibinfo {author} {\bibfnamefont
  {Z.}~\bibnamefont {Zuo}}, \bibinfo {author} {\bibfnamefont {Y.}~\bibnamefont
  {Liu}}, \bibinfo {author} {\bibfnamefont {B.}~\bibnamefont {Wang}},\ and\
  \bibinfo {author} {\bibfnamefont {R.-W.}\ \bibnamefont {Li}},\ }\bibfield
  {title} {\bibinfo {title} {Influence of {{Thermal Deformation}} on {{Exchange
  Bias}} in {{FeGa}}/{{IrMn Bilayers Grown}} on {{Flexible Polyvinylidene
  Fluoride Membranes}}},\ }\href {https://doi.org/10.1109/TMAG.2016.2529722}
  {\bibfield  {journal} {\bibinfo  {journal} {IEEE Transactions on Magnetics}\
  }\textbf {\bibinfo {volume} {52}},\ \bibinfo {pages} {1} (\bibinfo {year}
  {2016})}\BibitemShut {NoStop}%
\bibitem [{\citenamefont {Liu}\ \emph {et~al.}(2015{\natexlab{a}})\citenamefont
  {Liu}, \citenamefont {Wang}, \citenamefont {Zhan}, \citenamefont {Tang},
  \citenamefont {Yang}, \citenamefont {Liu}, \citenamefont {Zuo}, \citenamefont
  {Zhang}, \citenamefont {Xie}, \citenamefont {Zhu}, \citenamefont {Chen},
  \citenamefont {Wang},\ and\ \citenamefont {Li}}]{liu2014-thermal}%
  \BibitemOpen
  \bibfield  {author} {\bibinfo {author} {\bibfnamefont {Y.}~\bibnamefont
  {Liu}}, \bibinfo {author} {\bibfnamefont {B.}~\bibnamefont {Wang}}, \bibinfo
  {author} {\bibfnamefont {Q.}~\bibnamefont {Zhan}}, \bibinfo {author}
  {\bibfnamefont {Z.}~\bibnamefont {Tang}}, \bibinfo {author} {\bibfnamefont
  {H.}~\bibnamefont {Yang}}, \bibinfo {author} {\bibfnamefont {G.}~\bibnamefont
  {Liu}}, \bibinfo {author} {\bibfnamefont {Z.}~\bibnamefont {Zuo}}, \bibinfo
  {author} {\bibfnamefont {X.}~\bibnamefont {Zhang}}, \bibinfo {author}
  {\bibfnamefont {Y.}~\bibnamefont {Xie}}, \bibinfo {author} {\bibfnamefont
  {X.}~\bibnamefont {Zhu}}, \bibinfo {author} {\bibfnamefont {B.}~\bibnamefont
  {Chen}}, \bibinfo {author} {\bibfnamefont {J.}~\bibnamefont {Wang}},\ and\
  \bibinfo {author} {\bibfnamefont {R.-W.}\ \bibnamefont {Li}},\ }\bibfield
  {title} {\bibinfo {title} {Positive temperature coefficient of magnetic
  anisotropy in polyvinylidene fluoride ({{PVDF}})-based magnetic composites},\
  }\href {https://doi.org/10.1038/srep06615} {\bibfield  {journal} {\bibinfo
  {journal} {Sci Rep}\ }\textbf {\bibinfo {volume} {4}},\ \bibinfo {pages}
  {6615} (\bibinfo {year} {2015}{\natexlab{a}})}\BibitemShut {NoStop}%
\bibitem [{\citenamefont {Liu}\ \emph {et~al.}(2015{\natexlab{b}})\citenamefont
  {Liu}, \citenamefont {Zhan}, \citenamefont {Dai}, \citenamefont {Zhang},
  \citenamefont {Wang}, \citenamefont {Liu}, \citenamefont {Zuo}, \citenamefont
  {Rong}, \citenamefont {Yang}, \citenamefont {Zhu}, \citenamefont {Xie},
  \citenamefont {Chen},\ and\ \citenamefont {Li}}]{Liu2015_thermal}%
  \BibitemOpen
  \bibfield  {author} {\bibinfo {author} {\bibfnamefont {Y.}~\bibnamefont
  {Liu}}, \bibinfo {author} {\bibfnamefont {Q.}~\bibnamefont {Zhan}}, \bibinfo
  {author} {\bibfnamefont {G.}~\bibnamefont {Dai}}, \bibinfo {author}
  {\bibfnamefont {X.}~\bibnamefont {Zhang}}, \bibinfo {author} {\bibfnamefont
  {B.}~\bibnamefont {Wang}}, \bibinfo {author} {\bibfnamefont {G.}~\bibnamefont
  {Liu}}, \bibinfo {author} {\bibfnamefont {Z.}~\bibnamefont {Zuo}}, \bibinfo
  {author} {\bibfnamefont {X.}~\bibnamefont {Rong}}, \bibinfo {author}
  {\bibfnamefont {H.}~\bibnamefont {Yang}}, \bibinfo {author} {\bibfnamefont
  {X.}~\bibnamefont {Zhu}}, \bibinfo {author} {\bibfnamefont {Y.}~\bibnamefont
  {Xie}}, \bibinfo {author} {\bibfnamefont {B.}~\bibnamefont {Chen}},\ and\
  \bibinfo {author} {\bibfnamefont {R.-W.}\ \bibnamefont {Li}},\ }\bibfield
  {title} {\bibinfo {title} {Thermally assisted electric field control of
  magnetism in flexible multiferroic heterostructures},\ }\bibfield  {journal}
  {\bibinfo  {journal} {Scientific Reports}\ }\textbf {\bibinfo {volume} {4}},\
  \href {https://doi.org/10.1038/srep06925} {10.1038/srep06925} (\bibinfo
  {year} {2015}{\natexlab{b}})\BibitemShut {NoStop}%
\end{thebibliography}%


%
\end{document}